\documentclass[10ps]{article}
\usepackage{./packages/isabelle,./packages/isabellesym}
\usepackage{amsmath} 
\usepackage{amssymb} 
\usepackage{amsfonts} 
\usepackage{mathtools} 
\usepackage{mathrsfs} 
\usepackage{stmaryrd} 
\usepackage{mathpartir} 
\usepackage{graphicx} 
\usepackage{scalerel} 
\usepackage{enumerate} 
\usepackage{color}
\usepackage{appendix}
\usepackage[colorinlistoftodos]{todonotes} 
\usepackage{xparse} 
\usepackage{fullpage}

\makeatletter 
\NewDocumentCommand{\tagx}{om}{%
  \IfNoValueTF{#1}
   {
    \refstepcounter{equation}(\theequation)\label{#2}%
   }
   {
    (#1)\def\@currentlabel{#1}\label{#2}%
   }%
}

\newcommand*{\inlineequation}[2][]{%
  \begingroup
    \refstepcounter{equation}%
    \ifx\\#1\\%
    \else
      \label{#1}%
    \fi
    \relpenalty=10000 %
    \binoppenalty=10000 %
    \ensuremath{ 
      #2%
    }%
    ~\@eqnnum
  \endgroup
}
\makeatother

\usepackage[colorlinks=true]{hyperref} 

\hypersetup{
    colorlinks,
    linkcolor={red!50!black},
    citecolor={blue!50!black},
    urlcolor={blue!80!black}
  }
\graphicspath{ {./images/} }

\newcommand{\isalogo}{\includegraphics[width=9pt]{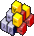}}
\newcommand{\isalink}[1]{\href{#1}{\isalogo}}
\definecolor{jjcolor}{cmyk}{0.36,0.44,0,0.07}

\newcommand{\leftAlignGather}[1]{\mathmakebox[0pt]{\mathmakebox[\textwidth][l]{#1}}}

\newcommand{\nats}{\mathbb{N}}

\DeclareMathOperator{\previously}{\mathrm{Y}}
\DeclareMathOperator{\nextly}{\mathrm{X}}
\newcommand{\since}{\mathbin{\mathrm{S}}}
\newcommand{\until}{\mathbin{\mathrm{U}}}
\newcommand{\trigger}{\mathbin{\mathrm{T}}}
\newcommand{\release}{\mathbin{\mathrm{R}}}
\newcommand{\pastly}{\mathop{\mathrm{P}}}
\newcommand{\eventually}{\mathop{\mathrm{F}}}
\newcommand{\historically}{\mathop{\mathrm{H}}}
\newcommand{\globally}{\mathop{\mathrm{G}}}
\newcommand{\sats}[1]{\llbracket#1\rrbracket}
\DeclareMathOperator{\fv}{fv}
\DeclareMathOperator*{\bigjoin}{\scalerel*{\bowtie}{\sum}}
\newcommand{\join}{\mathbin{\bowtie}}
\newcommand{\antijoin}{\mathbin{\scaleobj{1.2}{\triangleright}}}
\newcommand{\unit}[1]{\langle\rangle_{#1}}
\newcommand{\unitT}[1]{\mathbf{1}_{#1}}

\newcommand{\cons}{\mathbin{\texttt{\#}}}
\newcommand{\None}{\mathit{None}}
\DeclareMathOperator{\map}{map}
\DeclareMathOperator{\length}{length}
\newcommand{\teval}{\mathbin{!_{t}}}
\newcommand{\pred}{\mathbin{\dagger}}
\DeclareMathOperator{\var}{\mathbf{v}}
\DeclareMathOperator{\const}{\mathbf{c}}
\DeclareMathOperator{\wftuple}{\mathit{wf-tuple}}
\DeclareMathOperator{\vtable}{\mathit{table}}
\DeclareMathOperator{\qtable}{\mathit{qtable}}
\DeclareMathOperator{\issafe}{\mathit{is-safe}}
\DeclareMathOperator{\wfmformula}{\mathit{wf-mformula}}
\DeclareMathOperator{\nfv}{nfv}
\DeclareMathOperator{\dfv}{dfv}
\DeclareMathOperator{\ssfv}{ssfv}
\newcommand{\pairwu}{\mathbin{\uplus}}
\newcommand{\leftL}{\mathcal{L}}
\newcommand{\rightR}{\mathcal{R}}

\begin{document}
\isabellestyle{it}

\title{Relaxing safety for metric first-order temporal logic via dynamic free variables}
\author{Jonathan Juli\'an Huerta y Munive\\ University of Copenhagen, Denmark}
\date{\today}

\maketitle

\begin{abstract}
We define a fragment of metric first-order temporal logic formulas that 
guarantees the finiteness of their table-representations. We extend our
fragment's definition to cover the temporal dual operators 
\emph{trigger} and \emph{release} and show that our fragment 
is strictly larger than those previously used in the literature. We integrate 
these additions into an existing runtime verification tool and formally verify 
in Isabelle/HOL that the tool correctly outputs the table of constants that 
satisfy the monitored formula. Finally, we provide some 
example specifications that are now monitorable thanks to our contributions.
\end{abstract}

\section{Introduction}

Runtime verification (RV) complements other techniques for system quality-assurance
such as testing or model checking~\cite{LeuckerS09}. It allows monitoring 
properties during a system's execution by indicating when they are violated. 
Metric first-order temporal logic (MFOTL) is among the most expressive 
temporal and declarative specification languages for RV~\cite{FalconeKRT21}. 
It adds intervals to the logic's temporal operators to model quantitative 
descriptions of time~\cite{Chomicki95,Koymans90}. For these reasons, it is 
often used in monitor implementations~\cite{BasinKMZ15,HavelundPU17,SchneiderBKT19}.

Besides being expressive, monitors should also be efficient and trustworthy. 
Efficiency allows them to be deployed more invasively by course-correcting the 
evolution of a system~\cite{HublerBK22}, while trustworthiness makes them 
reliable in safety-critical applications. Recently, Verimon, an MFOTL-based 
RV-algorithm, has achieved high expressivity~\cite{ZinggKRST22} 
and efficiency~\cite{BasinDHKR0T20} while staying trustworthy because of 
its formally verified implementation in the Isabelle/HOL proof 
assistant~\cite{SchneiderBKT19}. It uses finite relations to represent
the set of valuations that make a specification true which contributes to its 
efficiency. Nevertheless, this feature also makes it inherit well-known issues 
from relational databases~\cite{GelderT91} forcing it to operate inside a fragment 
whose formula-evaluation guarantees finite outputs. This fragment is 
defined inductively on the structure of the formula via a predicate 
\isa{safe{\isacharunderscore}formula} which should not be confused with
the notion of safety property from model checking~\cite{KupfermanV01}. 
The set of \emph{safe} formulas that Verimon admits is rather restrictive.
The fact that well-known temporal operators such as \emph{historically} 
are immediately dubbed unsafe if they have free variables, is 
evidence of this. Here, we address this particular issue.


Our main contribution consists of the definition (\S\ref{sec-safe}) of a 
larger fragment of MFOTL-safe formulas, its formalisation in the 
Isabelle/HOL proof assistant, and its integration into Verimon's first 
implementation. This generalisation of safety enables us to monitor 
a wider variety of future and past operators including \emph{globally} and 
historically. To address safety of formulas involving these operators in 
conjunction with other connectives, we introduce the set of 
\emph{dynamic free variables} ($\dfv_i\, \alpha$) of a formula $\alpha$ 
at time-point $i$. It approximates the set of those free variables that 
contribute to the satisfiability of a formula at a specific point of a system's 
execution. Safety is then decided by computing a set of allowed sets of 
dynamic free variables and checking its nonemptiness. Essentially, we 
define a set of \emph{safe sets of free variables} ($\ssfv$) in such a way 
that whenever this set is nonempty, meaning that $\alpha$ is safe, 
$\dfv_i\, \alpha\in\ssfv\alpha$. 

Furthermore, with the view of integrating our generalisation into the latest 
optimised version of Verimon~\cite{ZinggKRST22}, we explicitly add 
conjunctions to the syntax of our implementation. This involves defining safety 
for specific cases where one of the conjuncts is an equality, a constraint, or the
negation of a safe formula. To go beyond the developments in the optimised 
version, we also add since and until's dual operators, \emph{trigger} and 
\emph{release} respectively, to Verimon's syntax. We therefore also extend 
Verimon's monitoring algorithm with functions to evaluate these operators 
(\S\ref{sec-dual}) and prove them correct.

Our formalisation and proof of correctness (\S\ref{sec-correct}) of the extended
monitoring algorithm is also a major contribution of this work. It involves 
redefining proof-invariants to accommodate dynamic free variables which 
largely reverberates in the proofs of correctness for each connective. The 
formalisation and proofs are available online, and corresponding definitions
are linked throughout the paper and indicated with the (clickable) Isabelle-logo
\isalink{https://bitbucket.org/jshs/monpoly/src/b4b63034eca0ccd5783085dececddb6c47cf6f52/thys/Relax_Safety/MFOTL/}.\footnote{Readers wishing to download and see the 
files in Isabelle, must ensure they download commit 
$b4b63034eca0ccd5783085dececddb6c47cf6f52$ of branch $\ssfv$s.} 
We add 
to the relevance of our safety relaxation by providing examples in 
\S\ref{sec-examples} that can now be monitored due to our contributions. 
We discuss future work and our conclusions in \S\ref{sec-conclusion}.

\paragraph{Related work.} In terms of expressivity of other monitors, our 
work directly extends the oldest Verimon version~\cite{SchneiderBKT19} 
by the above-mentioned additions. However, our generalisation has not been 
applied to the latest Verimon+~\cite{ZinggKRST22} as we do not deal with 
aggregators, recursive rules, or regular expressions. We foresee no issues to
adapt our approach to these other extensions. Verimon+ does not have an 
explicit version of dual operators and deems historically and globally unsafe.
A recent extension~\cite{HauserN21}, adds the dual operators and describes
lengthy encodings to monitor historically and globally. Yet, our work 
is complementary because of our larger safety fragment. Another past-only,
first-order monitor is DejaVu~\cite{HavelundPU17}. It uses 
binary decision diagrams which can model infinite relations and thus, does 
not require a notion of safety. In contrast, our work supports both past and 
bounded future operators and, in general, it is hard to compare performance
between both RV-approaches~\cite{HavelundPU17,ZinggKRST22}.

Safety has been well studied for relational
databases~\cite{AbiteboulHV95,Demolombe92,GelderT91,Kifer88,Ullman88}. 
Kifer~\cite{Kifer88} organises and relates various definitions of safety, 
and Avron and Hirshfeld~\cite{AvronH91} complement his work by answering
some of Kifer's conjectures at the end of his paper. More recently, evaluation of 
queries that are relatively safe has been explored~\cite{RaszykBKT22}. 
Thereby, there is an approach to translate any relational calculus query into a 
pair of safe queries where, if the second holds, then the original is unsafe 
(produces an infinite output). Otherwise, the original query's output is the same 
as the output of the first one. A recent extension to the temporal 
setting~\cite{Raszyk22} remains to be formally verified and integrated into
Verimon+. Arguably, using it as black boxes for the monitoring algorithm would
be less efficient than the direct integration we provide.


\section{Metric First-Order Temporal Logic}\label{sec-mfotl}
In this section, we briefly describe syntax and semantics of MFOTL for the 
Verimon implementation and introduce some concepts for later explanations. 
Following Isabelle/HOL notation and conventions, we use 
\isa{x\ {\isacharcolon}{\isacharcolon}\ {\isacharprime}a} to state that 
variable \isa{x} is of type \isa{{\isacharprime}a}. The type of lists and
the type of sets over \isa{{\isacharprime}a} are 
\isa{{\isacharprime}a\ list} and \isa{{\isacharprime}a\ set} respectively.
We implicitly use ``$s$'' to indicate a list of terms of some type, for 
instance, if $x::{\isacharprime}a$, then $\mathit{xs}::{\isacharprime}a\ list$.
The standard operations on lists that apply a function $f$ to each 
element ($\map\, f\, \mathit{xs}$), get the $n$th element ($\mathit{xs}\mathbin{!}n$), 
output the list's length ($\length \mathit{xs}$) or add an element to the 
left ($x\cons\, \mathit{xs}$) also appear throughout the paper. The expression
$f\ {\isacharprime}\ X$ denotes the set-image of $X$ under $f$. We freely 
use binary operations inside parenthesis as functions, that is 
we can write $x+y$ or $(+)\, x\, y$. Finally, the natural 
numbers have type $\nats$.

The type of terms ${\isacharprime}a\ \mathit{trm}$ simply consists of variables 
$\var x$ and constants $\const a$ with $x::\nats$ and 
\isa{a\ {\isacharcolon}{\isacharcolon}\ {\isacharprime}a}. The syntax
of MFOTL formulas ${\isacharprime}a\ \mathit{frm}$ that we use is\hfill\isalink{https://bitbucket.org/jshs/monpoly/src/b4b63034eca0ccd5783085dececddb6c47cf6f52/thys/Relax_Safety/MFOTL/MFOTL_Formula.thy\#lines-41}
\begin{align*}
\alpha\ ::=\ 
	& p\pred \mathit{ts} \mid t=_F t\mid \neg_F\, \alpha\mid \alpha\land_F\alpha\mid 
	\alpha\lor_F\alpha\mid \exists_F\, \alpha\mid\\
	& \previously_I\alpha\mid \nextly_I\alpha\mid \alpha\since_I\alpha\mid 
	\alpha\until_I\alpha\mid \alpha\trigger_I\alpha\mid \alpha\release_I\alpha,
\end{align*}
where $p::\mathit{string}$, $t::{\isacharprime}a\ \mathit{trm}$, 
$\mathit{ts}::{\isacharprime}a\ \mathit{trm}\ \mathit{list}$ and $I$ is a non-empty 
interval of natural numbers. We distinguish MFOTL connectives and quantifiers
from meta-statements via the subscript $F$ and freely use well-known interval 
notation, e.g. $[a,b)=\{n\mid a\leq n< b\}$ or $[a,b]=\{n\mid a\leq n\leq b\}$. 
Furthermore, we also write $m> I+n$ (and similar abbreviations) to state that 
$m$ is greater than $n$ plus any other element in $I$. Finally, the above syntax 
implies usage of De Bruijn indices, that is, $\exists_F\, p\pred [\var 1,\var 0]$
represents the formula $\exists\, y.\ p\, x\, y$.

Verimon encodes valuations as lists of natural numbers, $v::\nats\ list$ where 
the $x$th element of the list is the value of $\var x$. That is, the evaluation 
of terms ($\teval$) is given by $v \teval (\const a)=a$ and 
$v \teval (\var x)=v\mathbin{!}x$. For the semantics, a trace $\sigma$,
meaning an infinite time-stamped sequence of sets, models the input from the 
monitored system. The function $\tau\, \sigma\, i::\nats$ outputs the time-stamp 
at time-point $i::\nats$, whereas $\Gamma\, \sigma\, i$ outputs the 
corresponding set. When clear from context we use their abbreviated forms 
$\tau_i$ and $\Gamma_i$ respectively. The time-stamps are monotone 
$\forall i\leq j.\ \tau_i\leq\tau_j$ and eventually increasing 
$\forall n.\ \exists i.\ n\leq\tau_i$. The sets $\Gamma_i$ contain pairs 
$(p,\mathit{xs})$ where $p$ is a name for a predicate $p::\mathit{string}$ and 
$\mathit{xs}$ is a list of values ``satisfying'' $p$. Formally, the semantics 
are\hfill\isalink{https://bitbucket.org/jshs/monpoly/src/b4b63034eca0ccd5783085dececddb6c47cf6f52/thys/Relax_Safety/MFOTL/MFOTL_Formula.thy\#lines-211}
\small{
\begin{gather*} 
\leftAlignGather{
\langle\sigma, v, i\rangle\models p\pred \mathit{ts} \Leftrightarrow (p, \map\, ((\teval)\, v)\, \mathit{ts}) \in \Gamma_i,
\quad 
\langle\sigma, v, i\rangle\models \exists_F\, \alpha \Leftrightarrow \exists a::{\isacharprime}a.\ \langle\sigma, a\cons v, i\rangle\models\alpha
}\\ 
\leftAlignGather{\langle\sigma, v, i\rangle\models \alpha\land_F\beta \Leftrightarrow \langle\sigma, v, i\rangle\models\alpha\land \langle\sigma, v, i\rangle\models\beta, 
\quad 
\langle\sigma, v, i\rangle\models t_1=_Ft_2 \Leftrightarrow v\teval t_1 = v\teval t_2
}\\ 
\leftAlignGather{\langle\sigma, v, i\rangle\models \alpha\lor_F\beta \Leftrightarrow \langle\sigma, v, i\rangle\models\alpha\lor \langle\sigma, v, i\rangle\models\beta,
\qquad\qquad 
\langle\sigma, v, i\rangle\models \neg\, \alpha \Leftrightarrow \langle\sigma, v, i\rangle\not\models\alpha
}\\
\leftAlignGather{ 
\langle\sigma, v, i\rangle\models \nextly_I\alpha \Leftrightarrow \langle\sigma, v, i+1\rangle\models\alpha\land(\tau_{i+1} - \tau_{i})\in I\hspace{5em}
}\\
\leftAlignGather{
\langle\sigma, v, i\rangle\models \previously_I\alpha \Leftrightarrow \textit{ if }i=0\textit{ then }\text{false}\textit{ else }\langle\sigma, v, i-1\rangle\models\alpha\land(\tau_{i} - \tau_{i-1})\in I
}\\ 
\leftAlignGather{ 
\langle\sigma, v, i\rangle\models \alpha\since_I\beta \Leftrightarrow  \exists j\leq i.\ (\tau_i-\tau_j)\in I\land \langle\sigma, v, j\rangle\models\beta\land \left(\forall k\in\left(j,i\right].\ \langle\sigma, v, i\rangle\models\alpha\right)
}\\
\leftAlignGather{ 
\langle\sigma, v, i\rangle\models \alpha\until_I\beta \Leftrightarrow \exists j\geq i.\ (\tau_j-\tau_i)\in I\land \langle\sigma, v, j\rangle\models\beta\land \left(\forall k\in\left[i,j\right).\ \langle\sigma, v, i\rangle\models\alpha\right)
}\\
\leftAlignGather{ 
\langle\sigma, v, i\rangle\models \alpha\trigger_I\beta \Leftrightarrow \forall j\leq i.\ (\tau_i-\tau_j)\in I\Rightarrow \langle\sigma, v, j\rangle\models\beta\lor \left(\exists k\in\left(j,i\right].\ \langle\sigma, v, i\rangle\models\alpha\right)
}\\
\leftAlignGather{ 
\langle\sigma, v, i\rangle\models \alpha\release_I\beta \Leftrightarrow \forall j\geq i.\ (\tau_j-\tau_i)\in I\Rightarrow \langle\sigma, v, j\rangle\models\beta\lor \left(\exists k\in\left[i,j\right).\ \langle\sigma, v, i\rangle\models\alpha\right).
}
\end{gather*}
}

Other operators can be encoded, e.g. \emph{true} ($\top\equiv\const a=_F
\const a$), \emph{eventually} ($\eventually_I \alpha\equiv \top\until_I \alpha$), or 
\emph{historically} ($\historically_I\alpha\equiv \neg_F\,\eventually_I \neg_F\,\alpha$).
Moreover, trigger and release satisfy their dualities with since and until, that is,
$\langle\sigma, v, i\rangle\models \alpha\trigger_I\beta\Leftrightarrow 
\langle\sigma, v, i\rangle\models \neg_F\, ((\neg_F\, \alpha)\since_I(\neg_F\, \beta))$ 
and $\langle\sigma, v, i\rangle\models \alpha\release_I\beta\Leftrightarrow 
\langle\sigma, v, i\rangle\models \neg_F\, ((\neg_F\, \alpha)\until_I(\neg_F\, \beta))$. 
We do not encode them because it leads to hard-to-follow case distinctions 
in Isabelle proofs.

Intuitively, for the formula $\alpha$, a monitor outputs the set $\sats{\alpha}_i= 
\{v\mid  \langle \sigma, v,i\rangle\models\alpha\}$ at time-point $i::\nats$. 
However, these sets are redundant and infinite, e.g. $v_1=[4,5]$, 
$v_2=[4,5,1]$ and $v_3=[4,5,2,7]$ are all elements of 
$\sats{p\pred [\var 0,\var 1]}_i$ if $\Gamma\, \sigma\, i = \{(p,[4,5])\}$. 
For these reasons, Verimon makes valuations map variables $x::\nats$ that 
are not in the set of \emph{free variables} of $\alpha$, $x\notin \fv \alpha$, to 
$\None$ values of type ${\isacharprime}a\ \mathit{option}$. To address 
redundancy, Verimon also focuses on valuations with fixed length equal 
to $\nfv \alpha=\max\left(\{0\}\cup((+1)\ {\isacharprime}\ \fv\alpha)\right)$, that is, 
the least number $n$ such that if $x\in \fv \alpha$, then $x < n=\nfv \alpha$. For 
example, $v_4=[4,5,\None]$ satisfies $p\pred [\var 0,\var 1]$ at $i$ if it is a 
subformula of $\alpha$ with $\nfv \alpha=4$. Formally, the predicate 
$\wftuple n\, X\, v$ holds if $\length v=n$ and $\forall i<n.\ v\mathbin{!}i = 
\None \Leftrightarrow i\notin X$. Thus, the monitor 
outputs\hfill\isalink{https://bitbucket.org/jshs/monpoly/src/b4b63034eca0ccd5783085dececddb6c47cf6f52/thys/Relax_Safety/MFOTL/MFOTL_Table_Correct.thy\#lines-97}
\begin{equation*}
\sats{\alpha}_{i,n}^X=\{v::{\isacharprime}a\ \mathit{option}\ \mathit{list}\mid \langle \sigma, v,i\rangle
\models_M\alpha \land\wftuple n\, X\, v\},
\end{equation*}
where $\langle \sigma, v,i\rangle\models_M\alpha$ abbreviates $\langle \sigma, 
\map \mathit{the}\, v,i\rangle\models\alpha$ and $\mathit{the}$ is the standard 
function mapping optional values 
\isa{Some\ x\ {\isacharcolon}{\isacharcolon}{\isacharprime}a\ option} to their 
concrete counterparts \isa{x\ {\isacharcolon}{\isacharcolon}{\isacharprime}a} 
and $\None$ to an unspecified value of type \isa{{\isacharprime}a}. 

Outputs $\sats{\alpha}_{i,n}^X$ are relations or, from a database perspective, 
tables of values satisfying $\alpha$. For each subformula $\beta$ of the monitored 
formula $\alpha$, Verimon obtains the tables $\sats{\beta}_{i,\nfv\alpha}^{\fv\beta}$ 
and uses them to compute $\sats{\alpha}_{i,\nfv\alpha}^{\fv\alpha}$. For this, it 
includes common relational algebra operations like the (natural) join ($\join$), 
antijoin ($\antijoin$) and union ($\cup$) of tables. However, these outputs can 
quickly become infinite if not treated carefully, e.g. for datatypes with infinite carrier
sets, $\sats{\neg_F\, \alpha}_{i,n}^X$ is infinite when $\sats{\alpha}_{i,n}^X$ is finite.
Verimon uses the function \isa{safe{\isacharunderscore}formula} to define a 
fragment of MFOTL-formulas where finite outputs are guaranteed. For instance, 
tables for disjunctions $\sats{\alpha\lor_F\beta}_{i,n}^{(\fv\alpha)\cup(\fv\beta)}$ can 
only be computed as $\sats{\alpha}_{i,n}^{\fv\alpha}\cup\sats{\beta}_{i,n}^{\fv\beta}$ 
when they are \emph{union-compatible}, that is, they both have the same 
\emph{attributes} (columns) which \isa{safe{\isacharunderscore}formula} requires 
as $\fv\alpha=\fv\beta$. Similarly, negations are only allowed inside conjunctions 
$\sats{\alpha\land_F\neg\, \beta}_{i,n}^{\fv\alpha}$ with 
$\fv\beta\subseteq \fv\alpha$ to safely compute antijoins. 

This is why $\historically_I\, \alpha$ as encoded above is not generally 
considered safe by itself. Extending MFOTL's syntax to include trigger and 
release, encoding $\historically_I\alpha\equiv (\neg_F\,\top)\trigger_I\alpha$, 
and defining \isa{safe{\isacharunderscore}formula} for these 
cases~\cite{HauserN21} is still unsatisfactory. Following the semantics above,
formulas $\alpha\trigger_I\beta$ remain unsafe when $0\notin I$ because 
they could be vacuously true, i.e. if all $j\leq i$ satisfy $\tau_i-\tau_j\notin I$. 
Yet, crucially for our purposes, older Verimon versions deem some 
trivially true formulas like $\const a=_F \const a$ safe. They evaluate to 
\emph{unit tables} $\unitT{n}$, where 
$\unitT{n}=\{\unit{n}\}$ and $\unit{n}::{\isacharprime}a\ \mathit{option}\ \mathit{list}$ only has 
$\None$ repeated $n$ times. In fact, $\unit{n}$ is the only valuation $v$ that
satisfies $\wftuple n\, \emptyset\, v$. In the next section, we take advantage 
of this and define a notion of safety that allows an encoding of 
$\historically_I\, \alpha$ with $0\notin I$ to be safe.

The formalisation of syntax and semantics of the dual-operators is a 
technical contribution from our work. It involves routine extensions of 
definitions such as $\fv\alpha$ but it also requires adding properties
about satisfiability of both operators. We add more than 400 lines of
code to Verimon's formalisation of syntax and 
semantics~\cite{SchneiderBKT19,SchneiderT19}.

\section{Relaxation of Safety}\label{sec-safe}

If $\alpha\trigger_I\beta$ is vacuously true at $i$, then Verimon's output is correct if 
it is equal to $\sats{\alpha\trigger_I\beta}_{i,n}^{\emptyset}=\unitT{n}$. However, in
its current implementation, this would only be provable for formulas such that $\fv (\alpha\trigger_I\beta)=\emptyset$. 
Therefore, to define a larger fragment of evaluable formulas, it is convenient to 
choose the correct set of attributes at time-point $i$. Here, we generalise 
Verimon's fragment of safe formulas by introducing the 
\emph{dynamic free variables} $\dfv \sigma\, i\, \alpha$ of $\alpha$ at 
$i$~\isalink{https://bitbucket.org/jshs/monpoly/src/b4b63034eca0ccd5783085dececddb6c47cf6f52/thys/Relax_Safety/MFOTL/MFOTL_Safety.thy\#lines-816} and 
its set of \emph{safe sets of free variables} $\ssfv \alpha$~\isalink{https://bitbucket.org/jshs/monpoly/src/b4b63034eca0ccd5783085dececddb6c47cf6f52/thys/Relax_Safety/MFOTL/MFOTL_Safety.thy\#lines-349}. The function $\dfv\, \sigma\, i\, \alpha$ approximates 
the set of free variables that influence the satisfiability of $\alpha$ at $i$ in the trace 
$\sigma$. As with other trace-functions $\Gamma$ and $\tau$, we often use its 
abbreviated form $\dfv_i\, \alpha$. It is a semantic concept and we only need it to 
prove the algorithm's correctness: outputs are exactly the sets 
$\sats{\alpha}_{i,\nfv \alpha}^{\dfv_i\, \alpha}$ for the monitored formula $\alpha$ at 
each $i$. The function $\ssfv \alpha$ approximates all the possible combinations of 
attributes that tables for $\alpha$ might have at different time-points. It is recursively 
defined so that we can decide $\alpha$'s safety by checking $\ssfv\alpha\neq\emptyset$,
that is, we define $\issafe\alpha \Leftrightarrow \ssfv\alpha\neq\emptyset$. We describe 
our reasoning behind the definition for each connective below and enforce 
various properties with our definitions: on one hand, as they are sets of 
free variables, \tagx[$i$]{prpty:i} $\dfv\sigma\, i\, \alpha\subseteq\fv\alpha$ and 
\tagx[$ii$]{prpty:ii} $\bigcup\ssfv\alpha \subseteq \fv\alpha$. 
On the other hand, to prove correctness, if the formula is safe 
$\ssfv\alpha\neq\emptyset$, then our set of attributes should be a witness for it:
\tagx[$iii$]{prpty:iii} $\dfv\sigma\, i\, \alpha\in \ssfv\alpha$.  For the full formal 
definitions see also our Appendix~\ref{sec-appendix}.

\paragraph{Atomic formulas.} All atoms $p\pred \mathit{ts}$ are safe and their 
attributes do not change over time. Thus, we define $\ssfv (p\pred \mathit{ts}) = 
\{\fv (p\pred \mathit{ts})\}$ and $\dfv_i\, (p\pred \mathit{ts})=\fv (p\pred \mathit{ts})$. 
Following Verimon, we do not make $\var x=_F\var x$ safe as it is not practically
relevant for us. Therefore, $\ssfv$ maps equalities 
$\alpha\in\{\var x=_Ft,\ t=_F\var x,\ t_1=_Ft_2\}$ to $\{\fv \alpha\}$ whenever 
$\fv t=\emptyset$ (resp. $\fv t_1=\fv t_2=\emptyset$), and 
to $\emptyset$ otherwise. Similarly, we define $\dfv_i\, (t_1=_Ft_2) = 
\fv (t_1=_Ft_2)$ for all $t_1,t_2::{\isacharprime}a\ \mathit{trm}$ and $i::\nats$. 

\paragraph{Conjunctions.} If safe, each conjunct may have many combinations
of attributes. Moreover, a join ($\join$) outputs a table with all the attributes 
from its operands. Thus, if $\ssfv\alpha\neq\emptyset$ and $\ssfv\beta\neq\emptyset$, 
then $\ssfv(\alpha\land_F\beta)=\ssfv\alpha\pairwu\ssfv\beta$ where ($\pairwu$) is the 
pairwise union $A\pairwu B=\{a\cup b\mid a\in A\land b\in B\}$~\isalink{https://bitbucket.org/jshs/monpoly/src/b4b63034eca0ccd5783085dececddb6c47cf6f52/thys/Relax_Safety/MFOTL/MFOTL_Safety.thy\#lines-87}. We follow Verimon+ and define safety for cases when 
only the left conjunct $\alpha$ is safe.\footnote{Adding the symmetric case increases 
the number of proofs in the formalisation. It is easier to assume a formula rewriter can 
commute conjuncts if necessary.} If $\beta$ is an equality $t_1=_Ft_2$ with 
$\fv t_1\subseteq X$ or $\fv t_2\subseteq X$ for each $X\in\ssfv\alpha$, then we can 
safely add any single variable on the other side of the equation, possibly not in $\fv\alpha$,
to the elements of $\ssfv\alpha$. Therefore, in this case $\ssfv(\alpha\land_F\beta) = 
\left((\cup)\, (\fv\beta)\right)\ {\isacharprime}\ (\ssfv\alpha)$. Finally, if $\beta$ is a 
negation $\neg_F\, \beta'$ of a safe formula $\beta'$ and every $Y\in\ssfv\beta'$ 
satisfies $Y\subseteq X$ for each $X\in\ssfv\alpha$ then we can compute antijoins, 
and thus $\ssfv(\alpha\land_F\beta) = \ssfv\alpha$. If neither of these cases holds, 
then $\ssfv(\alpha\land_F\beta) = \emptyset$. Given the behaviour of join on columns, 
the dynamic free variables are simply $\dfv_i\, (\alpha\land_F\beta) = 
\dfv_i\, \alpha\cup \dfv_i\, \beta$. 

\paragraph{Disjunctions.} Due to union-compatibility, we can only take unions
of tables with the same attributes. Yet, we can generalise for cases when
formulas might be vacuously true at some time-points. To evaluate
disjunctions we use the function $\isa{eval{\isacharunderscore}or}\ n\, R_1\, R_2$ 
that outputs $\unitT{n}$ if either $R_1=\unitT{n}$ or $R_2=\unitT{n}$, and 
$R_1\cup R_2$ otherwise. Similarly, to ensure 
$\wftuple\, n\, (\dfv_i\, (\alpha\lor_F\beta))\, \unit{n}$, we state that 
there are no ``relevant'' variables ($\dfv_i\, (\alpha\lor_F\beta) = 
\emptyset$) for the satisfiability of $\alpha\lor_F\beta$ when either $\alpha$ 
or $\beta$ are logically valid at $i$. If the variables of $\alpha$ 
are ``irrelevant'' ($\dfv_i\, \alpha=\emptyset$) because $\alpha$ is unsatisfiable 
at $i$ ($\sats{\alpha}_i=\emptyset$), 
then we just need the variables of $\beta$: 
$\dfv_i\, (\alpha\lor_F\beta)= \dfv_i\, \beta$. 
The symmetric case also holds. If both disjuncts are relevant 
($\dfv_i\, \alpha\neq\emptyset\neq\dfv_i\, \beta$), we need both sets of variables: 
$\dfv_i\, (\alpha\lor_F\beta) = \dfv_i\alpha\cup\dfv_i\, \beta$. 

The behaviour of $\dfv$ on disjunctions means that if 
$\emptyset\in\ssfv\alpha$ or $\emptyset\in\ssfv\beta$, then $\emptyset$ 
should also be an element of $\ssfv(\alpha\lor_F\beta)$. This may happen in 
various ways. First, if $\fv\alpha=\emptyset$ or $\fv\beta=\emptyset$, then by 
$(ii)$ above, we know that $\ssfv\alpha=\{\emptyset\}$ or $\ssfv\beta=\{\emptyset\}$.
In this case, we can define $\ssfv(\alpha\lor_F\beta) = \ssfv\alpha\cup\ssfv\beta$ 
assuming both $\alpha$ and $\beta$ satisfy $\issafe$. Next, notice that if we allow 
attributes $X_\alpha\in\ssfv \alpha$ and $X_\beta\in\ssfv \beta$ such that 
$\emptyset\neq X_\alpha\neq X_\beta\neq\emptyset$, the corresponding table
for $\alpha\lor_F\beta$ would need to have infinite values. Therefore, at most 
we may allow $\ssfv\alpha\subseteq\{\emptyset,\fv \alpha\}$, 
$\ssfv\beta\subseteq\{\emptyset,\fv \beta\}$ and $\fv\alpha=\fv\beta$ with both 
$\alpha$ and $\beta$ having non-empty $\ssfv$s. In this case, if $\emptyset\in\ssfv\alpha$ or 
$\emptyset\in\ssfv\beta$ then $\ssfv(\alpha\lor_F\beta) = 
\{\emptyset\}\cup (\ssfv\alpha\pairwu\ssfv\beta)$, otherwise 
$\ssfv(\alpha\lor_F\beta)=\{\fv\alpha\}$. 

\paragraph{Negations.} If $\alpha$ is safe and closed ($\ssfv(\alpha)=\{\emptyset\}$ 
by $(ii)$), then we can safely evaluate its negation $\ssfv(\neg_F\, \alpha) =
\{\emptyset\}$.We also allow $\ssfv(\neg_F\, (t=_F t)) = \{\fv(t=_F t)\}$ for 
arbitrary term $t$ to encode the constantly false formula. Otherwise negations 
are unsafe $\ssfv(\neg_F\, \alpha)=\emptyset$. For dynamic free variables, we 
define $\dfv_i(\neg_F\, \alpha)=\dfv_i\alpha$. 

\paragraph{Quantifiers.} When interpreting De Bruijn indices, 
quantifiers remove $0$ from $\fv\alpha$ and subtract $1$ to all its elements. Thus, 
we define $\dfv_i\, (\exists_F\alpha)= (\lambda x.\ x-1)\ {\isacharprime}\ \left(\dfv_i\, \alpha-\{0\}\right)$
and $(\lambda x.\ x-1)\ {\isacharprime}\ \left(X-\{0\}\right)\in \ssfv(\exists_F\alpha)$
for each $X\in\ssfv\alpha$. 

\paragraph{Previous and next.} The definition of the dynamic free variables for 
one-step temporal operators follows that of their semantics: 
$\dfv_i\, (\nextly_I \alpha)=\dfv_{i+1} \alpha$ while 
$\dfv_i\, (\previously_I \alpha)=\dfv_{i-1} \alpha$ if $i>0$ and 
$\dfv_i\, (\previously_I \alpha)=\fv\alpha$ if $i=0$. For safety, all 
combinations of attributes of $\alpha$ might be used in its one-step temporal versions 
$\ssfv\, (\nextly_I \alpha) = \ssfv \alpha = \ssfv\, (\previously_I \alpha)$. 

\paragraph{Since and until.} Let us follow the semantics for since and until to 
define their $\dfv$s at $i$. The definition for one operator emerges by dualising 
the time-point order and flipping subtractions of time-stamps in the other's 
definition, hence we omit the description for until. We must collect all the 
dynamic free variables $\dfv_k\alpha$ and $\dfv_j\beta$ that influence the 
satisfiability of $\alpha\since_I\beta$. Start by defining $\downarrow_I i = 
\{j\mid j\leq i\land (\tau_i-\tau_j)\in I\}$ to identify the indices $j$ for $\beta$, and 
the predicate $\mathit{satisf{\isacharunderscore}at}\ j = \exists v.\ \langle\sigma, v, j\rangle\models\beta 
\land \left(\forall k\in\left(j,i\right].\ \langle\sigma, v, i\rangle\models\alpha\right)$ 
so that $\alpha\since_I\beta$ is unsatisfiable at $i$ if $\forall j\in\downarrow_I i.\ 
\neg\, \mathit{satisf{\isacharunderscore}at}\ j$. When this happens we let
$\dfv_i (\alpha\since_I\beta)=\fv (\alpha\since_I\beta)$. Otherwise, the indices $j$
for $\beta$ are $\mathcal{J} = \{j\in\downarrow_I i\mid \mathit{satisf{\isacharunderscore}at}\ j\}$
while those $k$ for $\alpha$ are $\mathcal{K}=\bigcup_{j\in\mathcal{J}} (j,i]$.
Having identified the indices, we obtain $\dfv_i (\alpha\since_I\beta)= 
{\big(\bigcup_{k\in\mathcal{K}}\dfv_k \alpha\big)\cup \big(\bigcup_{j\in\mathcal{J}}\dfv_j \beta\big)}$.

As we have seen for disjunctions, our definitions of safety depend on the operations 
in the formula-evaluation. In particular, to define $\ssfv$s for since, it is 
convenient to understand Verimon's~\cite{SchneiderBKT19} implementation 
roughly represented with the equations:\hfill\isalink{https://bitbucket.org/jshs/monpoly/src/b4b63034eca0ccd5783085dececddb6c47cf6f52/thys/Relax_Safety/MFOTL/MFOTL_Formula.thy\#lines-651}
\begin{equation}\label{eq-since}
\sats{\alpha\since_I\beta}_i{=} \bigcup_{j\in\downarrow_I i}\sats{\alpha\since_{\{\tau_i-\tau_j\}}\beta}_i
\text{ and } \sats{\alpha\since_{\{\tau_i-\tau_j\}}\beta}_i {=} \bigcup_{k\in\downarrow_{\{\tau_i-\tau_j\}} i}\sats{\beta}_k\cap\left(\bigcap_{l\in\left(k,i\right]}\sats{\alpha}_l\right).
\end{equation}
That is, the algorithm obtains the valuations in $\sats{\alpha\since_I\beta}_{i}$ 
by iteratively updating those in $\sats{\alpha\since_{\{\tau_i-\tau_j\}}\beta}_{i}$
until it has visited all time-points $j\in\downarrow_I i$, when it outputs their 
union. In the implementation, intersections are replaced with (anti)joins and sets
$\sats{\varphi}_l$, with tables $\sats{\varphi}_{l,n}^{X_l}$ having attributes $X_l$ 
at index $l$ for safe $\varphi$. If $0\in I$, one of the tables involved in the output
union is $\sats{\beta}^{\fv\beta}_{i,n}$. Hence, by union-compatibility, all the other 
table-operands (represented by $\sats{\alpha\since_{\{\tau_i-\tau_j\}}\beta}_{j}$)
must have the same attributes. To ensure this, we require
$\ssfv(\alpha\since_I\beta)$ to be non-empty only when $\ssfv\beta=\{\fv\beta\}$. 
Then, we define $\ssfv(\alpha\since_I\beta)=\{\fv\beta\}$ if $\fv\alpha\subseteq\fv\beta$ 
with $\ssfv\alpha\neq\emptyset$ because the joins in the construction of the tables
represented by $\sats{\alpha\since_{\{\tau_i-\tau_j\}}\beta}_{i}$ would always be 
guarded by the attributes $\fv\beta$ of $\sats{\beta}^{\fv\beta}_{k,n}$. Similarly, 
$\ssfv(\alpha\since_I\beta) = \{\fv\beta\}$ for $\alpha=\neg\, \alpha'$ with 
$\ssfv\alpha'\neq\emptyset$ and $\fv\alpha\subseteq\fv\beta$ because of the 
corresponding antijoins. 

The Verimon implementation for until is different and intuitively 
corresponds to\hfill\isalink{https://bitbucket.org/jshs/monpoly/src/b4b63034eca0ccd5783085dececddb6c47cf6f52/thys/Relax_Safety/MFOTL/MFOTL_Formula.thy\#lines-657}
\begin{equation*}
\sats{\alpha\until_I\beta}_i= \bigcup_{j\in\uparrow_I i}\sats{\beta}_j\cap\left(\bigcap_{k\in\left[i,j\right)}\sats{\alpha}_k\right).
\end{equation*}
As before, one of the operands may be $\sats{\beta}^{\fv\beta}_{i,n}$, 
therefore we require $\ssfv\beta=\{\fv\beta\}$. If $\fv\alpha\subseteq\fv\beta$ with
$\ssfv\alpha\neq\emptyset$, then $\ssfv(\alpha\since_I\beta)=\{\fv\beta\}$. But for 
$\alpha=\neg\alpha'$, the tables for $\alpha'$ 
are united separately and antijoined to each table for $\beta$ at $j$. Thus, we need to 
take union-compatibility into account when $\alpha=\neg\alpha'$. Hence, in this case,
we define $\ssfv(\alpha\since_I\beta)=\{\fv\beta\}$ if $\ssfv\beta=\{\fv\beta\}$ and 
$\ssfv\alpha'=\{\fv\alpha'\}$.

\paragraph{Trigger and release.} Our definition of $\dfv$s for these dual operators
is very similar to that of since and until. Assume $\mathbin{D}\in\{\trigger, \release\}$
and let $\mathit{idx}=\downarrow_I i$ and $\mathit{ivl}\, i\, j=\left(j,i\right]$ if 
$\mathbin{D}=\trigger$; otherwise, if $\mathbin{D}=\release$, then 
$\mathit{idx}=\uparrow_I i$ and $\mathit{ivl}\, i\, j=\left[i,j\right)$. The key difference 
in the definition of $\dfv$s for $\mathbin{D}$ is that if $\mathit{idx}=\emptyset$, then 
$\alpha\mathbin{D}_I\beta$ is vacuously true. Hence, we define 
$\dfv_i(\alpha\mathbin{D}_I\beta)=\emptyset$ because $\alpha\mathbin{D}_I\beta$ 
will evaluate to the unit table. As before, we also define a predicate 
$\mathit{satisf{\isacharunderscore}at}\ v\, j$ that makes the formula unsatisfiable 
whenever $\forall v.\ \exists j\in\mathit{idx}.\ 
\neg\, \mathit{satisf{\isacharunderscore}at}\ v\, j$. In this case, 
$\dfv_i(\alpha\mathbin{D}_I\beta)=\fv(\alpha\mathbin{D}_I\beta)$. The predicate \isa{satisf{\isacharunderscore}at} also allows us to define sets of indices $\mathcal{K}$ 
and $\mathcal{J}$ to collect $\dfv$s over time for the remaining cases: 
$\dfv_i (\alpha\mathbin{D}_I\beta) = {\big(\bigcup_{k\in\mathcal{K}}\dfv_k \alpha\big)\cup \big(\bigcup_{j\in\mathcal{J}}\dfv_j \beta\big)}$.

Our definition of safety for dual operators is intuitively understood by 
observing\hfill\isalink{https://bitbucket.org/jshs/monpoly/src/b4b63034eca0ccd5783085dececddb6c47cf6f52/thys/Relax_Safety/MFOTL/MFOTL_Formula.thy\#lines-680}
\begin{equation}\label{eq-dual}
\sats{\alpha\mathbin{D}_I\beta}_i= \bigcap_{j\in\mathit{idx}_I i}\sats{\beta}_j\cup\left(\bigcup_{k\in\mathit{ivl}\, i\, j}\mathit{rel}_k\cap\sats{\alpha}_k\right),
\end{equation}
where $\mathit{rel}_k=\sats{\beta}_k$ if $0\in I$, and $\mathit{rel}_k=\sats{\alpha}_k$ 
otherwise. As before, whenever $0\in I$, we define $\ssfv(\alpha\mathbin{D}_I\beta) 
= \{\fv\beta\}$ assuming $\ssfv\beta=\{\fv\beta\}$ and $\fv\alpha\subseteq\fv\beta$, 
both if $\ssfv\alpha\neq\emptyset$ or if $\ssfv\alpha=\emptyset$ but $\alpha=
\neg\, \alpha'$ with $\ssfv\alpha'\neq\emptyset$. When $0\notin I$, 
$\alpha\mathbin{D}_I\beta$ might be vacuously true and union-compatibility is 
relevant. Therefore, we define $\ssfv(\alpha\mathbin{D}_I\beta) = 
\{\emptyset,\fv\alpha\}$ whenever $\ssfv\alpha=\{\fv\alpha\}$, $\ssfv\beta=\{\fv\beta\}$ 
and $\fv\alpha=\fv\beta$. Otherwise $\ssfv(\alpha\mathbin{D}_I\beta)=\emptyset$. 

We formalise all cases above together with the definition 
$\issafe\alpha \Leftrightarrow \ssfv\alpha\neq\emptyset$ in Isabelle/HOL. Next, 
by induction on the definition of $\ssfv$s, we derive properties (\ref{prpty:i})~\isalink{https://bitbucket.org/jshs/monpoly/src/b4b63034eca0ccd5783085dececddb6c47cf6f52/thys/Relax_Safety/MFOTL/MFOTL_Safety.thy\#lines-853}, (\ref{prpty:ii})~\isalink{https://bitbucket.org/jshs/monpoly/src/b4b63034eca0ccd5783085dececddb6c47cf6f52/thys/Relax_Safety/MFOTL/MFOTL_Safety.thy\#lines-610} and (\ref{prpty:iii})~\isalink{https://bitbucket.org/jshs/monpoly/src/b4b63034eca0ccd5783085dececddb6c47cf6f52/thys/Relax_Safety/MFOTL/MFOTL_Safety.thy\#lines-1087} above. Additionally, we also prove that for 
any formula $\alpha$, $\issafe\alpha \Leftrightarrow \fv\alpha\in\ssfv\alpha$.

In classical logic, an important property for syntactic substitutions
of terms and formulas states that if two valuations $v$ and $v'$ coincide in $\fv\alpha$, 
then the value of $\alpha$ is the same under both valuations. 
Similarly, we have that if $v\mathbin{!}x=v'\mathbin{!}x$ for all 
$x\in\dfv\sigma\,i\,\alpha$, then $\langle\sigma,v,i\rangle\models\alpha\Leftrightarrow \langle\sigma,v',i\rangle\models\alpha$. This is useful for us because it ratifies 
that if $\langle\sigma,\unit{n},i\rangle\models_M\alpha$ for $\alpha$ with 
$\dfv_i\alpha=\emptyset$, then $\alpha$ is logically valid at $i$.

Let us compare our definition of safety with previous ones. To do this directly, 
we combine \isa{safe{\isacharunderscore}formula} predicates from 
Verimon~\cite{SchneiderBKT19} and Verimon+~\cite{HauserN21,ZinggKRST22},
restrict them to the syntax in \S\ref{sec-mfotl} and add them to our 
Appendix~\ref{sec-appendix}. Structurally, our definition of safety for conjunctions, 
negations, existential quantifiers, previous, next and until operators resembles 
that of Verimon+~\cite{ZinggKRST22} which is already more general than that of 
Verimon. However, in combination with other operators our definition deems 
more formulas safe (see \S\ref{sec-examples}). We discuss in~\S\ref{sec-conclusion}
possible generalisations for these cases that involve $\dfv$s and $\ssfv$s. 
For equalities, $\ssfv$ generalises \isa{safe{\isacharunderscore}formula} even 
when incorporating Verimon+'s more expressive term language that includes 
some arithmetic operations. By using \isa{eval{\isacharunderscore}or},
our definition of safety for disjunctions differs 
from \isa{safe{\isacharunderscore}formula} by admitting vacuously true formulas 
satisfying $\issafe$ in either disjunct. Additionally, $\issafe\, (\alpha\since_I\beta)$ 
allows $\alpha$ to have many $\ssfv$s which generalises 
$\isa{safe{\isacharunderscore}formula}\, (\alpha\since_I\beta)$. Dual operators are 
only safe in an unintegrated extension~\cite{HauserN21} of Verimon+; but we 
also generalise this work by making them safe when $0\notin I$, even 
when they are not part of a conjunction with a safe formula. In general, 
we show that $\issafe$ generalises previous monitoring fragments: on one 
hand we prove that $\issafe\alpha$ if $\isa{safe{\isacharunderscore}formula}\ 
\alpha$ for any $\alpha$~\isalink{https://bitbucket.org/jshs/monpoly/src/b4b63034eca0ccd5783085dececddb6c47cf6f52/thys/Relax_Safety/MFOTL/MFOTL_Safety.thy\#lines-1566}. On the other hand, the formula 
$\alpha\equiv\neg_F (\var x=_F\var x)\trigger_{[1,2]} \left(p\pred [\var x]\right)$, 
which is equivalent to $\historically_{[1,2]}p\pred [\var x]$, satisfies 
$\issafe\alpha$ but also $\neg\, \isa{safe{\isacharunderscore}formula}\ \alpha$. 




The definition and formalisation of safety for MFOTL-formulas is a major 
contribution of this work. It involves defining $\ssfv$s, $\dfv$s, pairwise 
unions and proving their corresponding properties. The developments on 
this section 
add more than 1600 lines of code to the Verimon formalisation~\cite{SchneiderBKT19,SchneiderT19}.

\section{Implementation of Dual Operators}\label{sec-dual}

Intuitively, the monitoring algorithm takes as inputs a safe MFOTL-formula 
$\alpha$ and an event trace $\sigma$ and outputs the table of satisfactions 
$\sats{\alpha}_{i,n}^{\dfv_i\, \alpha}$ at each time-point $i$. Concretely, 
Verimon provides two functions \isa{minit} and \isa{mstep} to initialise
and update the monitor's state respectively. The model for this monitor's 
\emph{state} at $i$ is a three-part record 
$\langle\rho^\alpha_i,\alpha^i_M,n\rangle$. Here, $n=\nfv\alpha$ and
$\rho^\alpha_i::nat$ is the \emph{progress} or the earliest time-point for 
which satisfactions of $\alpha$ cannot yet be evaluated for lack or information. 
Finally, $\alpha^i_M::{\isacharprime}a\ \mathit{mformula}$ is a recursively 
defined structure associated with $\alpha$ to store all the information needed 
to compute $\sats{\alpha}_{i,n}^{\dfv_i\, \alpha}$. We describe our extensions
to all these functions and structures in order to monitor dual operators trigger 
and release. In the sequel, we use $A_i=\sats{\alpha}_{i,n}^{\dfv_i\, \alpha}$, 
$B_i=\sats{\beta}_{i,n}^{\dfv_i\, \beta}$ and 
$C_i=\sats{\gamma}_{i,n}^{\dfv_i\, \gamma}$ to simplify notation.

The function \isa{minit} is just a wrapper calling $\mathit{minit}0$ to set the 
initial monitor's state to $\langle0,\alpha^0_M,\nfv\alpha\rangle$. Accordingly,
$\mathit{minit}0$ takes a safe MFOTL-formula 
$\alpha::{\isacharprime}a\ \mathit{frm}$ and transforms it into 
$\alpha^0_M::{\isacharprime}a\ \mathit{mformula}$. The datatype \isa{mformula} 
describes the information needed at each $i$ to compute $A_i$. For instance, 
if $\alpha$'s main connective is a binary operator with direct subformulas $\beta$
and $\gamma$, the set of satisfactions for either of them are only available up to 
$\rho^{\beta}_i$ and $\rho^\gamma_i$ respectively at $i$. This is why, $\alpha^i_M$
includes a \emph{buffer} \isa{buf{\isacharcolon}{\isacharcolon}{\isacharprime}a\ mbuf2} 
to store yet, unused tables $B_j$ (resp. $C_j$) such that $j\in[\rho_i^\gamma,\rho_i^\beta)$
(resp. $j\in[\rho_i^\beta,\rho_i^\gamma)$). Whenever both $B_j$ and $C_j$ are in the buffer, the 
algorithm operates them and either outputs the result or stores it in an 
\emph{auxiliary state} for future processing. To help the algorithm 
know whether to do joins or antijoins, $\alpha^i_M$ may also include a boolean
indicating whether one of its direct sub-formulas is not negated. More 
specifically, the state-representations of $\alpha\mathbin{D}_I\beta$ with 
$\mathbin{D}\in\{\trigger,\release\}$ include their sub-mformulas
$\alpha_M$ and $\beta_M$, a boolean indicating if $\alpha$ is not negated, the 
interval $I::\mathcal{I}$, the \emph{buffer} \isa{buf}, a corresponding list of unused 
time-stamps $\tau$s, and its auxiliary state. Formally, we add the last two 
lines below to the definition of the \emph{mformula} datatype:\hfill\isalink{https://bitbucket.org/jshs/monpoly/src/b4b63034eca0ccd5783085dececddb6c47cf6f52/thys/Relax_Safety/MFOTL/MFOTL_Monitor.thy\#lines-32}

\begin{isabellebody}\isanewline
\isacommand{datatype}\ {\isacharprime}a\ mformula\ {\isacharequal}\ MRel\ {\isachardoublequoteopen}
{\isacharprime}a\ option\ list\ set{\isachardoublequoteclose}\ {\isacharbar}\ {\isasymdots}\isanewline
\ \ {\isacharbar} MTrigger\ bool\ {\isachardoublequoteopen}{\isacharprime}a\ mformula{\isachardoublequoteclose}\ bool\ \ $\mathcal{I}$\ {\isachardoublequoteopen}{\isacharprime}a\ mformula{\isachardoublequoteclose}\ {\isachardoublequoteopen}{\isacharprime}a\ mbuf2{\isachardoublequoteclose}\ {\isachardoublequoteopen}\ $\nats$\ list{\isachardoublequoteclose}\ {\isachardoublequoteopen}{\isacharprime}a\ mtaux{\isachardoublequoteclose}\isanewline
\ \ {\isacharbar} MRelease\ bool\ {\isachardoublequoteopen}{\isacharprime}a\ mformula{\isachardoublequoteclose}\ bool\ \ $\mathcal{I}$\ {\isachardoublequoteopen}{\isacharprime}a\ mformula{\isachardoublequoteclose}\ {\isachardoublequoteopen}{\isacharprime}a\ mbuf2{\isachardoublequoteclose}\ {\isachardoublequoteopen}\ $\nats$\ list{\isachardoublequoteclose}\ {\isachardoublequoteopen}{\isacharprime}a\ mraux{\isachardoublequoteclose}\isanewline
\end{isabellebody}
\noindent where the second boolean indicates whether $0\in I$, and 
\isa{{\isacharprime}a\ mtaux} and \isa{{\isacharprime}a\ mraux} are the
type-abbreviations we use for trigger and release's auxiliary states 
respectively. Let us describe these in detail next.

Our implementation of the auxiliary state for trigger $\mathit{Ts}::{\isacharprime}a\ mtaux$ 
resembles Verimon's~\cite{SchneiderBKT19} auxiliary state for since. That is, it 
combines the intuitions in eqs.~(\ref{eq-since}) and~(\ref{eq-dual}): at 
time-point $i$, it is a list of time-stamped tables $\langle \tau_j,T_{\tau_j}^i\rangle$
that the monitor joins after it has passed all $j\in\downarrow_I i$. Abbreviating 
$\iota=(\min\rho^{\alpha}_{i}\rho^{\beta}_i)-1$ and using $\join^*$ to denote a join 
with non-negated subformula $\alpha$ and an antijoin with $\alpha$ for the 
direct subformula $\neg_F\, \alpha$, we intend
\begin{align}
T_{\tau_j}^i &=\bigjoin_{k\in\downarrow_{\{\tau_{\iota}-\tau_j\}} \iota}B_k\cup\left(\bigcup_{l\in\left(k, \iota\right]}B_l\join^* A_l\right), & \text{if }0\in I\text{ and }\label{eq-mtaux1}\\
T_{\tau_j}^i &=\sats{\alpha\trigger_{\{\tau_\iota-\tau_j\}}\beta}^{\fv\beta}_{\iota,n}, &\text{if }0\notin I.\label{eq-mtaux2}
\end{align}
The tables described by eq.~(\ref{eq-mtaux1}) would coincide with the tables for 
$\alpha\trigger_{\{\tau_\iota-\tau_j\}}\beta$ as in eq.~(\ref{eq-mtaux2}) if we remove 
the $B_l$s inside the union, but we use them to simplify our definition of safety 
as discussed after eq.~(\ref{eq-dual}). 

Likewise, following the intuition provided by eq.~(\ref{eq-dual}), the auxiliary state 
for release $\mathit{Rs}::{\isacharprime}a\ mtaux$ at $i$ is a list of ternary tuples 
$\langle \tau_j, R_\leftL^{j,i}, R_\rightR^j\rangle$ such that
\begin{align}
R_\leftL^{i,j} &=\mathit{if}\ 0\in I\ \mathit{ then }\ \bigcup_{k\in\left[j,\iota\right]}B_k\join^*A_k\ \mathit{ else }\ \bigcup_{k\in\left[j,\iota\right]}A_k\ \text{ and }\label{eq-mraux1} \\
R_\rightR^{i,j} &=
    \begin{cases}
    \mathit{if}\ 0\in I\ \mathit{ then }\ B_j\ \mathit{ else }\ \unitT{n}, &\text{if }\tau_\iota< \tau_j+I,\\
    \sats{\alpha\release_I\beta}^{\dfv_j(\alpha\release_I\beta)}_{j,n}, &\text{if }\tau_\iota>\tau_j+I,\\
    \bigjoin_{k\in [0,\iota]\cap\uparrow_I j}B_k\cup R^{k,i-1}_\leftL, &\text{if }(\tau_\iota-\tau_j)\in I.\\
    \end{cases}
\label{eq-mraux2}
\end{align}

The function that transforms $\alpha::{\isacharprime}a\ \mathit{frm}$ to 
$\alpha_M::{\isacharprime}a\ \mathit{mformula}$ is $\mathit{minit0}$. Our additions 
for this function on trigger and release follow our definition of safety. That is, 
$\mathit{minit0}$ maps $\alpha\mathbin{D}_I\beta$ with 
$\mathbin{D}\in\{\trigger,\release\}$ to 

\isa{MDual\ {\isacharparenleft}$\alpha\neq\neg_F\, \alpha'${\isacharparenright}\ {\isacharparenleft}minit0 n $\alpha${\isacharparenright}\ {\isacharparenleft}$0\in I${\isacharparenright}\ I\ {\isacharparenleft}minit0 n $\beta${\isacharparenright}\ {\isacharparenleft}{\isacharbrackleft}{\isacharbrackright}, {\isacharbrackleft}{\isacharbrackright}{\isacharparenright}\ {\isacharbrackleft}{\isacharbrackright}\ {\isacharbrackleft}{\isacharbrackright}}

\noindent for some $\alpha'$ 
and \isa{MDual\ $\in$ \{MTrigger, MRelease\}} accordingly.\hfill\isalink{https://bitbucket.org/jshs/monpoly/src/b4b63034eca0ccd5783085dececddb6c47cf6f52/thys/Relax_Safety/MFOTL/MFOTL_Monitor.thy\#lines-90}

When the monitor processes the $i$th time-point, the function \isa{mstep} 
outputs the new state $\langle\rho^\alpha_i,\alpha^i_M,n\rangle$. However,
it is also a wrapper for the function $\mathit{meval}$ in charge of updating 
$\alpha^i_M::{\isacharprime}a\ \mathit{mformula}$. Intuitively, $\mathit{meval}$ 
takes $\alpha^{i-1}_M$ and the trace information $\Gamma\, \sigma\, i$ and 
$\tau\, \sigma\, i$ and outputs the updated mformula $\alpha^{i}_M$ and the 
evaluated tables $A_j$ from $j=\rho^{i-1}_\alpha-1$ to $j=\rho^i_\alpha-1$. We
describe its behaviour on auxiliary states below.\hfill\isalink{https://bitbucket.org/jshs/monpoly/src/b4b63034eca0ccd5783085dececddb6c47cf6f52/thys/Relax_Safety/MFOTL/MFOTL_Monitor.thy\#lines-186}

Assuming $\mathit{meval}$ has produced tables $A_\iota$ and $B_\iota$, it uses 
the function \isa{update{\isacharunderscore}trigger} to update the auxiliary state 
$\mathit{Ts}::{\isacharprime}a\ mtaux$ and to possibly output 
$\sats{\alpha\trigger_I\beta}^{\fv\beta}_{\iota,n}$. This function first filters 
$\mathit{Ts}$ by removing all elements whose time-stamps $\tau_j$ are not relevant 
for future time-points, i.e. $\tau_j<\tau_\iota-I$. Then, following eqs.~(\ref{eq-mtaux1}) 
and~(\ref{eq-mtaux2}), it takes the union of the latest $B_\iota\join^*A_\iota$ or $A_\iota$ 
with all elements in $\mathit{Ts}$, depending on whether $0\in I$ or not respectively. Next, 
it adds $B_\iota$ either as a new element $\langle\tau_\iota,B_\iota\rangle\cons \mathit{Ts}$
if this is the first time $\tau_\iota$ is seen, or by joining it with the table in $\mathit{Ts}$ 
time-stamped with $\tau_{\iota-1}=\tau_\iota$, to obtain $T^i_{\tau_{\iota}}$. 
Finally, it joins all the tables $T^i_{\tau_{j}}$ 
in $\mathit{Ts}$ such that $j\in\downarrow_I \iota$.\hfill\isalink{https://bitbucket.org/jshs/monpoly/src/b4b63034eca0ccd5783085dececddb6c47cf6f52/thys/Relax_Safety/MFOTL/MFOTL_Monitor.thy\#lines-165}

The function \isa{update{\isacharunderscore}release} takes the union of 
the latest $B_\iota\join^*A_\iota$ or $A_\iota$ with each $R_\leftL^{j,i-1}$ in 
$\mathit{Rs}::{\isacharprime}a\ mraux$, 
depending on whether $0\in I$ (see eqs.~(\ref{eq-mraux1}) and~(\ref{eq-mraux2})). 
On the third entries $R_\rightR^{i,j}$, it joins $B_\iota\cup R_\leftL^{i-1,j}$ if $(\tau_i-\tau_j)\in I$, 
otherwise, it leaves $R_\rightR^{i,j}$ as it is. Finally, it adds at the end of $\mathit{Rs}$ the tuple 
$\langle \tau_\iota, B_\iota\join^*A_\iota, B_\iota\rangle$ if $0\in I$, or 
$\langle \tau_\iota, A_\iota,\unitT{n}\rangle$ if $0\notin I$. To output the final results
$\sats{\alpha\release_I\beta}^{\dfv_j\alpha\release_I\beta}_{j,n}$, $\mathit{meval}$ 
calls the function \isa{eval{\isacharunderscore}future} which traverses $\mathit{Rs}$ 
and outputs all $R_\rightR^{i,j}$ with $\tau_j+I<\tau_i$ and removes them from $\mathit{Rs}$.\hfill\isalink{https://bitbucket.org/jshs/monpoly/src/b4b63034eca0ccd5783085dececddb6c47cf6f52/thys/Relax_Safety/MFOTL/MFOTL_Monitor.thy\#lines-174}

The only other modification we do on $\mathit{meval}$ for the remaining logical 
connectives is in the disjunction case, where we have replaced every instance
of the traditional union $(\cup)$ with our function \isa{eval{\isacharunderscore}or}.
This is the only place where we perform this substitution.

For our purposes, this concludes the description of the monitoring algorithm. It
mainly consists of the initialisation function $\mathit{minit}$ and the single-step
function $\mathit{mstep}$. Together, they form Verimon's online interface with
the monitored system. Our additions in this regard are mostly conceptual since
we only implement the evaluation of trigger and release as functions 
\isa{update{\isacharunderscore}trigger} and \isa{update{\isacharunderscore}release}.

\section{Monitoring Examples}\label{sec-examples}

We describe various formulas that are safe according to our definition of 
$\ssfv$s but that are not safe in previous Verimon implementations. We 
present them through case-studies. Our examples occur not only in 
monitoring but also in relational databases. 

\paragraph{Quality assessment (operations on globally).} 
A company passes its products through sequential processes $p_1$, $p_2$ 
and $p_3$. Every item passes through all processes. Every minute, 
the company logs the time $\tau$ and the identification number ID of each 
item in process $p_i$. It uses an online monitor to classify its products 
according to their quality. The best ones are those that pass through 
process $p_i$ for exactly $n_i$ minutes and that move to $p_{i+1}$ 
immediately afterwards. The second-best ones are those that pass through 
at least one process $p_i$ in at least $n_i$ minutes. 
The remaining items need to be corrected after production. To identify the 
best ones, the company uses the following specification
\begin{align*}
\isa{best} &\equiv\Big(\globally\nolimits_{[0,n_1)}p_1\pred [\var x]\Big)\ 
\land_F \left(\globally\nolimits_{[n_1,n_2')}p_2\pred [\var x]\right)\ 
\land_F \left(\globally\nolimits_{[n_2',n_3')}p_3\pred [\var x]\right),
\end{align*}
where $n_2'=n_1+n_2$ and $n_3'=n_1+n_2+n_3$. Similarly, to identify
those products with just good quality, they use the specification
\begin{align*}
\isa{good}&\equiv\Big(\globally\nolimits_{[0,n_1)}p_1\pred [\var x]\Big)\ 
\lor_F \left(\globally\nolimits_{[n_1,n_2')}p_2\pred [\var x]\right)\ 
\lor_F\left(\globally\nolimits_{[n_2',n_3')}p_3\pred [\var x]\right).
\end{align*}
Our relaxation of safety, makes both formulas monitorable by encoding 
$\globally_I \left(p_i\pred [\var x]\right)$ as 
$\neg_F(\var\, x=_F\var\, x)\release_I \left(p_i\pred [\var x]\right)$. Specifically,
the safe sets of free variables are $\ssfv(\isa{best})=\{\{x\}\}\neq\emptyset$
and $\ssfv(\isa{good})=\{\emptyset,\{x\}\}\neq\emptyset$. We use the functions
\isa{minit} and \isa{mstep} to monitor \isa{best} through a manually made 
trace. We assume four products with IDs $0, 1, 2$ and $3$ and fix 
$n_1=n_2=n_3=2$. The table below represents said trace and shows the 
monitor's output at each time-point. 
\begin{center}
\begin{tabular}{r | c | c | c | c | c}
time & product\_id 0 & product\_id 1 & product\_id 2 & product\_id 3 & output for \isa{best} \\
\hline
0 & $p_1$ & $p_1$ & $p_1$ & $p_1$ & $\emptyset$\\
1 & $p_1$ & $p_1$ & $p_1$ & $p_1$ & $\emptyset$\\
2 & $p_2$ & $p_2$ & $p_1$ & $p_2$ & $\emptyset$\\
3 & $p_2$ & $p_2$ & $p_2$ & $p_2$ & $\emptyset$\\
4 & $p_3$ & $p_2$ & $p_2$ & $p_3$ & $\emptyset$\\
5 & $p_3$ & $p_3$ & $p_2$ & $p_3$ & $\emptyset$\\
6 & $-$ & $p_3$ & $p_3$ & $-$ & $\{[0],  [3]\}$ @$\tau=0$\\
\end{tabular}
\end{center}

The monitor correctly classifies the products with IDs $0$ and $3$
as the best ones after the first 6 minutes have passed. Before 
that, it outputs the empty set indicating that no product was
one of the best yet. Our Appendix~\ref{sec-appendix} includes the 
formalisation of this trace.
These encodings for \isa{best} and \isa{good} are not safe in any previous 
implementation. Assuming $0<a<b$, the equivalence~\cite{HauserN21}
\begin{align*}
\globally\nolimits_{[a,b)}\alpha \equiv (\eventually\nolimits_{[a,b)}\, \alpha) \land_F \neg_F\, (\eventually\nolimits_{[a,b)} ((\pastly\nolimits_{[0,b)}\alpha)\lor_F (\eventually\nolimits_{[0,b)}\alpha))\land_F \neg_F\, \alpha)
\end{align*}
also produces encodings of \isa{best} and \isa{good} that satisfy 
\isa{safe{\isacharunderscore}formula} but these are clearly longer than
simply using release $\release_I$ as above.

\paragraph{Vaccine refrigeration times (conjunction of negated historically).} 
A company that manufactures dry-ice thermal shipping containers has just 
reported a loss of their refrigeration effectiveness after $m$ hours. This means
that some vaccines transported in those containers are not effective because
the vaccines are very sensitive to thermal conditions. Vaccination centres need
to know which vaccines they can apply. They ask for help from the shipping company 
in charge of transporting the vaccines. This company needs to take into account its
unpacking time that consistently requires $n>0$ minutes. Fortunately, the shipping
company has a log, that among other things, registers ``@$\tau\ (\texttt{travelling},id)$''
if package with ID-number $id$ is travelling at time $\tau$, and annotates 
``@$\tau\ (\texttt{arrived},id)$'' when the package identified with $id$ arrives at a 
centre at time $\tau$. The company can deploy a monitor of the following specification 
over their log to know which packages contain vaccines that are safe to use:
\begin{equation*}
\left(\texttt{arrived}\pred [\var x]\right)\ 
\land_F\ \neg_F\,\historically\nolimits_{[n,m_\epsilon]}\left(\texttt{travelling}\pred [\var x]\right),
\end{equation*}
where $m_\epsilon$ is $m$ in minutes plus some margin of error $\epsilon$,
and $\historically_{I}\left(\texttt{travelling}\pred [\var x]\right)\equiv 
\neg_F(\var\, x=_F\var\, x)\trigger_I \left(\texttt{travelling}\pred [\var x]\right)$. 
Previous work on Verimon+ could tackle an equivalent specification, but it would 
require the less straightforward encoding~\cite{HauserN21}:
\begin{align*}
\historically\nolimits_{[a,b)}\alpha \equiv (\pastly\nolimits_{[a,b)}\, \alpha) \land_F \neg_F\, (\pastly\nolimits_{[a,b)} ((\pastly\nolimits_{[0,b)}\alpha)\lor_F (\eventually\nolimits_{[0,b)}\alpha))\land_F \neg_F\, \alpha),\text{ with }0<a<b.
\end{align*}

\paragraph{Financial crime investigation (historically with many variables).}
Data scientists suspect a vulnerability in the security system of their employer,
a new online bank. They notice the following pattern: various failed payment 
attempts of the same amount from one account to another for 5 consecutive 
minutes. Then, 30 minutes later, a successful payment of the same amount 
between the same accounts. One of the queries to the database that the 
scientists can issue to confirm their suspicions is
\begin{align*}
&(\texttt{approved\_trans\_from\_paid\_to}\pred [\var 0,\var 1,\var 2,\var 3])\\
&\land_F\historically\nolimits_{[30,34]}\, \left(\exists_F\ (\texttt{failed\_trans\_from\_paid\_to}\pred [\var 0,\var 1,\var 2,\var 3])\right).
\end{align*}
In both predicates, $\var 0$ represents the transaction ID, $\var 1$ is the 
account that pays, $\var 2$ corresponds to the amount of money transferred, 
and $\var 3$ denotes the receiving account. The query itself finds all transactions 
that were successful between two accounts $\var 1$ and $\var 3$ at a given 
time-point but that were attempted for 5 consecutive minutes, 30 minutes 
earlier. This not only provides all suspicious receivers, but also all possible
victims and the amount of money they lost per transaction. 

We can codify $\historically_I\, \alpha \equiv \left(\bot_F\, \alpha\right)\trigger_I\alpha$,
where the expression $\bot_F\, \alpha$ denotes the formula $\neg_F(\var 0=_F\var 0)\land_F\cdots\land_F\neg_F(\var n=_F\var n)$ and $n=\nfv\alpha-1$. Monitoring 
such a simple encoding of historically (with $0\notin I$) is possible due to the integration 
of our safety relaxation into the algorithm. Simplifying this further to 
$\bot_F\, \alpha\equiv\neg_F(\const 0=_F\const 0)$ produces an unsafe formula 
because the free variables on both sides of $\trigger_I$ do not coincide. 
We discuss generalisations involving $\ssfv$s and $\dfv$s to achieve this simplification 
in \S\ref{sec-conclusion}.

\paragraph{Monitoring piracy (release operator).} To deal with a recent 
increase in piracy, a shipping company integrates a monitor into 
its tanker tracking system. By standard, their vessels constantly broadcast 
a signal with their location through their automatic identification system, 
which in good conditions arrives every minute, but in adverse 
ones, can take more than 14 hours to update. This signalling system is one 
of the first turned off by pirates because it allows them to sell the tanker's 
contents in nearby unofficial ports. Thus, the company regularly registers the 
ID numbers of all moving ships whose signal is not being received via logs
``@$\tau\ (\texttt{no\_sign},id)$'' and those who are not in the correct
course as ``@$\tau\ (\texttt{off\_route},id)$'', where $id$ is the ID of a 
ship satisfying the respective status. The company decides it should monitor 
their tankers as
\begin{align*}
\isa{pirated} &\equiv\texttt{off\_route}\pred [\var x]\release_{[0,n)}\texttt{no\_sign}\pred [\var x].
\end{align*}
The monitor would send the company a warning after $n$ minutes if it 
observes either of two behaviours: not receiving a signal from the ship 
for the entire $n$ minutes, or not receiving a signal and suddenly receiving 
a position outside of its planned route. This lets the company collaborate with 
local authorities and try to locate their vessels through alternative means. 
Due to our explicit implementation of the bounded 
release operator, the specification above is now straightforwardly
monitorable. Indeed, running the monitor through a manual trace
illustrated in the table below and assuming for simplicity $n=2$
correctly identifies ships with IDs $1$ and $2$ as those possibly pirated.
The formalisation of the trace is available in our Appendix~\ref{sec-appendix}.
\begin{center}
\begin{tabular}{r | c | c | c | c}
time & ship\_id 1 & ship\_id 2 & ship\_id 3 & output for \isa{best} \\
\hline
0 & \texttt{no\_sign} & \texttt{no\_sign} & \texttt{sign} & $\emptyset$\\
1 & \texttt{no\_sign} & \texttt{no\_sign} & \texttt{sign} & $\emptyset$\\
2 & \texttt{no\_sign} & \texttt{no\_sign} & \texttt{sign} & $\emptyset$\\
3 & \texttt{off\_route} & \texttt{no\_sign} & \texttt{sign} & $\{[1],  [2]\}$ @$\tau=0$\\
4 & \texttt{off\_route} & \texttt{no\_sign} & \texttt{sign} & $\{[2]\}$ @$\tau=1$
\end{tabular}
\end{center}


\section{Correctness}\label{sec-correct}
In this section, we show that the integration of our relaxation of safety into
Verimon's monitoring algorithm is correct. We focus 
specifically on describing the proof of correctness for our implementation of 
dual operators. That is, we show that the algorithm described in previous
sections outputs exactly the tables $\sats{\alpha}^{\dfv_i\alpha}_{i,\nfv\alpha}$ 
at time-point $i$ for safe $\alpha$. We also comment on the overall additions 
and adaptations required in the proof of Verimon's correctness to accommodate 
our definition of safety.

First, an Isabelle predicate to describe that a given set of valuations $R$ is a 
proper table with $n$ attributes in $X$ is 
$\vtable n\, X\, R\Leftrightarrow (\forall v\in R.\ \wftuple n\, X\, R)$. The 
formalisation of Verimon~\cite{SchneiderT19}, uses the predicate 
$\qtable n\, X\, P\, Q\, R$ to state correctness of outputs, where $n::\nats$, 
$X::\nats\ set$, $P$ and $Q$ are predicates on valuations, and $R$ is a table. 
It is characterised by\hfill\isalink{https://bitbucket.org/jshs/monpoly/src/b4b63034eca0ccd5783085dececddb6c47cf6f52/thys/Relax_Safety/MFOTL/MFOTL_Table_Correct.thy\#lines-20}
\begin{equation*}
\qtable n\, X\, P\, Q\, R\Leftrightarrow(\vtable n\, X\, R)\land(\forall v.\ P\, v \Rightarrow (v\in R\leftrightarrow Q\, v \land \wftuple n\, X\, v)).
\end{equation*}
In our case, it is typically evaluated to $\qtable n\, (\dfv_i\alpha)\, P\, (\lambda v. \langle\sigma,v,i\rangle\models_M\alpha)\, R$ with $n\geq\nfv\alpha$. If $R$ is 
Verimon's output for $\alpha$ at $i$, it roughly states that 
$R=\sats{\alpha}^{\dfv_i\alpha}_{i,\nfv\alpha}$ modulo $P$ and that 
$\vtable (\nfv\alpha)\, (\dfv_i\alpha)\, R$. In Verimon's and our proof of 
correctness, $P$ is instantiated to a trivially true statement. We do not 
omit $P$ here because our general results require assumptions about it.

Given our use of empty $\dfv$s for vacuously true formulas, the behaviour of 
$\qtable$ on empty sets of variables is relevant for us. The only tables that 
satisfy $\qtable n\, \emptyset\, P\, Q\, R$ are $R=\unitT{n}$ and $R=\emptyset$
so that, assuming $P\, \unit{n}$, if $Q\, \unit{n}$ then $R=\unitT{n}$; 
otherwise, if $\neg\, (Q\, \unit{n})$ then $R=\emptyset$. In fact, the only way that
$\vtable n\, X\, \unitT{n}$ holds is if $X\subseteq\{x\mid x\geq n\}$. Given that we
only use sets $X$ such that $X\subseteq\{x\mid x< n\}$, only $X=\emptyset$ 
fits $\vtable n\, X\, \unitT{n}$. 
%

Since we must join all the tables in the auxiliary states for trigger 
and release, we also relate $\qtable$ and joins $(\join)$. Provided 
$\dfv\sigma\, i\, \alpha\subseteq X$
and $\dfv\sigma\, i\, \beta\subseteq Y$:\hfill~\isalink{https://bitbucket.org/jshs/monpoly/src/b4b63034eca0ccd5783085dececddb6c47cf6f52/thys/Relax_Safety/MFOTL/MFOTL_Table_Correct.thy\#lines-281}
\begin{align*}
\sats{\alpha\land_F\beta}_{i,n}^{X\cup Y} &=\sats{\alpha}^X_{i,n}\join\sats{\beta}^Y_{i,n},\text{ and }\\
\sats{\alpha\land_F\neg_F\, \beta}_{i,n}^{X\cup Y} &=\sats{\alpha}^X_{i,n}\antijoin\sats{\beta}^Y_{i,n},\text{ assuming }Y\subseteq X.
\end{align*}
More general results in terms of $\qtable$ are also available. For instance, if $\pi_X$ is 
the projection $\pi_X\, v = \map\ (\lambda i.\ \mathit{if}\ i\in X\ \mathit{then}\ v\mathbin{!}i\ 
\mathit{else}\, \None)\, [0,\dots,\length v-1]$,
then it holds that $\qtable n\, Z\, P\, Q\, (R_1\join R_2)$ if $\qtable n\, X\, P\, Q_1\, R_1$, 
$\qtable n\, Y\, P\, Q_2\, R_2$, $Z=X\cup Y$ and $\forall v.\, \wftuple n\, Z\, v\land P\, v\Rightarrow (Q\, v\Leftrightarrow Q_1\, (\pi_X\, v)\land Q_2\, (\pi_Y\, v))$. A similar statement is true
for antijoins. Moreover, the join of two tables with the same attributes is simply 
their intersection. By our definition of $\ssfv$s, our $n$-ary join on the auxiliary 
states is made on tables with the same attributes, thus we use the following fact: 
for a finite non-empty set of indices $\mathcal{I}$, $\qtable n\, X\, P\, Q\, \left(\bigcap_{i\in\mathcal{I}} R_i\right)$ holds if $\qtable n\, X\, P\, Q_i\, R_i$ and $\forall v.\, \wftuple n\, X\, v\land P\, v\Rightarrow (Q\, v\Leftrightarrow(\forall i\in\mathcal{I}.\ Q_i\, v))$.\hfill\isalink{https://bitbucket.org/jshs/monpoly/src/b4b63034eca0ccd5783085dececddb6c47cf6f52/thys/Relax_Safety/MFOTL/MFOTL_Table_Correct.thy\#lines-684}

The relationship between \isa{eval{\isacharunderscore}or} and $\qtable$ is also 
relevant for our correctness proof due to our modifications to the algorithm. We 
state here the specific statement: if $\qtable n\, (\dfv_i\alpha)\, P\, (\lambda v.\ \langle\sigma,v,i\rangle\models_M\alpha)\, R_1$ and $\qtable n\, (\dfv_i\beta)\, P\, (\lambda v.\ \langle\sigma,v,i\rangle\models_M\beta)\, R_2$ then $\qtable n\, (\dfv_i(\alpha\lor_F\beta))\, P\, (\lambda v.\ \langle\sigma,v,i\rangle\models_M\alpha\lor_F\beta)\, (\isa{eval{\isacharunderscore}or}\ n\, R_1\, R_2)$,
provided $P\, \unit{n}$ and $\dfv_i\alpha=\emptyset$, $\dfv_i\beta=\emptyset$ or
$\dfv_i\alpha=\dfv_i\beta$. See more results in the Isabelle code.\hfill\isalink{https://bitbucket.org/jshs/monpoly/src/b4b63034eca0ccd5783085dececddb6c47cf6f52/thys/Relax_Safety/MFOTL/MFOTL_Table_Correct.thy\#lines-537}

To state the correctness of the algorithm, Verimon defines an inductive predicate 
$\wfmformula \sigma\, i\, n\, U\, \alpha_M\, \alpha$ where $i,n::\nats$, 
$\alpha_M::{\isacharprime}a\ \mathit{mformula}$, $\alpha::{\isacharprime}a\ \mathit{frm}$ 
and $U$ is a set of valuations. The set $U$ is always instantiated
to the universal set $\mathit{UNIV}$, or the set of all terms of a given type. In fact, 
$P$ in $\qtable$ just checks for membership in $U=\mathit{UNIV}$. The predicate 
$\wfmformula$ is an invariant that holds after initialisation with $\mathit{minit}$ and 
that remains true after each application of $\mathit{mstep}$. It carries all the information
to prove correctness of outputs $R$ at $i$ via the predicate 
$\qtable n\, (\dfv_i\alpha)\, P\, (\lambda v. \langle\sigma,v,i\rangle\models_M\alpha)\, R$. 
We describe 
here only our additions for dual operators.\hfill\isalink{https://bitbucket.org/jshs/monpoly/src/b4b63034eca0ccd5783085dececddb6c47cf6f52/thys/Relax_Safety/MFOTL/MFOTL_Correctness.thy\#lines-401}

\begin{isabellebody}\isanewline
\isacommand{inductive}\ $\wfmformula$\ \isacommand{where}\ {\isasymdots}\isanewline 
\ \ {\isacharbar} Trigger: {\isachardoublequoteopen}$\wfmformula \sigma\ i\ n\ U\, \alpha_m\ \alpha$\ {\isasymLongrightarrow}\ $\wfmformula \sigma\ i\ n\ U\, \beta_m\ \beta$\isanewline
\ \ \ \ {\isasymLongrightarrow}\ $\alpha' = (\mathit{pos}\triangleright_+ \alpha)
\triangleleft \mathit{mem0} \triangleright \alpha$\ {\isasymLongrightarrow}\ $\mathit{mem0}$\ {\isasymleftrightarrow}\ $0\in I$\isanewline 
\ \ \ \ {\isasymLongrightarrow}\ $\issafe\ (\alpha'\trigger_I\beta)$\ {\isasymLongrightarrow}\ wf{\isacharunderscore}mbuf2'\ $\sigma\ i\ n\ U\ \alpha\ \beta$\,buf\ {\isasymLongrightarrow}\ wf{\isacharunderscore}ts\ $\sigma\ i\ \alpha\ \beta$\,nts \isanewline 
\ \ \ \ {\isasymLongrightarrow}\ wf{\isacharunderscore}trigger{\isacharunderscore}aux\ $\sigma\ n\ U$\ pos\ $\alpha$\ mem0\ I\ $\beta$\ aux\ {\isacharparenleft}progress\ $\sigma$\ $(\alpha'\trigger_I\beta)$\ i{\isacharparenright}\isanewline
\ \ \ \ {\isasymLongrightarrow}\ $\wfmformula \sigma\ i\ n\ U$\ {\isacharparenleft}MTrigger pos $\alpha_M$ mem0 I $\beta_M$ buf nts aux{\isacharparenright} $(\alpha'\trigger_I\beta)${\isachardoublequoteclose}\isanewline
\ \ {\isacharbar} Release: {\isachardoublequoteopen}$\wfmformula \sigma\ i\ n\ U\ \alpha_m\ \alpha$\ {\isasymLongrightarrow}\ $\wfmformula \sigma\ i\ n\ U\ \beta_m\ \beta$\isanewline
\ \ \ \ {\isasymLongrightarrow}\ $\alpha' = (\mathit{pos}\triangleright_+ \alpha)
\triangleleft \mathit{mem0} \triangleright \alpha$\ {\isasymLongrightarrow}\ $\mathit{mem0}$\ {\isasymleftrightarrow}\ $0\in I$\isanewline 
\ \ \ \ {\isasymLongrightarrow}\ $\issafe\ (\alpha'\release_I\beta)$\ {\isasymLongrightarrow}\ wf{\isacharunderscore}mbuf2'\ $\sigma\ i\ n\ U\ \alpha\ \beta$\,buf\ {\isasymLongrightarrow}\ wf{\isacharunderscore}ts\ $\sigma\ i\ \alpha\ \beta$\,nts \isanewline 
\ \ \ \ {\isasymLongrightarrow}\ wf{\isacharunderscore}release{\isacharunderscore}aux\ $\sigma\ n\ U$\ pos\ $\alpha$\ mem0\ I\ $\beta$\ aux\ {\isacharparenleft}progress\ $\sigma$\ $(\alpha'\trigger_I\beta)$\ i{\isacharparenright}\isanewline
\ \ \ \ {\isasymLongrightarrow}\ progress $\sigma\ (\alpha'\release_I\beta)\ i\ +$\ length\ aux\ {\isacharequal}\ min\ {\isacharparenleft}progress $\sigma\ \alpha\ i${\isacharparenright} {\isacharparenleft}progress $\sigma\ \beta\ i${\isacharparenright} \isanewline
\ \ \ \ {\isasymLongrightarrow}\ $\wfmformula \sigma\ i\ n\ U$\ {\isacharparenleft}MRelease pos $\alpha_M$ mem0 I $\beta_M$ buf nts aux{\isacharparenright} $(\alpha'\release_I\beta)${\isachardoublequoteclose}\isanewline
\end{isabellebody}
The code above states that if all of the conditions before the last arrow 
(\isa{\isasymLongrightarrow}) are satisfied, then we can assert $\wfmformula$ 
for trigger or release respectively. The function $\mathit{progress}\ \sigma\ \alpha\ i$
is $\rho^\alpha_i$, while $\mathit{pos}\mathbin{\triangleright}_+\alpha$ is our 
Isabelle abbreviation to state that $\alpha$ is not-negated according to the 
boolean $\mathit{pos}$. Similarly, $\alpha\triangleleft \mathit{test} \triangleright \beta$
is just $\alpha$ if $\mathit{test}$ is true, otherwise it is $\beta$. The predicates 
\isa{wf{\isacharunderscore}mbuf2'} and \isa{wf{\isacharunderscore}ts} check 
that the buffer and the corresponding list of time-stamps are well-formed in the 
sense that the buffer has every visited but yet unused table for $\alpha$ and 
$\beta$ while $\mathit{nts}$ has all the corresponding time-stamps $\tau_j$ with
$(\min \rho_i^\alpha\, \rho_i^\beta)\leq j<(\max \rho_i^\alpha\, \rho_i^\beta)$. 
Additionally, we describe below the corresponding invariants \isa{wf{\isacharunderscore}trigger{\isacharunderscore}aux} and \isa{wf{\isacharunderscore}release{\isacharunderscore}aux}
for the auxiliary states.

Recall from eqs.~(\ref{eq-mtaux1}) and~(\ref{eq-mtaux2}) that trigger's auxiliary 
state $\mathit{Ts}$ at time-point $i$ is a list of pairs $\langle \tau_j,T^i_{\tau_j}\rangle$. 
In the formalisation (see also Appendix~\ref{sec-appendix}), we split our definition 
of its invariant \isa{wf{\isacharunderscore}trigger{\isacharunderscore}aux} 
into two parts. First, we state the properties of the time-stamps $\tau_j$: 
they are strictly ordered, less than the latest $\tau_\iota$ and satisfy that 
$\tau_j\in \tau_\iota-I$ or $\tau_j>\tau_\iota-I$ for 
$\iota=(\min \rho_i^\alpha\, \rho_i^\beta)-1$. Conversely, it also affirms that 
every time-stamp satisfying these properties appears in $\mathit{Ts}$. The second 
part asserts correctness. That is, 
$\qtable\, n\, (\fv\beta)\, P\, Q^i_{\tau_j}\, T^i_{\tau_j}$ where $Q^i_{\tau_j}\, v 
\Leftrightarrow \langle\sigma,v,\iota\rangle\models_M\alpha\trigger_{\{\tau_\iota-\tau_j\}}\beta$ 
if $0\notin I$, and $Q^i_{\tau_j}\, v\Leftrightarrow(\forall k\leq \iota.\ \tau_k=\tau_j\Rightarrow\langle\sigma,v,k\rangle\models_M\beta\lor(\exists l\in\left(k,\iota\right].\ \langle\sigma,v,l\rangle\models_M))$ if $0\in I$.~\hfill\isalink{https://bitbucket.org/jshs/monpoly/src/b4b63034eca0ccd5783085dececddb6c47cf6f52/thys/Relax_Safety/MFOTL/MFOTL_Correctness.thy\#lines-218}

The invariant \isa{wf{\isacharunderscore}release{\isacharunderscore}aux} 
for release's auxiliary state $\mathit{Rs}$ is more verbose. Assuming $0\in I$, it 
asserts $\qtable n\, (\fv\beta)\, P\, Q_{\leftL,0\in I}^{i,j}\, R_\leftL^{i,j}$ and 
$\qtable n\, (\fv\beta)\, P\, Q_{\rightR,0\in I}^{i,j}\, R_\rightR^{i,j}$ where 
$Q_{\leftL,0\in I}^{i,j}$ and $Q_{\rightR,0\in I}^{i,j}$ describe the first parts 
of eqs.~(\ref{eq-mraux1}) and~(\ref{eq-mraux2}). That is,
\begin{align*}
Q_{\leftL,0\in I}^{i,j}\, v &\Leftrightarrow (\exists k\in\left[j,\iota\right).\ \langle\sigma,v,k\rangle\models_M\beta\land
\langle\sigma,v,k\rangle\models_M\alpha)\text{ and }\\
Q_{\rightR,0\in I}^{i,j}\, v &\Leftrightarrow (\forall k\in\left[j,\iota\right).\ (\tau_k-\tau_j)\in I\Rightarrow \langle\sigma,v,k\rangle\models_M\beta\lor Q_{\leftL,0\in I}^{i,k}\, v).
\end{align*}
However, when $0\notin I$, the invariant asserts $\qtable n\, (\fv\beta)\, P\, Q_{\leftL,0\notin I}^{i,j}\, R_\leftL^{i,j}$ where
\begin{equation*}
Q_{\leftL,0\notin I}^{i,j}\, v \Leftrightarrow (\exists k\in\left[j,\iota\right).\ \langle\sigma,v,k\rangle\models_M\alpha).
\end{equation*}
For the right table, if $\tau_\iota<\tau_j+I$, it simply asserts $R_R^{i,j}=\unitT{n}$. 
However, if $(\tau_\iota-\tau_j)\in I$, the invariant states that $\qtable n\, (\fv\beta)\, P\, Q_{\rightR,0\notin I}^{i,j}\, R_\rightR^{i,j}$ where
\begin{equation*}
Q_{\rightR,0\notin I}^{i,j}\, v \Leftrightarrow (\forall k\in\left[j,\iota\right).\ (\tau_k-\tau_j)\in I\Rightarrow \langle\sigma,v,k\rangle\models_M\beta\lor Q_{\leftL,0\notin I}^{i,k}\, v).
\end{equation*}
Finally, the case when $\tau_j+I<\tau_i$ asserts $\qtable n\, (\dfv_j(\alpha\release_I\beta))\, P\, Q_{\rightR,0\notin I}^{i,j}\, R_\rightR^{i,j}$.\hfill\isalink{https://bitbucket.org/jshs/monpoly/src/b4b63034eca0ccd5783085dececddb6c47cf6f52/thys/Relax_Safety/MFOTL/MFOTL_Correctness.thy\#lines-258}

We then adapt Verimon's proof of correctness~\cite{SchneiderBKT19} for the
monitored formula $\alpha$. It consists of two facts: \tagx[$a$]{prpty:a} after 
initialisation, $\alpha_M^0$ satisfies $\wfmformula$, and \tagx[$b$]{prpty:b} 
whenever $\alpha_M^{i-1}$ satisfies $\wfmformula$, then after an execution 
of $\mathit{meval}$, the new $\alpha_M^i$ also satisfies $\wfmformula$ and
all the outputs of $\mathit{meval}$ are correct. 

 
At initialisation (\ref{prpty:a}), our relaxation of safety allows us to replace
\isa{safe{\isacharunderscore}formula} with $\issafe$. Also, due to the 
condition $P\, \unit{n}$ in our results about $\qtable$ and 
\isa{eval{\isacharunderscore}or}, we need to assume $\unit{n}\in U$ 
where $U$ is the set referred in $\wfmformula$. Formally, our correctness 
of initialisation states that if $\issafe\alpha$, $\unit{n}$ is an element of the 
set $U$, and the free variables of $\alpha$ are all less than $n$, then 
$\wfmformula\sigma\, 0\, n\, U\, (\mathit{minit0}\, n\, \alpha)\, \alpha$.
The proof is a typical application of inductive reasoning but not fully 
automatic since we need case distinctions for negations, conjunctions 
and dual operators.\hfill\isalink{https://bitbucket.org/jshs/monpoly/src/b4b63034eca0ccd5783085dececddb6c47cf6f52/thys/Relax_Safety/MFOTL/MFOTL_Correctness.thy\#lines-497}

The addition of $\dfv$s and $\ssfv$s produces more changes in Verimon's 
invariant preservation proof (\ref{prpty:b}) than in the initialisation proof 
(\ref{prpty:a}). In many preliminary definitions and lemmas, 
including that of \isa{wf{\isacharunderscore}mbuf2'}, we replace the argument 
$\fv\alpha$ with $\dfv_i\, \alpha$. In others, a less straightforward substitution
is necessary. For instance, in the auxiliary state for until, we do not simply use
$\dfv_i\, \alpha$ but the union of various $\dfv$s. This reverberates in the proof
of correctness of the auxiliary state which quintuples its size from 28 to 140 lines 
of code due to the various cases generated by both $\dfv$s and $\ssfv$s.\hfill\isalink{https://bitbucket.org/jshs/monpoly/src/b4b63034eca0ccd5783085dececddb6c47cf6f52/thys/Relax_Safety/MFOTL/MFOTL_Correctness.thy\#lines-769}

Our proof of correctness for trigger's auxiliary state consists of a step-wise
decomposition of \isa{update{\isacharunderscore}trigger} and stating, at 
each step, what the tables in the auxiliary state satisfy in terms of $\qtable$.
It is 440 lines of code long, double the size of the proof for since due to the 
case distinctions $0\in I$ and $0\notin I$.~\isalink{https://bitbucket.org/jshs/monpoly/src/b4b63034eca0ccd5783085dececddb6c47cf6f52/thys/Relax_Safety/MFOTL/MFOTL_Correctness.thy\#lines-1101} These also appear in the corresponding 
proof of correctness for release and dictate its main structure. On one 
hand, the case $0\in I$ for release is further split into whether the auxiliary state 
was previously an empty list or not. On the other hand, the assumption $0\notin I$ 
requires analysing the different cases $\tau_\iota<\tau_j+I$, $(\tau_\iota-\tau_j)\in I$ 
and $\tau_j+I<\tau_i$ as above. Each of these also considers the emptiness of 
the auxiliary state at the previous time-point.\hfill\isalink{https://bitbucket.org/jshs/monpoly/src/b4b63034eca0ccd5783085dececddb6c47cf6f52/thys/Relax_Safety/MFOTL/MFOTL_Correctness.thy\#lines-1554}

Finally, the theorem that uses all of these correctness results and modifications
is the invariant preservation proof (\ref{prpty:b}) above. In more detail, it states that 
if we start with $\alpha_M$ such that $\wfmformula\sigma\, i\, n\, U\, \alpha_M\, \alpha$ 
and $\mathit{meval}\, n\, \tau_i\, \Gamma_i\, \alpha_M=(\mathit{outputs},\alpha_{M'})$, 
then $\wfmformula\sigma\, (i+1)\, n\, U\, \alpha_{M'}\, \alpha$ and 
$\qtable n\, (\dfv_j\alpha)\, P\, (\lambda v.\ \langle\sigma,v,j\rangle\models_M\alpha) R_j$
for each $R_j\in\mathit{outputs}$ with $j\in [\rho_i^\alpha,\rho_{i+1}^\alpha)$. 
We do its proof over the structure of $\alpha_M::{\isacharprime}a\ \isa{mformula}$.
The base step for equalities and inequalities requires some case distinctions
and our results about $\qtable$ and $\unitT{n}$. The inductive steps require
mostly the same argument: from $\wfmformula$ we know most of the information
to prove $\qtable$, we supply it to our preliminary lemmas like
the correctness of auxiliary states and buffers, finally we use these results
and the inductive definition of $\wfmformula$ to obtain our desired conclusion.
Our separation of preliminary lemmas from the main body of the proof of (\ref{prpty:b})
highly increases readability of this long argument.\hfill\isalink{https://bitbucket.org/jshs/monpoly/src/b4b63034eca0ccd5783085dececddb6c47cf6f52/thys/Relax_Safety/MFOTL/MFOTL_Correctness.thy\#lines-2048}

This concludes our description of the correctness argument from definitions
to explanations on the proof structure. Their formalisation is one of our major 
contributions. Our additions on properties about $\qtable$ and $\dfv$s consists 
of approximately 350 lines of code, while those to the proof of correctness are 
more than 1500. This still does not take into account the additions on other 
already existing results, like the modifications to the proof of correctness of 
until's auxiliary state. In total, the correctness argument changed from 
approximately 1000 lines of code to more than 3000.


\section{Conclusion}\label{sec-conclusion}

We defined a fragment of MFOTL-formulas guaranteeing their 
relational-algebra representations to be computed through well-known
table operations. For this, we introduced the set of safe sets of free 
variables ($\ssfv$) of a formula which collects all possible allowed 
attributes of the formula's table-representations over time. The fragment 
required this set to be non-empty. We argued that this 
\emph{safe} fragment is larger than others from previous work on temporal 
properties and pointed to our Isabelle/HOL proof of this fact. We integrated 
our relaxation of safety into a monitoring algorithm. The formal 
verification of this integration was possible due to our newly introduced 
concept of dynamic free variables ($\dfv$) of the 
monitored specification. We also extended the algorithm with
concrete syntax and functions to monitor MFOTL dual operators
\emph{trigger} and \emph{release}. The combination of $\ssfv$s, $\dfv$s and dual 
operators enabled the algorithm to monitor more specifications, some of which we illustrated via examples.

\paragraph{Future work.} Our relaxation of safety can be generalised in
various ways. The simplest of these add cases to our definition of $\ssfv$s.
For instance, asserting $\ssfv(t=_Ft)=\{\emptyset\}$ is possible since we 
can map it to $\unitT{n}$. However, doing this has unintended consequences
that forces us to rethink other cases, e.g. the conjunction 
$\var x=_F\var x\land_F\var x=_F\var y$ would become safe under the current
definition. Furthermore, we lose some ``nice'' properties like 
$\issafe\alpha\Leftrightarrow\fv\alpha\in\ssfv\alpha$. It is also unsatisfactory
that safety for $(\neg_F\, \alpha)\since_I\beta$ only requires $\ssfv\beta=\{\beta\}$
and $\fv\alpha\subseteq\fv\beta$ while that for $(\neg_F\, \alpha)\until_I\beta$ needs 
the stronger condition $\ssfv\beta=\{\beta\}$ and $\ssfv\alpha=\{\fv\alpha\}$. A 
reimplementation of the monitoring functions for until would alleviate this situation. 

An orthogonal development replaces every instance of union, $(\cup)$ or 
$(\bigcup)$, in the implementation with our generalised \isa{eval{\isacharunderscore}or}. 
This would allow us to change our definition of safety so that more 
attributes are available for the right-hand-side formula in temporal operators 
(i.e. $\ssfv\beta\subseteq\{\emptyset,\fv\beta\}$). Consequently, this would allow 
us to write combinations of them, e.g. 
$\pastly_I \left(p\pred \mathit{xs}\land_F\historically_Jq\pred ys\right)$,
where $\pastly_I \alpha\equiv\top\since_I\alpha$.

A different avenue of research follows the standard approach in logic 
and the database community and defines a series of transformations that
determine if a formula is equivalent to a safe one~\cite{AbiteboulHV95,GelderT91,Raszyk22,RaszykBKT22}. 
If such a transformation is obtained, formally verified and implemented, 
its integration into Verimon would mean that many more future-bounded 
formulas would be monitorable.

With the long-term view of developing a more trustworthy, expressive and 
efficient monitor than other non-verified tools, we intend to integrate our 
relaxation of safety into Verimon+~\cite{ZinggKRST22}. This 
requires adding more complex terms inside equalities and inequalities 
that contain additions, multiplications, divisions and type castings. Additionally, 
safety would need to be defined for aggregations like sum or average, dynamic 
operators from metric first-order dynamic logic, and recursive let operations. A 
first attempt and its not-yet complete integration into an old Verimon+ 
version~\cite{BasinDHKR0T20} are available online.\hfill\isalink{https://bitbucket.org/jshs/monpoly/src/b4b63034eca0ccd5783085dececddb6c47cf6f52/thys/Relax_Safety/MFODL/MFODL_Safety.thy\#lines-428}

\paragraph{Acknowledgements} I would like to thank Dmitriy Traytel
and Joshua Schneider for discussions and pointers. Martin Raszyk came
up with the idea of using a set of sets representing tables' attributes to 
define safety and coined the term ``dynamic free variables'', I highly 
appreciate his feedback throughout the implementation 
of this project. I also thank Leonardo Lima, Rafael Castro G. Silva and Phebe
L. Bonilla Prado for their comments on early drafts of the paper. The work 
itself is funded by a Novo Nordisk Fonden Start Package Grant~(NNF20OC0063462).

\newpage
\bibliographystyle{abbrv}
\bibliography{main}

\newpage
\appendix
\section{Appendix: Formal Definitions}\label{sec-appendix}

We supply our Isabelle/HOL definitions of dynamic free variables, 
safe sets of free variables, a combination of Verimon's 
\isa{safe{\isacharunderscore}{\kern0pt}formula} 
predicates~\cite{HauserN21,SchneiderBKT19,ZinggKRST22}, and trigger 
and release's invariants for auxiliary states. We also provide the formalisation
of the traces in \S\ref{sec-examples} to showcase the working monitoring algorithm.

\paragraph{Dynamic free variables.}  In the code below, the notation 
\isa{mem\ I\ {\isadigit0}} represents $0\in I$, for interval $I$. Also, \isa{Suc} is 
the successor function on natural numbers. The first two definitions correspond 
to the sets $\downarrow_I i$ and $\uparrow_I i$ of \S\ref{sec-safe} respectively.

\begin{isabellebody}\scriptsize\isanewline
\isacommand{definition}\isamarkupfalse%
\ {\isachardoublequoteopen}down{\isacharunderscore}{\kern0pt}cl{\isacharunderscore}{\kern0pt}ivl\ {\isasymsigma}\ I\ i\ {\isasymequiv}\ {\isacharbraceleft}{\kern0pt}j\ {\isacharbar}{\kern0pt}j{\isachardot}{\kern0pt}\ j\ {\isasymle}\ i\ {\isasymand}\ mem\ I\ {\isacharparenleft}{\kern0pt}{\isacharparenleft}{\kern0pt}{\isasymtau}\ {\isasymsigma}\ i\ {\isacharminus}{\kern0pt}\ {\isasymtau}\ {\isasymsigma}\ j{\isacharparenright}{\kern0pt}{\isacharparenright}{\kern0pt}{\isacharbraceright}{\kern0pt}{\isachardoublequoteclose}
\end{isabellebody}

\begin{isabellebody}\scriptsize\isanewline
\isacommand{definition}\isamarkupfalse%
\ {\isachardoublequoteopen}up{\isacharunderscore}{\kern0pt}cl{\isacharunderscore}{\kern0pt}ivl\ {\isasymsigma}\ I\ i\ {\isasymequiv}\ {\isacharbraceleft}{\kern0pt}j\ {\isacharbar}{\kern0pt}j{\isachardot}{\kern0pt}\ i\ {\isasymle}\ j\ {\isasymand}\ mem\ I\ {\isacharparenleft}{\kern0pt}{\isacharparenleft}{\kern0pt}{\isasymtau}\ {\isasymsigma}\ j\ {\isacharminus}{\kern0pt}\ {\isasymtau}\ {\isasymsigma}\ i{\isacharparenright}{\kern0pt}{\isacharparenright}{\kern0pt}{\isacharbraceright}{\kern0pt}{\isachardoublequoteclose}
\end{isabellebody}

\begin{isabellebody}\scriptsize\isanewline
\isacommand{fun}\isamarkupfalse%
\ dfv\ {\isacharcolon}{\kern0pt}{\isacharcolon}{\kern0pt}\ {\isachardoublequoteopen}{\isacharparenleft}{\kern0pt}char\ list\ {\isasymtimes}\ {\isacharprime}{\kern0pt}a\ list{\isacharparenright}{\kern0pt}\ set\ trace\ {\isasymRightarrow}\ nat\ {\isasymRightarrow}\ {\isacharprime}{\kern0pt}a\ MFOTL{\isacharunderscore}{\kern0pt}Formula{\isachardot}{\kern0pt}formula\ {\isasymRightarrow}\ nat\ set{\isachardoublequoteclose}\ \isanewline
\ \ \isakeyword{where}\ {\isachardoublequoteopen}dfv\ {\isasymsigma}\ i\ {\isacharparenleft}{\kern0pt}p\ {\isasymdagger}\ ts{\isacharparenright}{\kern0pt}\ {\isacharequal}{\kern0pt}\ FV\ {\isacharparenleft}{\kern0pt}p\ {\isasymdagger}\ ts{\isacharparenright}{\kern0pt}{\isachardoublequoteclose}\isanewline
\ \ {\isacharbar}{\kern0pt}\ {\isachardoublequoteopen}dfv\ {\isasymsigma}\ i\ {\isacharparenleft}{\kern0pt}t{\isadigit{1}}\ {\isacharequal}{\kern0pt}\isactrlsub F\ t{\isadigit{2}}{\isacharparenright}{\kern0pt}\ {\isacharequal}{\kern0pt}\ FV\ {\isacharparenleft}{\kern0pt}t{\isadigit{1}}\ {\isacharequal}{\kern0pt}\isactrlsub F\ t{\isadigit{2}}{\isacharparenright}{\kern0pt}{\isachardoublequoteclose}\isanewline
\ \ {\isacharbar}{\kern0pt}\ {\isachardoublequoteopen}dfv\ {\isasymsigma}\ i\ {\isacharparenleft}{\kern0pt}{\isasymnot}\isactrlsub F\ {\isasymalpha}{\isacharparenright}{\kern0pt}\ {\isacharequal}{\kern0pt}\ dfv\ {\isasymsigma}\ i\ {\isasymalpha}{\isachardoublequoteclose}\isanewline
\ \ {\isacharbar}{\kern0pt}\ {\isachardoublequoteopen}dfv\ {\isasymsigma}\ i\ {\isacharparenleft}{\kern0pt}{\isasymalpha}\ {\isasymand}\isactrlsub F\ {\isasymbeta}{\isacharparenright}{\kern0pt}\ {\isacharequal}{\kern0pt}\ dfv\ {\isasymsigma}\ i\ {\isasymalpha}\ {\isasymunion}\ dfv\ {\isasymsigma}\ i\ {\isasymbeta}{\isachardoublequoteclose}\isanewline
\ \ {\isacharbar}{\kern0pt}\ {\isachardoublequoteopen}dfv\ {\isasymsigma}\ i\ {\isacharparenleft}{\kern0pt}{\isasymalpha}\ {\isasymor}\isactrlsub F\ {\isasymbeta}{\isacharparenright}{\kern0pt}\ {\isacharequal}{\kern0pt}\ {\isacharparenleft}{\kern0pt}if\ dfv\ {\isasymsigma}\ i\ {\isasymalpha}\ {\isacharequal}{\kern0pt}\ {\isacharbraceleft}{\kern0pt}{\isacharbraceright}{\kern0pt}\ \isanewline
\ \ \ \ \ \ then\ if\ {\isacharbraceleft}{\kern0pt}v{\isachardot}{\kern0pt}\ {\isasymlangle}{\isasymsigma}{\isacharcomma}{\kern0pt}\ v{\isacharcomma}{\kern0pt}\ i{\isasymrangle}\ {\isasymTurnstile}\ {\isasymalpha}{\isacharbraceright}{\kern0pt}\ {\isacharequal}{\kern0pt}\ {\isacharbraceleft}{\kern0pt}{\isacharbraceright}{\kern0pt}\ then\ dfv\ {\isasymsigma}\ i\ {\isasymbeta}\ else\ {\isacharbraceleft}{\kern0pt}{\isacharbraceright}{\kern0pt}\isanewline
\ \ \ \ \ \ else\ if\ dfv\ {\isasymsigma}\ i\ {\isasymbeta}\ {\isacharequal}{\kern0pt}\ {\isacharbraceleft}{\kern0pt}{\isacharbraceright}{\kern0pt}\ then\ if\ {\isacharbraceleft}{\kern0pt}v{\isachardot}{\kern0pt}\ {\isasymlangle}{\isasymsigma}{\isacharcomma}{\kern0pt}\ v{\isacharcomma}{\kern0pt}\ i{\isasymrangle}\ {\isasymTurnstile}\ {\isasymbeta}{\isacharbraceright}{\kern0pt}\ {\isacharequal}{\kern0pt}\ {\isacharbraceleft}{\kern0pt}{\isacharbraceright}{\kern0pt}\ then\ dfv\ {\isasymsigma}\ i\ {\isasymalpha}\ else\ {\isacharbraceleft}{\kern0pt}{\isacharbraceright}{\kern0pt}\isanewline
\ \ \ \ \ \ else\ dfv\ {\isasymsigma}\ i\ {\isasymalpha}\ {\isasymunion}\ dfv\ {\isasymsigma}\ i\ {\isasymbeta}{\isacharparenright}{\kern0pt}{\isachardoublequoteclose}\isanewline
\ \ {\isacharbar}{\kern0pt}\ {\isachardoublequoteopen}dfv\ {\isasymsigma}\ i\ {\isacharparenleft}{\kern0pt}{\isasymexists}\isactrlsub F\ {\isasymalpha}{\isacharparenright}{\kern0pt}\ {\isacharequal}{\kern0pt}\ {\isacharparenleft}{\kern0pt}{\isasymlambda}x{\isacharcolon}{\kern0pt}{\isacharcolon}{\kern0pt}nat{\isachardot}{\kern0pt}\ x\ {\isacharminus}{\kern0pt}\ {\isadigit{1}}{\isacharparenright}{\kern0pt}\ {\isacharbackquote}{\kern0pt}\ {\isacharparenleft}{\kern0pt}{\isacharparenleft}{\kern0pt}dfv\ {\isasymsigma}\ i\ {\isasymalpha}{\isacharparenright}{\kern0pt}\ {\isacharminus}{\kern0pt}\ {\isacharbraceleft}{\kern0pt}{\isadigit{0}}{\isacharbraceright}{\kern0pt}{\isacharparenright}{\kern0pt}{\isachardoublequoteclose}\isanewline
\ \ {\isacharbar}{\kern0pt}\ {\isachardoublequoteopen}dfv\ {\isasymsigma}\ i\ {\isacharparenleft}{\kern0pt}\isactrlbold Y\ I\ {\isasymalpha}{\isacharparenright}{\kern0pt}\ {\isacharequal}{\kern0pt}\ {\isacharparenleft}{\kern0pt}if\ i{\isacharequal}{\kern0pt}{\isadigit{0}}\ then\ FV\ {\isasymalpha}\ else\ dfv\ {\isasymsigma}\ {\isacharparenleft}{\kern0pt}i{\isacharminus}{\kern0pt}{\isadigit{1}}{\isacharparenright}{\kern0pt}\ {\isasymalpha}{\isacharparenright}{\kern0pt}{\isachardoublequoteclose}\isanewline
\ \ {\isacharbar}{\kern0pt}\ {\isachardoublequoteopen}dfv\ {\isasymsigma}\ i\ {\isacharparenleft}{\kern0pt}\isactrlbold X\ I\ {\isasymalpha}{\isacharparenright}{\kern0pt}\ {\isacharequal}{\kern0pt}\ dfv\ {\isasymsigma}\ {\isacharparenleft}{\kern0pt}Suc\ i{\isacharparenright}{\kern0pt}\ {\isasymalpha}{\isachardoublequoteclose}\isanewline
\ \ {\isacharbar}{\kern0pt}\ {\isachardoublequoteopen}dfv\ {\isasymsigma}\ i\ {\isacharparenleft}{\kern0pt}{\isasymalpha}\ \isactrlbold S\ I\ {\isasymbeta}{\isacharparenright}{\kern0pt}\ {\isacharequal}{\kern0pt}\ \isanewline
\ \ \ \ \ \ {\isacharparenleft}{\kern0pt}let\ satisf{\isacharunderscore}{\kern0pt}at\ {\isacharequal}{\kern0pt}\ {\isasymlambda}j{\isachardot}{\kern0pt}\ {\isasymexists}v{\isachardot}{\kern0pt}\ {\isasymlangle}{\isasymsigma}{\isacharcomma}{\kern0pt}\ v{\isacharcomma}{\kern0pt}\ j{\isasymrangle}\ {\isasymTurnstile}\ {\isasymbeta}\ {\isasymand}\ {\isacharparenleft}{\kern0pt}{\isasymforall}k{\isasymin}{\isacharbraceleft}{\kern0pt}j{\isacharless}{\kern0pt}{\isachardot}{\kern0pt}{\isachardot}{\kern0pt}i{\isacharbraceright}{\kern0pt}{\isachardot}{\kern0pt}\ {\isasymlangle}{\isasymsigma}{\isacharcomma}{\kern0pt}\ v{\isacharcomma}{\kern0pt}\ k{\isasymrangle}\ {\isasymTurnstile}\ {\isasymalpha}{\isacharparenright}{\kern0pt}\ in\isanewline
\ \ \ \ \ \ \ \ {\isacharparenleft}{\kern0pt}if\ {\isacharparenleft}{\kern0pt}{\isasymforall}j{\isasymin}down{\isacharunderscore}{\kern0pt}cl{\isacharunderscore}{\kern0pt}ivl\ {\isasymsigma}\ I\ i{\isachardot}{\kern0pt}\ {\isasymnot}\ satisf{\isacharunderscore}{\kern0pt}at\ j{\isacharparenright}{\kern0pt}\ then\ FV\ {\isacharparenleft}{\kern0pt}{\isasymalpha}\ \isactrlbold S\ I\ {\isasymbeta}{\isacharparenright}{\kern0pt}\isanewline
\ \ \ \ \ \ \ \ \ else\ {\isacharparenleft}{\kern0pt}let\ J\ {\isacharequal}{\kern0pt}\ {\isacharbraceleft}{\kern0pt}j{\isasymin}down{\isacharunderscore}{\kern0pt}cl{\isacharunderscore}{\kern0pt}ivl\ {\isasymsigma}\ I\ i{\isachardot}{\kern0pt}\ satisf{\isacharunderscore}{\kern0pt}at\ j{\isacharbraceright}{\kern0pt}{\isacharsemicolon}{\kern0pt}\ K\ {\isacharequal}{\kern0pt}\ {\isasymUnion}{\isacharbraceleft}{\kern0pt}{\isacharbraceleft}{\kern0pt}j{\isacharless}{\kern0pt}{\isachardot}{\kern0pt}{\isachardot}{\kern0pt}i{\isacharbraceright}{\kern0pt}{\isacharbar}{\kern0pt}j{\isachardot}{\kern0pt}\ j\ {\isasymin}\ J{\isacharbraceright}{\kern0pt}\ in\isanewline
\ \ \ \ \ \ \ \ \ \ {\isacharparenleft}{\kern0pt}{\isasymUnion}{\isacharbraceleft}{\kern0pt}{\isacharparenleft}{\kern0pt}dfv\ {\isasymsigma}\ k\ {\isasymalpha}{\isacharparenright}{\kern0pt}{\isacharbar}{\kern0pt}k{\isachardot}{\kern0pt}\ k\ {\isasymin}\ K{\isacharbraceright}{\kern0pt}{\isacharparenright}{\kern0pt}\ {\isasymunion}\ {\isacharparenleft}{\kern0pt}{\isasymUnion}{\isacharbraceleft}{\kern0pt}{\isacharparenleft}{\kern0pt}dfv\ {\isasymsigma}\ j\ {\isasymbeta}{\isacharparenright}{\kern0pt}{\isacharbar}{\kern0pt}j{\isachardot}{\kern0pt}\ j\ {\isasymin}\ J{\isacharbraceright}{\kern0pt}{\isacharparenright}{\kern0pt}{\isacharparenright}{\kern0pt}{\isacharparenright}{\kern0pt}{\isacharparenright}{\kern0pt}{\isachardoublequoteclose}\isanewline
\ \ {\isacharbar}{\kern0pt}\ {\isachardoublequoteopen}dfv\ {\isasymsigma}\ i\ {\isacharparenleft}{\kern0pt}{\isasymalpha}\ \isactrlbold U\ I\ {\isasymbeta}{\isacharparenright}{\kern0pt}\ {\isacharequal}{\kern0pt}\ \isanewline
\ \ \ \ \ \ {\isacharparenleft}{\kern0pt}let\ satisf{\isacharunderscore}{\kern0pt}at\ {\isacharequal}{\kern0pt}\ {\isasymlambda}j{\isachardot}{\kern0pt}{\isasymexists}v{\isachardot}{\kern0pt}\ {\isasymlangle}{\isasymsigma}{\isacharcomma}{\kern0pt}\ v{\isacharcomma}{\kern0pt}\ j{\isasymrangle}\ {\isasymTurnstile}\ {\isasymbeta}\ {\isasymand}\ {\isacharparenleft}{\kern0pt}{\isasymforall}k{\isasymin}{\isacharbraceleft}{\kern0pt}i{\isachardot}{\kern0pt}{\isachardot}{\kern0pt}{\isacharless}{\kern0pt}j{\isacharbraceright}{\kern0pt}{\isachardot}{\kern0pt}\ {\isasymlangle}{\isasymsigma}{\isacharcomma}{\kern0pt}\ v{\isacharcomma}{\kern0pt}\ k{\isasymrangle}\ {\isasymTurnstile}\ {\isasymalpha}{\isacharparenright}{\kern0pt}\ in\isanewline
\ \ \ \ \ \ \ \ {\isacharparenleft}{\kern0pt}if\ {\isacharparenleft}{\kern0pt}{\isasymforall}j{\isasymin}up{\isacharunderscore}{\kern0pt}cl{\isacharunderscore}{\kern0pt}ivl\ {\isasymsigma}\ I\ i{\isachardot}{\kern0pt}\ {\isasymnot}\ satisf{\isacharunderscore}{\kern0pt}at\ j{\isacharparenright}{\kern0pt}\ then\ FV\ {\isacharparenleft}{\kern0pt}{\isasymalpha}\ \isactrlbold U\ I\ {\isasymbeta}{\isacharparenright}{\kern0pt}\isanewline
\ \ \ \ \ \ \ \ \ else\ {\isacharparenleft}{\kern0pt}let\ J\ {\isacharequal}{\kern0pt}\ {\isacharbraceleft}{\kern0pt}j{\isasymin}up{\isacharunderscore}{\kern0pt}cl{\isacharunderscore}{\kern0pt}ivl\ {\isasymsigma}\ I\ i{\isachardot}{\kern0pt}\ satisf{\isacharunderscore}{\kern0pt}at\ j{\isacharbraceright}{\kern0pt}{\isacharsemicolon}{\kern0pt}\ K\ {\isacharequal}{\kern0pt}\ {\isasymUnion}{\isacharbraceleft}{\kern0pt}{\isacharbraceleft}{\kern0pt}i{\isachardot}{\kern0pt}{\isachardot}{\kern0pt}{\isacharless}{\kern0pt}j{\isacharbraceright}{\kern0pt}{\isacharbar}{\kern0pt}j{\isachardot}{\kern0pt}\ j\ {\isasymin}\ J{\isacharbraceright}{\kern0pt}\ in\isanewline
\ \ \ \ \ \ \ \ \ \ {\isacharparenleft}{\kern0pt}{\isasymUnion}{\isacharbraceleft}{\kern0pt}{\isacharparenleft}{\kern0pt}dfv\ {\isasymsigma}\ k\ {\isasymalpha}{\isacharparenright}{\kern0pt}{\isacharbar}{\kern0pt}k{\isachardot}{\kern0pt}\ k\ {\isasymin}\ K{\isacharbraceright}{\kern0pt}{\isacharparenright}{\kern0pt}\ {\isasymunion}\ {\isacharparenleft}{\kern0pt}{\isasymUnion}{\isacharbraceleft}{\kern0pt}{\isacharparenleft}{\kern0pt}dfv\ {\isasymsigma}\ j\ {\isasymbeta}{\isacharparenright}{\kern0pt}{\isacharbar}{\kern0pt}j{\isachardot}{\kern0pt}\ j\ {\isasymin}\ J{\isacharbraceright}{\kern0pt}{\isacharparenright}{\kern0pt}{\isacharparenright}{\kern0pt}{\isacharparenright}{\kern0pt}{\isacharparenright}{\kern0pt}{\isachardoublequoteclose}\isanewline
\ \ {\isacharbar}{\kern0pt}\ {\isachardoublequoteopen}dfv\ {\isasymsigma}\ i\ {\isacharparenleft}{\kern0pt}{\isasymalpha}\ \isactrlbold T\ I\ {\isasymbeta}{\isacharparenright}{\kern0pt}\ {\isacharequal}{\kern0pt}\ \isanewline
\ \ \ \ \ \ {\isacharparenleft}{\kern0pt}let\ satisf{\isacharunderscore}{\kern0pt}at\ {\isacharequal}{\kern0pt}\ {\isasymlambda}v\ j{\isachardot}{\kern0pt}\ {\isasymlangle}{\isasymsigma}{\isacharcomma}{\kern0pt}\ v{\isacharcomma}{\kern0pt}\ j{\isasymrangle}\ {\isasymTurnstile}\ {\isasymbeta}\ {\isasymor}\ {\isacharparenleft}{\kern0pt}{\isasymexists}k{\isasymin}{\isacharbraceleft}{\kern0pt}j{\isacharless}{\kern0pt}{\isachardot}{\kern0pt}{\isachardot}{\kern0pt}i{\isacharbraceright}{\kern0pt}{\isachardot}{\kern0pt}\ {\isasymlangle}{\isasymsigma}{\isacharcomma}{\kern0pt}\ v{\isacharcomma}{\kern0pt}\ k{\isasymrangle}\ {\isasymTurnstile}\ {\isasymalpha}{\isacharparenright}{\kern0pt}\ in\isanewline
\ \ \ \ \ \ \ \ {\isacharparenleft}{\kern0pt}if\ down{\isacharunderscore}{\kern0pt}cl{\isacharunderscore}{\kern0pt}ivl\ {\isasymsigma}\ I\ i\ {\isacharequal}{\kern0pt}\ {\isacharbraceleft}{\kern0pt}{\isacharbraceright}{\kern0pt}\ then\ {\isacharbraceleft}{\kern0pt}{\isacharbraceright}{\kern0pt}\isanewline
\ \ \ \ \ \ \ \ else\ if\ {\isacharparenleft}{\kern0pt}{\isasymforall}v{\isachardot}{\kern0pt}\ {\isasymexists}j{\isasymin}down{\isacharunderscore}{\kern0pt}cl{\isacharunderscore}{\kern0pt}ivl\ {\isasymsigma}\ I\ i{\isachardot}{\kern0pt}\ {\isasymnot}\ satisf{\isacharunderscore}{\kern0pt}at\ v\ j{\isacharparenright}{\kern0pt}\ then\ FV\ {\isacharparenleft}{\kern0pt}{\isasymalpha}\ \isactrlbold T\ I\ {\isasymbeta}{\isacharparenright}{\kern0pt}\isanewline
\ \ \ \ \ \ \ \ else\ {\isacharparenleft}{\kern0pt}let\ J\ {\isacharequal}{\kern0pt}\ {\isacharbraceleft}{\kern0pt}j{\isasymin}down{\isacharunderscore}{\kern0pt}cl{\isacharunderscore}{\kern0pt}ivl\ {\isasymsigma}\ I\ i{\isachardot}{\kern0pt}\ {\isasymexists}v{\isachardot}{\kern0pt}\ {\isasymlangle}{\isasymsigma}{\isacharcomma}{\kern0pt}\ v{\isacharcomma}{\kern0pt}\ j{\isasymrangle}\ {\isasymTurnstile}\ {\isasymbeta}{\isacharbraceright}{\kern0pt}{\isacharsemicolon}{\kern0pt}\ \isanewline
\ \ \ \ \ \ \ \ \ \ K\ {\isacharequal}{\kern0pt}\ {\isacharbraceleft}{\kern0pt}k{\isachardot}{\kern0pt}\ {\isasymexists}v{\isachardot}{\kern0pt}\ {\isasymexists}j{\isasymin}down{\isacharunderscore}{\kern0pt}cl{\isacharunderscore}{\kern0pt}ivl\ {\isasymsigma}\ I\ i{\isachardot}{\kern0pt}\ j\ {\isacharless}{\kern0pt}\ k\ {\isasymand}\ k\ {\isasymle}\ i\ {\isasymand}\ {\isasymlangle}{\isasymsigma}{\isacharcomma}{\kern0pt}\ v{\isacharcomma}{\kern0pt}\ k{\isasymrangle}\ {\isasymTurnstile}\ {\isasymalpha}{\isacharbraceright}{\kern0pt}\ in\isanewline
\ \ \ \ \ \ \ \ \ \ {\isacharparenleft}{\kern0pt}{\isasymUnion}{\isacharbraceleft}{\kern0pt}{\isacharparenleft}{\kern0pt}dfv\ {\isasymsigma}\ k\ {\isasymalpha}{\isacharparenright}{\kern0pt}{\isacharbar}{\kern0pt}k{\isachardot}{\kern0pt}\ k\ {\isasymin}\ K{\isacharbraceright}{\kern0pt}{\isacharparenright}{\kern0pt}\ {\isasymunion}\ {\isacharparenleft}{\kern0pt}{\isasymUnion}{\isacharbraceleft}{\kern0pt}{\isacharparenleft}{\kern0pt}dfv\ {\isasymsigma}\ j\ {\isasymbeta}{\isacharparenright}{\kern0pt}{\isacharbar}{\kern0pt}j{\isachardot}{\kern0pt}\ j\ {\isasymin}\ J{\isacharbraceright}{\kern0pt}{\isacharparenright}{\kern0pt}{\isacharparenright}{\kern0pt}{\isacharparenright}{\kern0pt}{\isacharparenright}{\kern0pt}{\isachardoublequoteclose}\isanewline
\ \ {\isacharbar}{\kern0pt}\ {\isachardoublequoteopen}dfv\ {\isasymsigma}\ i\ {\isacharparenleft}{\kern0pt}{\isasymalpha}\ \isactrlbold R\ I\ {\isasymbeta}{\isacharparenright}{\kern0pt}\ {\isacharequal}{\kern0pt}\ \isanewline
\ \ \ \ \ \ {\isacharparenleft}{\kern0pt}let\ satisf{\isacharunderscore}{\kern0pt}at\ {\isacharequal}{\kern0pt}\ {\isasymlambda}v\ j{\isachardot}{\kern0pt}\ {\isasymlangle}{\isasymsigma}{\isacharcomma}{\kern0pt}\ v{\isacharcomma}{\kern0pt}\ j{\isasymrangle}\ {\isasymTurnstile}\ {\isasymbeta}\ {\isasymor}\ {\isacharparenleft}{\kern0pt}{\isasymexists}k{\isasymin}{\isacharbraceleft}{\kern0pt}i{\isachardot}{\kern0pt}{\isachardot}{\kern0pt}{\isacharless}{\kern0pt}j{\isacharbraceright}{\kern0pt}{\isachardot}{\kern0pt}\ {\isasymlangle}{\isasymsigma}{\isacharcomma}{\kern0pt}\ v{\isacharcomma}{\kern0pt}\ k{\isasymrangle}\ {\isasymTurnstile}\ {\isasymalpha}{\isacharparenright}{\kern0pt}\ in\isanewline
\ \ \ \ \ \ \ \ {\isacharparenleft}{\kern0pt}if\ up{\isacharunderscore}{\kern0pt}cl{\isacharunderscore}{\kern0pt}ivl\ {\isasymsigma}\ I\ i\ {\isacharequal}{\kern0pt}\ {\isacharbraceleft}{\kern0pt}{\isacharbraceright}{\kern0pt}\ then\ {\isacharbraceleft}{\kern0pt}{\isacharbraceright}{\kern0pt}\isanewline
\ \ \ \ \ \ \ \ else\ if\ {\isacharparenleft}{\kern0pt}{\isasymforall}v{\isachardot}{\kern0pt}\ {\isasymexists}j{\isasymin}up{\isacharunderscore}{\kern0pt}cl{\isacharunderscore}{\kern0pt}ivl\ {\isasymsigma}\ I\ i{\isachardot}{\kern0pt}\ {\isasymnot}\ satisf{\isacharunderscore}{\kern0pt}at\ v\ j{\isacharparenright}{\kern0pt}\ then\ FV\ {\isacharparenleft}{\kern0pt}{\isasymalpha}\ \isactrlbold R\ I\ {\isasymbeta}{\isacharparenright}{\kern0pt}\isanewline
\ \ \ \ \ \ \ \ else\ {\isacharparenleft}{\kern0pt}let\ J\ {\isacharequal}{\kern0pt}\ {\isacharbraceleft}{\kern0pt}j{\isasymin}up{\isacharunderscore}{\kern0pt}cl{\isacharunderscore}{\kern0pt}ivl\ {\isasymsigma}\ I\ i{\isachardot}{\kern0pt}\ {\isasymexists}v{\isachardot}{\kern0pt}\ {\isasymlangle}{\isasymsigma}{\isacharcomma}{\kern0pt}\ v{\isacharcomma}{\kern0pt}\ j{\isasymrangle}\ {\isasymTurnstile}\ {\isasymbeta}{\isacharbraceright}{\kern0pt}{\isacharsemicolon}{\kern0pt}\ \isanewline
\ \ \ \ \ \ \ \ \ \ K\ {\isacharequal}{\kern0pt}\ {\isacharbraceleft}{\kern0pt}k{\isachardot}{\kern0pt}\ {\isasymexists}v{\isachardot}{\kern0pt}\ {\isasymexists}j{\isasymin}up{\isacharunderscore}{\kern0pt}cl{\isacharunderscore}{\kern0pt}ivl\ {\isasymsigma}\ I\ i{\isachardot}{\kern0pt}\ i\ {\isasymle}\ k\ {\isasymand}\ k\ {\isacharless}{\kern0pt}\ j\ {\isasymand}\ {\isasymlangle}{\isasymsigma}{\isacharcomma}{\kern0pt}\ v{\isacharcomma}{\kern0pt}\ k{\isasymrangle}\ {\isasymTurnstile}\ {\isasymalpha}{\isacharbraceright}{\kern0pt}\ in\isanewline
\ \ \ \ \ \ \ \ \ \ {\isacharparenleft}{\kern0pt}{\isasymUnion}{\isacharbraceleft}{\kern0pt}{\isacharparenleft}{\kern0pt}dfv\ {\isasymsigma}\ k\ {\isasymalpha}{\isacharparenright}{\kern0pt}{\isacharbar}{\kern0pt}k{\isachardot}{\kern0pt}\ k\ {\isasymin}\ K{\isacharbraceright}{\kern0pt}{\isacharparenright}{\kern0pt}\ {\isasymunion}\ {\isacharparenleft}{\kern0pt}{\isasymUnion}{\isacharbraceleft}{\kern0pt}{\isacharparenleft}{\kern0pt}dfv\ {\isasymsigma}\ j\ {\isasymbeta}{\isacharparenright}{\kern0pt}{\isacharbar}{\kern0pt}j{\isachardot}{\kern0pt}\ j\ {\isasymin}\ J{\isacharbraceright}{\kern0pt}{\isacharparenright}{\kern0pt}{\isacharparenright}{\kern0pt}{\isacharparenright}{\kern0pt}{\isacharparenright}{\kern0pt}{\isachardoublequoteclose}\isanewline
\end{isabellebody}

\paragraph{Safe sets of free variables.} 

\begin{isabellebody}\scriptsize\isanewline
\isanewline
\isacommand{fun}\isamarkupfalse%
\ {\isachardoublequoteopen}is{\isacharunderscore}{\kern0pt}constraint\ {\isacharparenleft}{\kern0pt}t{\isadigit{1}}\ {\isacharequal}{\kern0pt}\isactrlsub F\ t{\isadigit{2}}{\isacharparenright}{\kern0pt}\ {\isacharequal}{\kern0pt}\ True{\isachardoublequoteclose}\isanewline
{\isacharbar}{\kern0pt}\ {\isachardoublequoteopen}is{\isacharunderscore}{\kern0pt}constraint\ {\isacharparenleft}{\kern0pt}{\isasymnot}\isactrlsub F\ {\isacharparenleft}{\kern0pt}t{\isadigit{1}}\ {\isacharequal}{\kern0pt}\isactrlsub F\ t{\isadigit{2}}{\isacharparenright}{\kern0pt}{\isacharparenright}{\kern0pt}\ {\isacharequal}{\kern0pt}\ True{\isachardoublequoteclose}\isanewline
{\isacharbar}{\kern0pt}\ {\isachardoublequoteopen}is{\isacharunderscore}{\kern0pt}constraint\ {\isacharunderscore}{\kern0pt}\ {\isacharequal}{\kern0pt}\ False{\isachardoublequoteclose}\isanewline

\isacommand{definition}\isamarkupfalse%
\ {\isachardoublequoteopen}safe{\isacharunderscore}{\kern0pt}assignment\ X\ {\isasymalpha}\ {\isacharequal}{\kern0pt}\ {\isacharparenleft}{\kern0pt}case\ {\isasymalpha}\ of\isanewline
\ \ \ \ \ \ \ \isactrlbold v\ x\ {\isacharequal}{\kern0pt}\isactrlsub F\ \isactrlbold v\ y\ {\isasymRightarrow}\ {\isacharparenleft}{\kern0pt}x\ {\isasymnotin}\ X\ {\isasymlongleftrightarrow}\ y\ {\isasymin}\ X{\isacharparenright}{\kern0pt}\isanewline
\ \ \ \ \ {\isacharbar}{\kern0pt}\ \isactrlbold v\ x\ {\isacharequal}{\kern0pt}\isactrlsub F\ t\ {\isasymRightarrow}\ {\isacharparenleft}{\kern0pt}x\ {\isasymnotin}\ X\ {\isasymand}\ fv{\isacharunderscore}{\kern0pt}trm\ t\ {\isasymsubseteq}\ X{\isacharparenright}{\kern0pt}\isanewline
\ \ \ \ \ {\isacharbar}{\kern0pt}\ t\ {\isacharequal}{\kern0pt}\isactrlsub F\ \isactrlbold v\ x\ {\isasymRightarrow}\ {\isacharparenleft}{\kern0pt}x\ {\isasymnotin}\ X\ {\isasymand}\ fv{\isacharunderscore}{\kern0pt}trm\ t\ {\isasymsubseteq}\ X{\isacharparenright}{\kern0pt}\isanewline
\ \ \ \ \ {\isacharbar}{\kern0pt}\ {\isacharunderscore}{\kern0pt}\ {\isasymRightarrow}\ False{\isacharparenright}{\kern0pt}{\isachardoublequoteclose}\isanewline

\isacommand{fun}\isamarkupfalse%
\ ssfv\ {\isacharcolon}{\kern0pt}{\isacharcolon}{\kern0pt}\ {\isachardoublequoteopen}{\isacharprime}{\kern0pt}a\ MFOTL{\isacharunderscore}{\kern0pt}Formula{\isachardot}{\kern0pt}formula\ {\isasymRightarrow}\ nat\ set\ set{\isachardoublequoteclose}\ \isanewline
\ \ \isakeyword{where}\ {\isachardoublequoteopen}ssfv\ {\isacharparenleft}{\kern0pt}p\ {\isasymdagger}\ trms{\isacharparenright}{\kern0pt}\ {\isacharequal}{\kern0pt}\ {\isacharbraceleft}{\kern0pt}FV\ {\isacharparenleft}{\kern0pt}p\ {\isasymdagger}\ trms{\isacharparenright}{\kern0pt}{\isacharbraceright}{\kern0pt}{\isachardoublequoteclose}\isanewline
\ \ {\isacharbar}{\kern0pt}\ {\isachardoublequoteopen}ssfv\ {\isacharparenleft}{\kern0pt}\isactrlbold v\ x\ {\isacharequal}{\kern0pt}\isactrlsub F\ t{\isacharparenright}{\kern0pt}\ {\isacharequal}{\kern0pt}\ {\isacharparenleft}{\kern0pt}if\ FV\isactrlsub t\ t\ {\isacharequal}{\kern0pt}\ {\isacharbraceleft}{\kern0pt}{\isacharbraceright}{\kern0pt}\ then\ {\isacharbraceleft}{\kern0pt}{\isacharbraceleft}{\kern0pt}x{\isacharbraceright}{\kern0pt}{\isacharbraceright}{\kern0pt}\ else\ {\isacharbraceleft}{\kern0pt}{\isacharbraceright}{\kern0pt}{\isacharparenright}{\kern0pt}{\isachardoublequoteclose}\isanewline
\ \ {\isacharbar}{\kern0pt}\ {\isachardoublequoteopen}ssfv\ {\isacharparenleft}{\kern0pt}t\ {\isacharequal}{\kern0pt}\isactrlsub F\ \isactrlbold v\ x{\isacharparenright}{\kern0pt}\ {\isacharequal}{\kern0pt}\ {\isacharparenleft}{\kern0pt}if\ FV\isactrlsub t\ t\ {\isacharequal}{\kern0pt}\ {\isacharbraceleft}{\kern0pt}{\isacharbraceright}{\kern0pt}\ then\ {\isacharbraceleft}{\kern0pt}{\isacharbraceleft}{\kern0pt}x{\isacharbraceright}{\kern0pt}{\isacharbraceright}{\kern0pt}\ else\ {\isacharbraceleft}{\kern0pt}{\isacharbraceright}{\kern0pt}{\isacharparenright}{\kern0pt}{\isachardoublequoteclose}\isanewline
\ \ {\isacharbar}{\kern0pt}\ {\isachardoublequoteopen}ssfv\ {\isacharparenleft}{\kern0pt}t{\isadigit{1}}\ {\isacharequal}{\kern0pt}\isactrlsub F\ t{\isadigit{2}}{\isacharparenright}{\kern0pt}\ {\isacharequal}{\kern0pt}\ {\isacharparenleft}{\kern0pt}if\ FV\isactrlsub t\ t{\isadigit{1}}\ {\isasymunion}\ FV\isactrlsub t\ t{\isadigit{2}}\ {\isacharequal}{\kern0pt}\ {\isacharbraceleft}{\kern0pt}{\isacharbraceright}{\kern0pt}\ then\ {\isacharbraceleft}{\kern0pt}{\isacharbraceleft}{\kern0pt}{\isacharbraceright}{\kern0pt}{\isacharbraceright}{\kern0pt}\ else\ {\isacharbraceleft}{\kern0pt}{\isacharbraceright}{\kern0pt}{\isacharparenright}{\kern0pt}{\isachardoublequoteclose}\isanewline
\ \ {\isacharbar}{\kern0pt}\ {\isachardoublequoteopen}ssfv\ {\isacharparenleft}{\kern0pt}{\isasymnot}\isactrlsub F\ {\isacharparenleft}{\kern0pt}t{\isadigit{1}}\ {\isacharequal}{\kern0pt}\isactrlsub F\ t{\isadigit{2}}{\isacharparenright}{\kern0pt}{\isacharparenright}{\kern0pt}\ {\isacharequal}{\kern0pt}\ {\isacharparenleft}{\kern0pt}let\ X\ {\isacharequal}{\kern0pt}\ FV\ {\isacharparenleft}{\kern0pt}t{\isadigit{1}}\ {\isacharequal}{\kern0pt}\isactrlsub F\ t{\isadigit{2}}{\isacharparenright}{\kern0pt}\ in\ if\ t{\isadigit{1}}\ {\isacharequal}{\kern0pt}\ t{\isadigit{2}}\ {\isasymor}\ X\ {\isacharequal}{\kern0pt}\ {\isacharbraceleft}{\kern0pt}{\isacharbraceright}{\kern0pt}\ then\ {\isacharbraceleft}{\kern0pt}X{\isacharbraceright}{\kern0pt}\ else\ {\isacharbraceleft}{\kern0pt}{\isacharbraceright}{\kern0pt}{\isacharparenright}{\kern0pt}{\isachardoublequoteclose}\isanewline
\ \ {\isacharbar}{\kern0pt}\ {\isachardoublequoteopen}ssfv\ {\isacharparenleft}{\kern0pt}{\isasymalpha}\ {\isasymand}\isactrlsub F\ {\isasymbeta}{\isacharparenright}{\kern0pt}\ {\isacharequal}{\kern0pt}\ {\isacharparenleft}{\kern0pt}let\ {\isasymA}\ {\isacharequal}{\kern0pt}\ ssfv\ {\isasymalpha}{\isacharsemicolon}{\kern0pt}\ {\isasymB}\ {\isacharequal}{\kern0pt}\ ssfv\ {\isasymbeta}\ in\ \isanewline
\ \ \ \ \ \ if\ {\isasymA}\ {\isasymnoteq}\ {\isacharbraceleft}{\kern0pt}{\isacharbraceright}{\kern0pt}\ then\ \isanewline
\ \ \ \ \ \ \ \ if\ {\isasymB}\ {\isasymnoteq}\ {\isacharbraceleft}{\kern0pt}{\isacharbraceright}{\kern0pt}\ then\ {\isasymA}\ {\isasymuplus}\ {\isasymB}\isanewline
\ \ \ \ \ \ \ \ else\ if\ {\isasymforall}X{\isasymin}{\isasymA}{\isachardot}{\kern0pt}\ safe{\isacharunderscore}{\kern0pt}assignment\ X\ {\isasymbeta}\ then\ {\isacharparenleft}{\kern0pt}{\isacharparenleft}{\kern0pt}{\isasymunion}{\isacharparenright}{\kern0pt}\ {\isacharparenleft}{\kern0pt}FV\ {\isasymbeta}{\isacharparenright}{\kern0pt}{\isacharparenright}{\kern0pt}\ {\isacharbackquote}{\kern0pt}\ {\isasymA}\ \isanewline
\ \ \ \ \ \ \ \ else\ if\ is{\isacharunderscore}{\kern0pt}constraint\ {\isasymbeta}\ {\isasymand}\ {\isacharparenleft}{\kern0pt}{\isasymforall}X{\isasymin}{\isasymA}{\isachardot}{\kern0pt}\ FV\ {\isasymbeta}\ {\isasymsubseteq}\ X{\isacharparenright}{\kern0pt}\ then\ {\isasymA}\isanewline
\ \ \ \ \ \ \ \ else\ {\isacharparenleft}{\kern0pt}case\ {\isasymbeta}\ of\ {\isasymnot}\isactrlsub F\ {\isasymbeta}{\isacharprime}{\kern0pt}\ {\isasymRightarrow}\ {\isacharparenleft}{\kern0pt}let\ {\isasymB}{\isacharprime}{\kern0pt}\ {\isacharequal}{\kern0pt}\ ssfv\ {\isasymbeta}{\isacharprime}{\kern0pt}\ in\isanewline
\ \ \ \ \ \ \ \ \ \ {\isacharparenleft}{\kern0pt}if\ {\isasymB}{\isacharprime}{\kern0pt}\ {\isasymnoteq}\ {\isacharbraceleft}{\kern0pt}{\isacharbraceright}{\kern0pt}\ {\isasymand}\ {\isacharparenleft}{\kern0pt}{\isasymforall}Y{\isasymin}{\isasymB}{\isacharprime}{\kern0pt}{\isachardot}{\kern0pt}\ {\isasymforall}X{\isasymin}{\isasymA}{\isachardot}{\kern0pt}\ Y\ {\isasymsubseteq}\ X{\isacharparenright}{\kern0pt}\ then\ {\isasymA}\ else\ {\isacharbraceleft}{\kern0pt}{\isacharbraceright}{\kern0pt}{\isacharparenright}{\kern0pt}{\isacharparenright}{\kern0pt}\ {\isacharbar}{\kern0pt}\ {\isacharunderscore}{\kern0pt}\ {\isasymRightarrow}\ {\isacharbraceleft}{\kern0pt}{\isacharbraceright}{\kern0pt}{\isacharparenright}{\kern0pt}\isanewline
\ \ \ \ \ \ else\ {\isacharbraceleft}{\kern0pt}{\isacharbraceright}{\kern0pt}{\isacharparenright}{\kern0pt}{\isachardoublequoteclose}\isanewline
\ \ {\isacharbar}{\kern0pt}\ {\isachardoublequoteopen}ssfv\ {\isacharparenleft}{\kern0pt}{\isasymalpha}\ {\isasymor}\isactrlsub F\ {\isasymbeta}{\isacharparenright}{\kern0pt}\ {\isacharequal}{\kern0pt}\ {\isacharparenleft}{\kern0pt}let\ {\isasymA}\ {\isacharequal}{\kern0pt}\ ssfv\ {\isasymalpha}{\isacharsemicolon}{\kern0pt}\ {\isasymB}\ {\isacharequal}{\kern0pt}\ ssfv\ {\isasymbeta}{\isacharsemicolon}{\kern0pt}\ X\ {\isacharequal}{\kern0pt}\ FV\ {\isasymalpha}{\isacharsemicolon}{\kern0pt}\ Y\ {\isacharequal}{\kern0pt}\ FV\ {\isasymbeta}\ in\isanewline
\ \ \ \ \ \ if\ {\isacharparenleft}{\kern0pt}{\isasymA}\ {\isasymnoteq}\ {\isacharbraceleft}{\kern0pt}{\isacharbraceright}{\kern0pt}\ {\isasymand}\ {\isasymB}\ {\isasymnoteq}\ {\isacharbraceleft}{\kern0pt}{\isacharbraceright}{\kern0pt}{\isacharparenright}{\kern0pt}\ then\ \isanewline
\ \ \ \ \ \ \ \ if\ X\ {\isacharequal}{\kern0pt}\ Y\ {\isasymand}\ {\isasymA}\ {\isasymsubseteq}\ {\isacharbraceleft}{\kern0pt}{\isacharbraceleft}{\kern0pt}{\isacharbraceright}{\kern0pt}{\isacharcomma}{\kern0pt}X{\isacharbraceright}{\kern0pt}\ {\isasymand}\ {\isasymB}\ {\isasymsubseteq}\ {\isacharbraceleft}{\kern0pt}{\isacharbraceleft}{\kern0pt}{\isacharbraceright}{\kern0pt}{\isacharcomma}{\kern0pt}Y{\isacharbraceright}{\kern0pt}\ then\ \isanewline
\ \ \ \ \ \ \ \ \ \ {\isacharparenleft}{\kern0pt}if\ {\isacharbraceleft}{\kern0pt}{\isacharbraceright}{\kern0pt}\ {\isasymin}\ {\isasymA}\ {\isasymor}\ {\isacharbraceleft}{\kern0pt}{\isacharbraceright}{\kern0pt}\ {\isasymin}\ {\isasymB}\ then\ {\isacharbraceleft}{\kern0pt}{\isacharbraceleft}{\kern0pt}{\isacharbraceright}{\kern0pt}{\isacharbraceright}{\kern0pt}\ {\isasymunion}\ {\isacharparenleft}{\kern0pt}{\isasymA}\ {\isasymuplus}\ {\isasymB}{\isacharparenright}{\kern0pt}\ else\ {\isasymA}\ {\isasymuplus}\ {\isasymB}{\isacharparenright}{\kern0pt}\isanewline
\ \ \ \ \ \ \ \ else\ \isanewline
\ \ \ \ \ \ \ \ \ \ {\isacharparenleft}{\kern0pt}if\ X\ {\isacharequal}{\kern0pt}\ {\isacharbraceleft}{\kern0pt}{\isacharbraceright}{\kern0pt}\ {\isasymor}\ Y\ {\isacharequal}{\kern0pt}\ {\isacharbraceleft}{\kern0pt}{\isacharbraceright}{\kern0pt}\ then\ {\isasymA}\ {\isasymunion}\ {\isasymB}\ else\ {\isacharbraceleft}{\kern0pt}{\isacharbraceright}{\kern0pt}{\isacharparenright}{\kern0pt}\isanewline
\ \ \ \ \ \ else\ {\isacharbraceleft}{\kern0pt}{\isacharbraceright}{\kern0pt}{\isacharparenright}{\kern0pt}{\isachardoublequoteclose}\isanewline
\ \ {\isacharbar}{\kern0pt}\ {\isachardoublequoteopen}ssfv\ {\isacharparenleft}{\kern0pt}{\isasymexists}\isactrlsub F\ {\isasymalpha}{\isacharparenright}{\kern0pt}\ {\isacharequal}{\kern0pt}\ {\isacharparenleft}{\kern0pt}{\isacharparenleft}{\kern0pt}{\isacharparenleft}{\kern0pt}{\isacharbackquote}{\kern0pt}{\isacharparenright}{\kern0pt}\ {\isacharparenleft}{\kern0pt}{\isasymlambda}x{\isacharcolon}{\kern0pt}{\isacharcolon}{\kern0pt}nat{\isachardot}{\kern0pt}\ x\ {\isacharminus}{\kern0pt}\ {\isadigit{1}}{\isacharparenright}{\kern0pt}{\isacharparenright}{\kern0pt}\ {\isasymcirc}\ {\isacharparenleft}{\kern0pt}{\isasymlambda}X{\isachardot}{\kern0pt}\ X\ {\isacharminus}{\kern0pt}\ {\isacharbraceleft}{\kern0pt}{\isadigit{0}}{\isacharbraceright}{\kern0pt}{\isacharparenright}{\kern0pt}{\isacharparenright}{\kern0pt}\ {\isacharbackquote}{\kern0pt}\ ssfv\ {\isasymalpha}{\isachardoublequoteclose}\isanewline
\ \ {\isacharbar}{\kern0pt}\ {\isachardoublequoteopen}ssfv\ {\isacharparenleft}{\kern0pt}\isactrlbold Y\ I\ {\isasymalpha}{\isacharparenright}{\kern0pt}\ {\isacharequal}{\kern0pt}\ ssfv\ {\isasymalpha}{\isachardoublequoteclose}\isanewline
\ \ {\isacharbar}{\kern0pt}\ {\isachardoublequoteopen}ssfv\ {\isacharparenleft}{\kern0pt}\isactrlbold X\ I\ {\isasymalpha}{\isacharparenright}{\kern0pt}\ {\isacharequal}{\kern0pt}\ ssfv\ {\isasymalpha}{\isachardoublequoteclose}\isanewline
\ \ {\isacharbar}{\kern0pt}\ {\isachardoublequoteopen}ssfv\ {\isacharparenleft}{\kern0pt}{\isasymalpha}\ \isactrlbold S\ I\ {\isasymbeta}{\isacharparenright}{\kern0pt}\ {\isacharequal}{\kern0pt}\ {\isacharparenleft}{\kern0pt}let\ {\isasymA}\ {\isacharequal}{\kern0pt}\ ssfv\ {\isasymalpha}{\isacharsemicolon}{\kern0pt}\ {\isasymB}\ {\isacharequal}{\kern0pt}\ ssfv\ {\isasymbeta}{\isacharsemicolon}{\kern0pt}\ X\ {\isacharequal}{\kern0pt}\ FV\ {\isasymalpha}{\isacharsemicolon}{\kern0pt}\ Y\ {\isacharequal}{\kern0pt}\ FV\ {\isasymbeta}\ in\ \isanewline
\ \ \ \ \ \ if\ {\isacharparenleft}{\kern0pt}{\isasymB}\ {\isacharequal}{\kern0pt}\ {\isacharbraceleft}{\kern0pt}Y{\isacharbraceright}{\kern0pt}{\isacharparenright}{\kern0pt}\ then\isanewline
\ \ \ \ \ \ \ \ if\ {\isasymA}\ {\isasymnoteq}\ {\isacharbraceleft}{\kern0pt}{\isacharbraceright}{\kern0pt}\ {\isasymand}\ X\ {\isasymsubseteq}\ Y\ then\ {\isacharbraceleft}{\kern0pt}Y{\isacharbraceright}{\kern0pt}\isanewline
\ \ \ \ \ \ \ \ else\ {\isacharparenleft}{\kern0pt}case\ {\isasymalpha}\ of\ \isanewline
\ \ \ \ \ \ \ \ \ \ \ \ {\isasymnot}\isactrlsub F\ {\isasymalpha}{\isacharprime}{\kern0pt}\ {\isasymRightarrow}\ {\isacharparenleft}{\kern0pt}let\ {\isasymA}{\isacharprime}{\kern0pt}\ {\isacharequal}{\kern0pt}\ ssfv\ {\isasymalpha}{\isacharprime}{\kern0pt}\ in\ if\ {\isasymA}{\isacharprime}{\kern0pt}\ {\isasymnoteq}\ {\isacharbraceleft}{\kern0pt}{\isacharbraceright}{\kern0pt}\ {\isasymand}\ X\ {\isasymsubseteq}\ Y\ then\ {\isacharbraceleft}{\kern0pt}Y{\isacharbraceright}{\kern0pt}\ else\ {\isacharbraceleft}{\kern0pt}{\isacharbraceright}{\kern0pt}{\isacharparenright}{\kern0pt}\ \isanewline
\ \ \ \ \ \ \ \ \ \ {\isacharbar}{\kern0pt}\ \ \ \ {\isacharunderscore}{\kern0pt}\ {\isasymRightarrow}\ {\isacharbraceleft}{\kern0pt}{\isacharbraceright}{\kern0pt}{\isacharparenright}{\kern0pt}\isanewline
\ \ \ \ \ \ else\ {\isacharbraceleft}{\kern0pt}{\isacharbraceright}{\kern0pt}{\isacharparenright}{\kern0pt}{\isachardoublequoteclose}\isanewline
\ \ {\isacharbar}{\kern0pt}\ {\isachardoublequoteopen}ssfv\ {\isacharparenleft}{\kern0pt}{\isasymalpha}\ \isactrlbold U\ I\ {\isasymbeta}{\isacharparenright}{\kern0pt}\ {\isacharequal}{\kern0pt}\ {\isacharparenleft}{\kern0pt}let\ {\isasymA}\ {\isacharequal}{\kern0pt}\ ssfv\ {\isasymalpha}{\isacharsemicolon}{\kern0pt}\ {\isasymB}\ {\isacharequal}{\kern0pt}\ ssfv\ {\isasymbeta}{\isacharsemicolon}{\kern0pt}\ X\ {\isacharequal}{\kern0pt}\ FV\ {\isasymalpha}{\isacharsemicolon}{\kern0pt}\ Y\ {\isacharequal}{\kern0pt}\ FV\ {\isasymbeta}\ in\ \isanewline
\ \ \ \ \ \ if\ {\isacharparenleft}{\kern0pt}{\isasymB}\ {\isacharequal}{\kern0pt}\ {\isacharbraceleft}{\kern0pt}Y{\isacharbraceright}{\kern0pt}{\isacharparenright}{\kern0pt}\ then\isanewline
\ \ \ \ \ \ \ \ if\ {\isasymA}\ {\isasymnoteq}\ {\isacharbraceleft}{\kern0pt}{\isacharbraceright}{\kern0pt}\ {\isasymand}\ X\ {\isasymsubseteq}\ Y\ then\ {\isacharbraceleft}{\kern0pt}Y{\isacharbraceright}{\kern0pt}\isanewline
\ \ \ \ \ \ \ \ else\ {\isacharparenleft}{\kern0pt}case\ {\isasymalpha}\ of\ \isanewline
\ \ \ \ \ \ \ \ \ \ \ \ {\isasymnot}\isactrlsub F\ {\isasymalpha}{\isacharprime}{\kern0pt}\ {\isasymRightarrow}\ {\isacharparenleft}{\kern0pt}let\ {\isasymA}{\isacharprime}{\kern0pt}\ {\isacharequal}{\kern0pt}\ ssfv\ {\isasymalpha}{\isacharprime}{\kern0pt}\ in\ if\ X\ {\isasymsubseteq}\ Y\ {\isasymand}\ {\isasymA}{\isacharprime}{\kern0pt}\ {\isacharequal}{\kern0pt}\ {\isacharbraceleft}{\kern0pt}X{\isacharbraceright}{\kern0pt}\ then\ {\isacharbraceleft}{\kern0pt}Y{\isacharbraceright}{\kern0pt}\ else\ {\isacharbraceleft}{\kern0pt}{\isacharbraceright}{\kern0pt}{\isacharparenright}{\kern0pt}\ \isanewline
\ \ \ \ \ \ \ \ \ \ {\isacharbar}{\kern0pt}\ \ \ \ {\isacharunderscore}{\kern0pt}\ \ {\isasymRightarrow}\ {\isacharbraceleft}{\kern0pt}{\isacharbraceright}{\kern0pt}{\isacharparenright}{\kern0pt}\isanewline
\ \ \ \ \ \ else\ {\isacharbraceleft}{\kern0pt}{\isacharbraceright}{\kern0pt}{\isacharparenright}{\kern0pt}{\isachardoublequoteclose}\isanewline
\ \ {\isacharbar}{\kern0pt}\ {\isachardoublequoteopen}ssfv\ {\isacharparenleft}{\kern0pt}{\isasymalpha}\ \isactrlbold T\ I\ {\isasymbeta}{\isacharparenright}{\kern0pt}\ {\isacharequal}{\kern0pt}\ {\isacharparenleft}{\kern0pt}let\ {\isasymA}\ {\isacharequal}{\kern0pt}\ ssfv\ {\isasymalpha}{\isacharsemicolon}{\kern0pt}\ {\isasymB}\ {\isacharequal}{\kern0pt}\ ssfv\ {\isasymbeta}{\isacharsemicolon}{\kern0pt}\ X\ {\isacharequal}{\kern0pt}\ FV\ {\isasymalpha}{\isacharsemicolon}{\kern0pt}\ Y\ {\isacharequal}{\kern0pt}\ FV\ {\isasymbeta}\ in\isanewline
\ \ \ \ \ \ if\ mem\ I\ {\isadigit{0}}\ then\ \isanewline
\ \ \ \ \ \ \ \ if\ {\isacharparenleft}{\kern0pt}{\isasymB}\ {\isacharequal}{\kern0pt}\ {\isacharbraceleft}{\kern0pt}Y{\isacharbraceright}{\kern0pt}{\isacharparenright}{\kern0pt}\ then\isanewline
\ \ \ \ \ \ \ \ \ \ if\ {\isasymA}\ {\isasymnoteq}\ {\isacharbraceleft}{\kern0pt}{\isacharbraceright}{\kern0pt}\ {\isasymand}\ X\ {\isasymsubseteq}\ Y\ then\ {\isacharbraceleft}{\kern0pt}Y{\isacharbraceright}{\kern0pt}\isanewline
\ \ \ \ \ \ \ \ \ \ else\ {\isacharparenleft}{\kern0pt}case\ {\isasymalpha}\ of\ \isanewline
\ \ \ \ \ \ \ \ \ \ \ \ \ \ {\isasymnot}\isactrlsub F\ {\isasymalpha}{\isacharprime}{\kern0pt}\ {\isasymRightarrow}\ {\isacharparenleft}{\kern0pt}let\ {\isasymA}{\isacharprime}{\kern0pt}\ {\isacharequal}{\kern0pt}\ ssfv\ {\isasymalpha}{\isacharprime}{\kern0pt}\ in\ if\ {\isasymA}{\isacharprime}{\kern0pt}\ {\isasymnoteq}\ {\isacharbraceleft}{\kern0pt}{\isacharbraceright}{\kern0pt}\ {\isasymand}\ X\ {\isasymsubseteq}\ Y\ then\ {\isacharbraceleft}{\kern0pt}Y{\isacharbraceright}{\kern0pt}\ else\ {\isacharbraceleft}{\kern0pt}{\isacharbraceright}{\kern0pt}{\isacharparenright}{\kern0pt}\ \isanewline
\ \ \ \ \ \ \ \ \ \ \ \ {\isacharbar}{\kern0pt}\ \ \ \ {\isacharunderscore}{\kern0pt}\ {\isasymRightarrow}\ {\isacharbraceleft}{\kern0pt}{\isacharbraceright}{\kern0pt}{\isacharparenright}{\kern0pt}\isanewline
\ \ \ \ \ \ \ \ else\ {\isacharbraceleft}{\kern0pt}{\isacharbraceright}{\kern0pt}\isanewline
\ \ \ \ \ \ else\isanewline
\ \ \ \ \ \ \ \ if\ X\ {\isacharequal}{\kern0pt}\ Y\ {\isasymand}\ {\isasymA}\ {\isacharequal}{\kern0pt}\ {\isacharbraceleft}{\kern0pt}X{\isacharbraceright}{\kern0pt}\ {\isasymand}\ {\isasymB}\ {\isacharequal}{\kern0pt}\ {\isacharbraceleft}{\kern0pt}Y{\isacharbraceright}{\kern0pt}\ then\ {\isacharbraceleft}{\kern0pt}{\isacharbraceleft}{\kern0pt}{\isacharbraceright}{\kern0pt}{\isacharcomma}{\kern0pt}X{\isacharbraceright}{\kern0pt}\ else\ {\isacharbraceleft}{\kern0pt}{\isacharbraceright}{\kern0pt}{\isacharparenright}{\kern0pt}{\isachardoublequoteclose}\isanewline
\ \ {\isacharbar}{\kern0pt}\ {\isachardoublequoteopen}ssfv\ {\isacharparenleft}{\kern0pt}{\isasymalpha}\ \isactrlbold R\ I\ {\isasymbeta}{\isacharparenright}{\kern0pt}\ {\isacharequal}{\kern0pt}\ {\isacharparenleft}{\kern0pt}let\ {\isasymA}\ {\isacharequal}{\kern0pt}\ ssfv\ {\isasymalpha}{\isacharsemicolon}{\kern0pt}\ {\isasymB}\ {\isacharequal}{\kern0pt}\ ssfv\ {\isasymbeta}{\isacharsemicolon}{\kern0pt}\ X\ {\isacharequal}{\kern0pt}\ FV\ {\isasymalpha}{\isacharsemicolon}{\kern0pt}\ Y\ {\isacharequal}{\kern0pt}\ FV\ {\isasymbeta}\ in\isanewline
\ \ \ \ \ \ if\ mem\ I\ {\isadigit{0}}\ then\ \isanewline
\ \ \ \ \ \ \ \ if\ {\isacharparenleft}{\kern0pt}{\isasymB}\ {\isacharequal}{\kern0pt}\ {\isacharbraceleft}{\kern0pt}Y{\isacharbraceright}{\kern0pt}{\isacharparenright}{\kern0pt}\ then\isanewline
\ \ \ \ \ \ \ \ \ \ if\ {\isasymA}\ {\isasymnoteq}\ {\isacharbraceleft}{\kern0pt}{\isacharbraceright}{\kern0pt}\ {\isasymand}\ X\ {\isasymsubseteq}\ Y\ then\ {\isacharbraceleft}{\kern0pt}Y{\isacharbraceright}{\kern0pt}\isanewline
\ \ \ \ \ \ \ \ \ \ else\ {\isacharparenleft}{\kern0pt}case\ {\isasymalpha}\ of\ \isanewline
\ \ \ \ \ \ \ \ \ \ \ \ \ \ {\isasymnot}\isactrlsub F\ {\isasymalpha}{\isacharprime}{\kern0pt}\ {\isasymRightarrow}\ {\isacharparenleft}{\kern0pt}let\ {\isasymA}{\isacharprime}{\kern0pt}\ {\isacharequal}{\kern0pt}\ ssfv\ {\isasymalpha}{\isacharprime}{\kern0pt}\ in\ if\ {\isasymA}{\isacharprime}{\kern0pt}\ {\isasymnoteq}\ {\isacharbraceleft}{\kern0pt}{\isacharbraceright}{\kern0pt}\ {\isasymand}\ X\ {\isasymsubseteq}\ Y\ then\ {\isacharbraceleft}{\kern0pt}Y{\isacharbraceright}{\kern0pt}\ else\ {\isacharbraceleft}{\kern0pt}{\isacharbraceright}{\kern0pt}{\isacharparenright}{\kern0pt}\ \isanewline
\ \ \ \ \ \ \ \ \ \ \ \ {\isacharbar}{\kern0pt}\ \ \ \ {\isacharunderscore}{\kern0pt}\ {\isasymRightarrow}\ {\isacharbraceleft}{\kern0pt}{\isacharbraceright}{\kern0pt}{\isacharparenright}{\kern0pt}\isanewline
\ \ \ \ \ \ \ \ else\ {\isacharbraceleft}{\kern0pt}{\isacharbraceright}{\kern0pt}\isanewline
\ \ \ \ \ \ else\isanewline
\ \ \ \ \ \ \ \ if\ X\ {\isacharequal}{\kern0pt}\ Y\ {\isasymand}\ {\isasymA}\ {\isacharequal}{\kern0pt}\ {\isacharbraceleft}{\kern0pt}X{\isacharbraceright}{\kern0pt}\ {\isasymand}\ {\isasymB}\ {\isacharequal}{\kern0pt}\ {\isacharbraceleft}{\kern0pt}Y{\isacharbraceright}{\kern0pt}\ then\ {\isacharbraceleft}{\kern0pt}{\isacharbraceleft}{\kern0pt}{\isacharbraceright}{\kern0pt}{\isacharcomma}{\kern0pt}X{\isacharbraceright}{\kern0pt}\ else\ {\isacharbraceleft}{\kern0pt}{\isacharbraceright}{\kern0pt}{\isacharparenright}{\kern0pt}{\isachardoublequoteclose}\isanewline
\ \ {\isacharbar}{\kern0pt}\ {\isachardoublequoteopen}ssfv\ {\isacharparenleft}{\kern0pt}{\isasymnot}\isactrlsub F\ {\isasymalpha}{\isacharparenright}{\kern0pt}\ {\isacharequal}{\kern0pt}\ {\isacharparenleft}{\kern0pt}if\ ssfv\ {\isasymalpha}\ {\isacharequal}{\kern0pt}\ {\isacharbraceleft}{\kern0pt}{\isacharbraceleft}{\kern0pt}{\isacharbraceright}{\kern0pt}{\isacharbraceright}{\kern0pt}\ then\ {\isacharbraceleft}{\kern0pt}{\isacharbraceleft}{\kern0pt}{\isacharbraceright}{\kern0pt}{\isacharbraceright}{\kern0pt}\ else\ {\isacharbraceleft}{\kern0pt}{\isacharbraceright}{\kern0pt}{\isacharparenright}{\kern0pt}{\isachardoublequoteclose}
\end{isabellebody}

\paragraph{Verimon's \isa{safe}{\isacharunderscore}formula predicate.} Below 
we provide the definition of \isa{safe{\isacharunderscore}{\kern0pt}formula} used
in our proof that $\issafe$ defines a larger fragment. This is also the predicate that
does not hold for our examples in \S\ref{sec-examples}.

\begin{isabellebody}\scriptsize\isanewline\isanewline
\isacommand{definition}\isamarkupfalse%
\ safe{\isacharunderscore}{\kern0pt}dual\ \isakeyword{where}\ {\isachardoublequoteopen}safe{\isacharunderscore}{\kern0pt}dual\ conjoined\ safe{\isacharunderscore}{\kern0pt}formula\ {\isasymalpha}\ I\ {\isasymbeta}\ {\isacharequal}{\kern0pt}\ {\isacharparenleft}{\kern0pt}\isanewline
\ \ if\ {\isacharparenleft}{\kern0pt}mem\ I\ {\isadigit{0}}{\isacharparenright}{\kern0pt}\ then\isanewline
\ \ \ \ {\isacharparenleft}{\kern0pt}safe{\isacharunderscore}{\kern0pt}formula\ {\isasymbeta}\ {\isasymand}\ fv\ {\isasymalpha}\ {\isasymsubseteq}\ fv\ {\isasymbeta}\ \isanewline
\ \ \ \ \ \ {\isasymand}\ {\isacharparenleft}{\kern0pt}safe{\isacharunderscore}{\kern0pt}formula\ {\isasymalpha}\isanewline
\ \ \ \ \ \ \ \ {\isasymor}\ {\isacharparenleft}{\kern0pt}case\ {\isasymalpha}\ of\ {\isasymnot}\isactrlsub F\ {\isasymalpha}{\isacharprime}{\kern0pt}\ {\isasymRightarrow}\ safe{\isacharunderscore}{\kern0pt}formula\ {\isasymalpha}{\isacharprime}{\kern0pt}\ {\isacharbar}{\kern0pt}\ {\isacharunderscore}{\kern0pt}\ {\isasymRightarrow}\ False{\isacharparenright}{\kern0pt}{\isacharparenright}{\kern0pt}{\isacharparenright}{\kern0pt}\ \isanewline
\ \ else\isanewline
\ \ \ \ conjoined\ {\isasymand}\ {\isacharparenleft}{\kern0pt}safe{\isacharunderscore}{\kern0pt}formula\ {\isasymalpha}\ {\isasymand}\ safe{\isacharunderscore}{\kern0pt}formula\ {\isasymbeta}\ {\isasymand}\ fv\ {\isasymalpha}\ {\isacharequal}{\kern0pt}\ fv\ {\isasymbeta}{\isacharparenright}{\kern0pt}{\isacharparenright}{\kern0pt}{\isachardoublequoteclose}\isanewline

\isacommand{function}\isamarkupfalse%
\ safe{\isacharunderscore}{\kern0pt}formula\ {\isacharcolon}{\kern0pt}{\isacharcolon}{\kern0pt}\ {\isachardoublequoteopen}{\isacharprime}{\kern0pt}a\ MFOTL{\isacharunderscore}{\kern0pt}Formula{\isachardot}{\kern0pt}formula\ {\isasymRightarrow}\ bool{\isachardoublequoteclose}\ \isanewline
\ \ \isakeyword{where}\ {\isachardoublequoteopen}safe{\isacharunderscore}{\kern0pt}formula\ {\isacharparenleft}{\kern0pt}t{\isadigit{1}}\ {\isacharequal}{\kern0pt}\isactrlsub F\ t{\isadigit{2}}{\isacharparenright}{\kern0pt}\ {\isacharequal}{\kern0pt}\ {\isacharparenleft}{\kern0pt}{\isacharparenleft}{\kern0pt}trm{\isachardot}{\kern0pt}is{\isacharunderscore}{\kern0pt}Const\ t{\isadigit{1}}\ {\isasymand}\ {\isacharparenleft}{\kern0pt}trm{\isachardot}{\kern0pt}is{\isacharunderscore}{\kern0pt}Const\ t{\isadigit{2}}\ {\isasymor}\ trm{\isachardot}{\kern0pt}is{\isacharunderscore}{\kern0pt}Var\ t{\isadigit{2}}{\isacharparenright}{\kern0pt}{\isacharparenright}{\kern0pt}\ \isanewline
\ \ \ \ {\isasymor}\ {\isacharparenleft}{\kern0pt}trm{\isachardot}{\kern0pt}is{\isacharunderscore}{\kern0pt}Var\ t{\isadigit{1}}\ {\isasymand}\ trm{\isachardot}{\kern0pt}is{\isacharunderscore}{\kern0pt}Const\ t{\isadigit{2}}{\isacharparenright}{\kern0pt}{\isacharparenright}{\kern0pt}{\isachardoublequoteclose}\isanewline
\ \ {\isacharbar}{\kern0pt}\ {\isachardoublequoteopen}safe{\isacharunderscore}{\kern0pt}formula\ {\isacharparenleft}{\kern0pt}{\isasymnot}\isactrlsub F\ {\isacharparenleft}{\kern0pt}\isactrlbold v\ x\ {\isacharequal}{\kern0pt}\isactrlsub F\ \isactrlbold v\ y{\isacharparenright}{\kern0pt}{\isacharparenright}{\kern0pt}\ {\isacharequal}{\kern0pt}\ {\isacharparenleft}{\kern0pt}x\ {\isacharequal}{\kern0pt}\ y{\isacharparenright}{\kern0pt}{\isachardoublequoteclose}\isanewline
\ \ {\isacharbar}{\kern0pt}\ {\isachardoublequoteopen}safe{\isacharunderscore}{\kern0pt}formula\ {\isacharparenleft}{\kern0pt}p\ {\isasymdagger}\ ts{\isacharparenright}{\kern0pt}\ {\isacharequal}{\kern0pt}\ {\isacharparenleft}{\kern0pt}{\isasymforall}t{\isasymin}set\ ts{\isachardot}{\kern0pt}\ trm{\isachardot}{\kern0pt}is{\isacharunderscore}{\kern0pt}Var\ t\ {\isasymor}\ trm{\isachardot}{\kern0pt}is{\isacharunderscore}{\kern0pt}Const\ t{\isacharparenright}{\kern0pt}{\isachardoublequoteclose}\isanewline
\ \ {\isacharbar}{\kern0pt}\ {\isachardoublequoteopen}safe{\isacharunderscore}{\kern0pt}formula\ {\isacharparenleft}{\kern0pt}{\isasymnot}\isactrlsub F\ {\isasymalpha}{\isacharparenright}{\kern0pt}\ {\isacharequal}{\kern0pt}\ {\isacharparenleft}{\kern0pt}fv\ {\isasymalpha}\ {\isacharequal}{\kern0pt}\ {\isacharbraceleft}{\kern0pt}{\isacharbraceright}{\kern0pt}\ {\isasymand}\ safe{\isacharunderscore}{\kern0pt}formula\ {\isasymalpha}{\isacharparenright}{\kern0pt}{\isachardoublequoteclose}\isanewline
\ \ {\isacharbar}{\kern0pt}\ {\isachardoublequoteopen}safe{\isacharunderscore}{\kern0pt}formula\ {\isacharparenleft}{\kern0pt}{\isasymalpha}\ {\isasymor}\isactrlsub F\ {\isasymbeta}{\isacharparenright}{\kern0pt}\ {\isacharequal}{\kern0pt}\ {\isacharparenleft}{\kern0pt}fv\ {\isasymbeta}\ {\isacharequal}{\kern0pt}\ fv\ {\isasymalpha}\ {\isasymand}\ safe{\isacharunderscore}{\kern0pt}formula\ {\isasymalpha}\ {\isasymand}\ safe{\isacharunderscore}{\kern0pt}formula\ {\isasymbeta}{\isacharparenright}{\kern0pt}{\isachardoublequoteclose}\isanewline
\ \ {\isacharbar}{\kern0pt}\ {\isachardoublequoteopen}safe{\isacharunderscore}{\kern0pt}formula\ {\isacharparenleft}{\kern0pt}{\isasymalpha}\ {\isasymand}\isactrlsub F\ {\isasymbeta}{\isacharparenright}{\kern0pt}\ {\isacharequal}{\kern0pt}\ {\isacharparenleft}{\kern0pt}safe{\isacharunderscore}{\kern0pt}formula\ {\isasymalpha}\ {\isasymand}\ \isanewline
\ \ \ \ \ \ \ \ {\isacharparenleft}{\kern0pt}safe{\isacharunderscore}{\kern0pt}assignment\ {\isacharparenleft}{\kern0pt}fv\ {\isasymalpha}{\isacharparenright}{\kern0pt}\ {\isasymbeta}\ \isanewline
\ \ \ \ \ \ \ \ {\isasymor}\ safe{\isacharunderscore}{\kern0pt}formula\ {\isasymbeta}\ \isanewline
\ \ \ \ \ \ \ \ {\isasymor}\ {\isacharparenleft}{\kern0pt}fv\ {\isasymbeta}\ {\isasymsubseteq}\ fv\ {\isasymalpha}\ {\isasymand}\ {\isacharparenleft}{\kern0pt}is{\isacharunderscore}{\kern0pt}constraint\ {\isasymbeta}\ \isanewline
\ \ \ \ \ \ \ \ \ \ {\isasymor}\ {\isacharparenleft}{\kern0pt}case\ {\isasymbeta}\ of\ \isanewline
\ \ \ \ \ \ \ \ \ \ \ \ {\isasymnot}\isactrlsub F\ {\isasymbeta}{\isacharprime}{\kern0pt}\ {\isasymRightarrow}\ safe{\isacharunderscore}{\kern0pt}formula\ {\isasymbeta}{\isacharprime}{\kern0pt}\ \isanewline
\ \ \ \ \ \ \ \ \ \ \ \ {\isacharbar}{\kern0pt}\ {\isasymalpha}{\isacharprime}{\kern0pt}\ \isactrlbold T\ I\ {\isasymbeta}{\isacharprime}{\kern0pt}\ {\isasymRightarrow}\ safe{\isacharunderscore}{\kern0pt}dual\ True\ safe{\isacharunderscore}{\kern0pt}formula\ {\isasymalpha}{\isacharprime}{\kern0pt}\ I\ {\isasymbeta}{\isacharprime}{\kern0pt}\isanewline
\ \ \ \ \ \ \ \ \ \ \ \ {\isacharbar}{\kern0pt}\ {\isasymalpha}{\isacharprime}{\kern0pt}\ \isactrlbold R\ I\ {\isasymbeta}{\isacharprime}{\kern0pt}\ {\isasymRightarrow}\ safe{\isacharunderscore}{\kern0pt}dual\ True\ safe{\isacharunderscore}{\kern0pt}formula\ {\isasymalpha}{\isacharprime}{\kern0pt}\ I\ {\isasymbeta}{\isacharprime}{\kern0pt}\isanewline
\ \ \ \ \ \ \ \ \ \ \ \ {\isacharbar}{\kern0pt}\ {\isacharunderscore}{\kern0pt}\ {\isasymRightarrow}\ False{\isacharparenright}{\kern0pt}{\isacharparenright}{\kern0pt}{\isacharparenright}{\kern0pt}{\isacharparenright}{\kern0pt}{\isacharparenright}{\kern0pt}{\isachardoublequoteclose}\isanewline
\ \ {\isacharbar}{\kern0pt}\ {\isachardoublequoteopen}safe{\isacharunderscore}{\kern0pt}formula\ {\isacharparenleft}{\kern0pt}{\isasymexists}\isactrlsub F\ {\isasymalpha}{\isacharparenright}{\kern0pt}\ {\isacharequal}{\kern0pt}\ {\isacharparenleft}{\kern0pt}safe{\isacharunderscore}{\kern0pt}formula\ {\isasymalpha}{\isacharparenright}{\kern0pt}{\isachardoublequoteclose}\isanewline
\ \ {\isacharbar}{\kern0pt}\ {\isachardoublequoteopen}safe{\isacharunderscore}{\kern0pt}formula\ {\isacharparenleft}{\kern0pt}\isactrlbold Y\ I\ {\isasymalpha}{\isacharparenright}{\kern0pt}\ {\isacharequal}{\kern0pt}\ {\isacharparenleft}{\kern0pt}safe{\isacharunderscore}{\kern0pt}formula\ {\isasymalpha}{\isacharparenright}{\kern0pt}{\isachardoublequoteclose}\isanewline
\ \ {\isacharbar}{\kern0pt}\ {\isachardoublequoteopen}safe{\isacharunderscore}{\kern0pt}formula\ {\isacharparenleft}{\kern0pt}\isactrlbold X\ I\ {\isasymalpha}{\isacharparenright}{\kern0pt}\ {\isacharequal}{\kern0pt}\ {\isacharparenleft}{\kern0pt}safe{\isacharunderscore}{\kern0pt}formula\ {\isasymalpha}{\isacharparenright}{\kern0pt}{\isachardoublequoteclose}\isanewline
\ \ {\isacharbar}{\kern0pt}\ {\isachardoublequoteopen}safe{\isacharunderscore}{\kern0pt}formula\ {\isacharparenleft}{\kern0pt}{\isasymalpha}\ \isactrlbold S\ I\ {\isasymbeta}{\isacharparenright}{\kern0pt}\ {\isacharequal}{\kern0pt}\ {\isacharparenleft}{\kern0pt}safe{\isacharunderscore}{\kern0pt}formula\ {\isasymbeta}\ {\isasymand}\ fv\ {\isasymalpha}\ {\isasymsubseteq}\ fv\ {\isasymbeta}\ {\isasymand}\isanewline
\ \ \ \ {\isacharparenleft}{\kern0pt}safe{\isacharunderscore}{\kern0pt}formula\ {\isasymalpha}\ {\isasymor}\ {\isacharparenleft}{\kern0pt}case\ {\isasymalpha}\ of\ {\isasymnot}\isactrlsub F\ {\isasymalpha}{\isacharprime}{\kern0pt}\ {\isasymRightarrow}\ safe{\isacharunderscore}{\kern0pt}formula\ {\isasymalpha}{\isacharprime}{\kern0pt}\ {\isacharbar}{\kern0pt}\ {\isacharunderscore}{\kern0pt}\ {\isasymRightarrow}\ False{\isacharparenright}{\kern0pt}{\isacharparenright}{\kern0pt}{\isacharparenright}{\kern0pt}{\isachardoublequoteclose}\isanewline
\ \ {\isacharbar}{\kern0pt}\ {\isachardoublequoteopen}safe{\isacharunderscore}{\kern0pt}formula\ {\isacharparenleft}{\kern0pt}{\isasymalpha}\ \isactrlbold U\ I\ {\isasymbeta}{\isacharparenright}{\kern0pt}\ {\isacharequal}{\kern0pt}\ {\isacharparenleft}{\kern0pt}safe{\isacharunderscore}{\kern0pt}formula\ {\isasymbeta}\ {\isasymand}\ fv\ {\isasymalpha}\ {\isasymsubseteq}\ fv\ {\isasymbeta}\ {\isasymand}\isanewline
\ \ \ \ {\isacharparenleft}{\kern0pt}safe{\isacharunderscore}{\kern0pt}formula\ {\isasymalpha}\ {\isasymor}\ {\isacharparenleft}{\kern0pt}case\ {\isasymalpha}\ of\ {\isasymnot}\isactrlsub F\ {\isasymalpha}{\isacharprime}{\kern0pt}\ {\isasymRightarrow}\ safe{\isacharunderscore}{\kern0pt}formula\ {\isasymalpha}{\isacharprime}{\kern0pt}\ {\isacharbar}{\kern0pt}\ {\isacharunderscore}{\kern0pt}\ {\isasymRightarrow}\ False{\isacharparenright}{\kern0pt}{\isacharparenright}{\kern0pt}{\isacharparenright}{\kern0pt}{\isachardoublequoteclose}\isanewline
\ \ {\isacharbar}{\kern0pt}\ {\isachardoublequoteopen}safe{\isacharunderscore}{\kern0pt}formula\ {\isacharparenleft}{\kern0pt}{\isasymalpha}\ \isactrlbold T\ I\ {\isasymbeta}{\isacharparenright}{\kern0pt}\ {\isacharequal}{\kern0pt}\ safe{\isacharunderscore}{\kern0pt}dual\ False\ safe{\isacharunderscore}{\kern0pt}formula\ {\isasymalpha}\ I\ {\isasymbeta}{\isachardoublequoteclose}\isanewline
\ \ {\isacharbar}{\kern0pt}\ {\isachardoublequoteopen}safe{\isacharunderscore}{\kern0pt}formula\ {\isacharparenleft}{\kern0pt}{\isasymalpha}\ \isactrlbold R\ I\ {\isasymbeta}{\isacharparenright}{\kern0pt}\ {\isacharequal}{\kern0pt}\ safe{\isacharunderscore}{\kern0pt}dual\ False\ safe{\isacharunderscore}{\kern0pt}formula\ {\isasymalpha}\ I\ {\isasymbeta}{\isachardoublequoteclose}
\end{isabellebody}

\paragraph{Auxiliary states for trigger and release.} The predicate 
\isa{wf{\isacharunderscore}{\kern0pt}past{\isacharunderscore}{\kern0pt}aux} 
below is the first part described in \S\ref{sec-correct} of the invariant for \isa{wf{\isacharunderscore}{\kern0pt}trigger{\isacharunderscore}{\kern0pt}aux}. We also abuse notation here and 
use the predicate $Q^{i-1}_{t}$ from \S\ref{sec-correct} instead of the 
names \isa{mem0{\isacharunderscore}{\kern0pt}taux{\isacharunderscore}{\kern0pt}sat} 
and \isa{nmem0{\isacharunderscore}{\kern0pt}taux{\isacharunderscore}{\kern0pt}sat} 
in the formalisation.\hfill\isalink{https://bitbucket.org/jshs/monpoly/src/b4b63034eca0ccd5783085dececddb6c47cf6f52/thys/Relax_Safety/MFOTL/MFOTL_Correctness.thy\#lines-218}

\begin{isabellebody}\scriptsize\isanewline
\isacommand{definition}\isamarkupfalse%
\ {\isachardoublequoteopen}wf{\isacharunderscore}{\kern0pt}past{\isacharunderscore}{\kern0pt}aux\ {\isasymsigma}\ I\ i\ aux\ {\isasymlongleftrightarrow}\ {\isacharparenleft}{\kern0pt}sorted{\isacharunderscore}{\kern0pt}wrt\ {\isacharparenleft}{\kern0pt}{\isasymlambda}x\ y{\isachardot}{\kern0pt}\ fst\ x\ {\isachargreater}{\kern0pt}\ fst\ y{\isacharparenright}{\kern0pt}\ aux{\isacharparenright}{\kern0pt}\isanewline
\ \ {\isasymand}\ {\isacharparenleft}{\kern0pt}{\isasymforall}t\ R{\isachardot}{\kern0pt}\ {\isacharparenleft}{\kern0pt}t{\isacharcomma}{\kern0pt}\ R{\isacharparenright}{\kern0pt}\ {\isasymin}\ set\ aux\ {\isasymlongrightarrow}\ i\ {\isasymnoteq}\ {\isadigit{0}}\ {\isasymand}\ t\ {\isasymle}\ {\isasymtau}\ {\isasymsigma}\ {\isacharparenleft}{\kern0pt}i{\isacharminus}{\kern0pt}{\isadigit{1}}{\isacharparenright}{\kern0pt}\ {\isasymand}\ memR\ I\ {\isacharparenleft}{\kern0pt}{\isasymtau}\ {\isasymsigma}\ {\isacharparenleft}{\kern0pt}i{\isacharminus}{\kern0pt}{\isadigit{1}}{\isacharparenright}{\kern0pt}\ {\isacharminus}{\kern0pt}\ t{\isacharparenright}{\kern0pt}\isanewline 
\ \ \ \ {\isasymand}\ {\isacharparenleft}{\kern0pt}{\isasymexists}j{\isachardot}{\kern0pt}\ t\ {\isacharequal}{\kern0pt}\ {\isasymtau}\ {\isasymsigma}\ j{\isacharparenright}{\kern0pt}{\isacharparenright}{\kern0pt}\isanewline
\ \ {\isasymand}\ {\isacharparenleft}{\kern0pt}{\isasymforall}t{\isachardot}{\kern0pt}\ i\ {\isasymnoteq}\ {\isadigit{0}}\ {\isasymand}\ t\ {\isasymle}\ {\isasymtau}\ {\isasymsigma}\ {\isacharparenleft}{\kern0pt}i{\isacharminus}{\kern0pt}{\isadigit{1}}{\isacharparenright}{\kern0pt}\ {\isasymand}\ memR\ I\ {\isacharparenleft}{\kern0pt}{\isasymtau}\ {\isasymsigma}\ {\isacharparenleft}{\kern0pt}i{\isacharminus}{\kern0pt}{\isadigit{1}}{\isacharparenright}{\kern0pt}\ {\isacharminus}{\kern0pt}\ t{\isacharparenright}{\kern0pt}\ {\isasymand}\ {\isacharparenleft}{\kern0pt}{\isasymexists}j{\isachardot}{\kern0pt}\ {\isasymtau}\ {\isasymsigma}\ j\ {\isacharequal}{\kern0pt}\ t{\isacharparenright}{\kern0pt}\ \isanewline
\ \ \ \ {\isasymlongrightarrow}\ {\isacharparenleft}{\kern0pt}{\isasymexists}X{\isachardot}{\kern0pt}\ {\isacharparenleft}{\kern0pt}t{\isacharcomma}{\kern0pt}\ X{\isacharparenright}{\kern0pt}\ {\isasymin}\ set\ aux{\isacharparenright}{\kern0pt}{\isacharparenright}{\kern0pt}{\isachardoublequoteclose}\isanewline

\isacommand{definition}\isamarkupfalse%
\ {\isachardoublequoteopen}wf{\isacharunderscore}{\kern0pt}trigger{\isacharunderscore}{\kern0pt}aux\ {\isasymsigma}\ n\ U\ pos\ {\isasymalpha}\ mem{\isadigit{0}}\ I\ {\isasymbeta}\ aux\ i\ {\isasymlongleftrightarrow}\ {\isacharparenleft}{\kern0pt}wf{\isacharunderscore}{\kern0pt}past{\isacharunderscore}{\kern0pt}aux\ {\isasymsigma}\ I\ i\ aux\isanewline
\ \ {\isasymand}\ {\isacharparenleft}{\kern0pt}{\isasymforall}t\ R{\isachardot}{\kern0pt}\ {\isacharparenleft}{\kern0pt}t{\isacharcomma}{\kern0pt}\ R{\isacharparenright}{\kern0pt}\ {\isasymin}\ set\ aux\ {\isasymlongrightarrow}\isanewline
\ \ \ {\isacharparenleft}{\kern0pt}mem{\isadigit{0}}\ {\isasymlongrightarrow}\ qtable\ n\ {\isacharparenleft}{\kern0pt}FV\ {\isasymbeta}{\isacharparenright}{\kern0pt}\ {\isacharparenleft}{\kern0pt}mem{\isacharunderscore}{\kern0pt}restr\ U{\isacharparenright}{\kern0pt}\ {\isacharparenleft}{\kern0pt}$Q^{i-1}_{t}${\isacharparenright}{\kern0pt}\ R{\isacharparenright}{\kern0pt}\isanewline
\ \ \ {\isasymand}\ {\isacharparenleft}{\kern0pt}{\isasymnot}\ mem{\isadigit{0}}\ {\isasymlongrightarrow}\ qtable\ n\ {\isacharparenleft}{\kern0pt}FV\ {\isasymbeta}{\isacharparenright}{\kern0pt}\ {\isacharparenleft}{\kern0pt}mem{\isacharunderscore}{\kern0pt}restr\ U{\isacharparenright}{\kern0pt}\ {\isacharparenleft}{\kern0pt}$Q^{i-1}_{t}${\isacharparenright}{\kern0pt}\ R{\isacharparenright}{\kern0pt}{\isacharparenright}{\kern0pt}{\isacharparenright}{\kern0pt}{\isachardoublequoteclose}\isanewline
\end{isabellebody}

The notation \isa{list{\isacharunderscore}{\kern0pt}all{\isadigit{2}}} below 
indicates universal pairwise quantification over its two list-arguments 
\isa{aux} and \isa{{\isacharbrackleft}{\kern0pt}ne{\isachardot}{\kern0pt}{\isachardot}{\kern0pt}{\isacharless}{\kern0pt}ne{\isacharplus}{\kern0pt}length\ aux{\isacharbrackright}}. 
The first argument is the auxiliary state, while the notation $[a..{<}b]$
represents the list of all natural numbers greater or equal than $a$ and 
less than $b$. As before, we use notation $Q^{i, ne+\length aux}_{\leftL,0\in I}$
and $Q^{i, ne+\length aux}_{\rightR,0\notin I}$ from \S\ref{sec-correct} 
instead of that in the formalisation.\hfill\isalink{https://bitbucket.org/jshs/monpoly/src/b4b63034eca0ccd5783085dececddb6c47cf6f52/thys/Relax_Safety/MFOTL/MFOTL_Correctness.thy\#lines-258}

\begin{isabellebody}\scriptsize\isanewline\isanewline\isanewline
\isacommand{definition}\ {\isachardoublequoteopen}wf{\isacharunderscore}{\kern0pt}release{\isacharunderscore}{\kern0pt}aux\ {\isasymsigma}\ n\ U\ pos\ {\isasymalpha}\ mem{\isadigit{0}}\ I\ {\isasymbeta}\ aux\ ne\ {\isasymlongleftrightarrow}\ \isanewline
\ \ \ \ {\isacharparenleft}{\kern0pt}if\ mem{\isadigit{0}}\ then\isanewline
\ \ \ \ \ \ {\isacharparenleft}{\kern0pt}list{\isacharunderscore}{\kern0pt}all{\isadigit{2}}\ {\isacharparenleft}{\kern0pt}{\isasymlambda}x\ i{\isachardot}{\kern0pt}\ case\ x\ of\ {\isacharparenleft}{\kern0pt}t{\isacharcomma}{\kern0pt}\ r{\isadigit{1}}{\isacharcomma}{\kern0pt}\ r{\isadigit{2}}{\isacharparenright}{\kern0pt}\ {\isasymRightarrow}\ t\ {\isacharequal}{\kern0pt}\ {\isasymtau}\ {\isasymsigma}\ i\ {\isasymand}\ {\isacharparenleft}{\kern0pt}\isanewline
\ \ \ \ \ \ \ \ \ \ qtable\ n\ {\isacharparenleft}{\kern0pt}FV\ {\isasymbeta}{\isacharparenright}{\kern0pt}\ {\isacharparenleft}{\kern0pt}mem{\isacharunderscore}{\kern0pt}restr\ U{\isacharparenright}{\kern0pt}\ {\isacharparenleft}{\kern0pt}$Q^{i, ne+\length aux}_{\leftL,0\in I}${\isacharparenright}{\kern0pt}\ r{\isadigit{1}}\ \isanewline
\ \ \ \ \ \ \ \ \ \ {\isasymand}\ qtable\ n\ {\isacharparenleft}{\kern0pt}FV\ {\isasymbeta}{\isacharparenright}{\kern0pt}\ {\isacharparenleft}{\kern0pt}mem{\isacharunderscore}{\kern0pt}restr\ U{\isacharparenright}{\kern0pt}\ {\isacharparenleft}{\kern0pt}$Q^{i, ne+\length aux}_{\rightR,0\in I}${\isacharparenright}{\kern0pt}\ r{\isadigit{2}}{\isacharparenright}{\kern0pt}{\isacharparenright}{\kern0pt}{\isacharparenright}{\kern0pt}\isanewline
\ \ \ \ \ \ aux\ {\isacharbrackleft}{\kern0pt}ne{\isachardot}{\kern0pt}{\isachardot}{\kern0pt}{\isacharless}{\kern0pt}ne{\isacharplus}{\kern0pt}length\ aux{\isacharbrackright}{\kern0pt}\isanewline
\ \ \ \ else\ \isanewline
\ \ \ \ \ \ list{\isacharunderscore}{\kern0pt}all{\isadigit{2}}\ {\isacharparenleft}{\kern0pt}{\isasymlambda}x\ i{\isachardot}{\kern0pt}\ case\ x\ of\ {\isacharparenleft}{\kern0pt}t{\isacharcomma}{\kern0pt}\ r{\isadigit{1}}{\isacharcomma}{\kern0pt}\ r{\isadigit{2}}{\isacharparenright}{\kern0pt}\ {\isasymRightarrow}\ t\ {\isacharequal}{\kern0pt}\ {\isasymtau}\ {\isasymsigma}\ i\ \isanewline
\ \ \ \ \ \ \ \ {\isasymand}\ qtable\ n\ {\isacharparenleft}{\kern0pt}FV\ {\isasymbeta}{\isacharparenright}{\kern0pt}\ {\isacharparenleft}{\kern0pt}mem{\isacharunderscore}{\kern0pt}restr\ U{\isacharparenright}{\kern0pt}\ {\isacharparenleft}{\kern0pt}$Q^{i, ne+\length aux}_{\leftL,0\notin I}${\isacharparenright}{\kern0pt}\ r{\isadigit{1}}\isanewline
\ \ \ \ \ \ \ \ {\isasymand}\ {\isacharparenleft}{\kern0pt}if\ {\isasymnot}\ memL\ I\ {\isacharparenleft}{\kern0pt}{\isasymtau}\ {\isasymsigma}\ {\isacharparenleft}{\kern0pt}ne\ {\isacharplus}{\kern0pt}\ length\ aux\ {\isacharminus}{\kern0pt}\ {\isadigit{1}}{\isacharparenright}{\kern0pt}\ {\isacharminus}{\kern0pt}\ {\isasymtau}\ {\isasymsigma}\ i{\isacharparenright}{\kern0pt}\ then\isanewline
\ \ \ \ \ \ \ \ \ \ \ \ \ \ r{\isadigit{2}}\ {\isacharequal}{\kern0pt}\ unit{\isacharunderscore}{\kern0pt}table\ n\isanewline
\ \ \ \ \ \ \ \ \ \ else\ if\ memR\ I\ {\isacharparenleft}{\kern0pt}{\isasymtau}\ {\isasymsigma}\ {\isacharparenleft}{\kern0pt}ne\ {\isacharplus}{\kern0pt}\ length\ aux\ {\isacharminus}{\kern0pt}\ {\isadigit{1}}{\isacharparenright}{\kern0pt}\ {\isacharminus}{\kern0pt}\ {\isasymtau}\ {\isasymsigma}\ i{\isacharparenright}{\kern0pt}\ then\isanewline
\ \ \ \ \ \ \ \ \ \ \ \ qtable\ n\ {\isacharparenleft}{\kern0pt}FV\ {\isasymbeta}{\isacharparenright}{\kern0pt}\ {\isacharparenleft}{\kern0pt}mem{\isacharunderscore}{\kern0pt}restr\ U{\isacharparenright}{\kern0pt}\ {\isacharparenleft}{\kern0pt}$Q^{i, ne+\length aux}_{\rightR,0\notin I}${\isacharparenright}{\kern0pt}\ r{\isadigit{2}}\isanewline
\ \ \ \ \ \ \ \ \ \ else\ \isanewline
\ \ \ \ \ \ \ \ \ \ \ \ qtable\ n\ {\isacharparenleft}{\kern0pt}if\ {\isasymforall}j{\isasymin}{\isacharbraceleft}{\kern0pt}i{\isachardot}{\kern0pt}{\isachardot}{\kern0pt}{\isacharless}{\kern0pt}ne{\isacharplus}{\kern0pt}length\ aux{\isacharbraceright}{\kern0pt}{\isachardot}{\kern0pt}\ {\isasymnot}\ mem\ I\ {\isacharparenleft}{\kern0pt}{\isasymtau}\ {\isasymsigma}\ j\ {\isacharminus}{\kern0pt}\ {\isasymtau}\ {\isasymsigma}\ i{\isacharparenright}{\kern0pt}\ then\ {\isacharbraceleft}{\kern0pt}{\isacharbraceright}{\kern0pt}\ else\ FV\ {\isasymbeta}{\isacharparenright}{\kern0pt}\ \isanewline
\ \ \ \ \ \ \ \ \ \ \ \ \ \ {\isacharparenleft}{\kern0pt}mem{\isacharunderscore}{\kern0pt}restr\ U{\isacharparenright}{\kern0pt}\ \isanewline
\ \ \ \ \ \ \ \ \ \ \ \ \ \ {\isacharparenleft}{\kern0pt}$Q^{i, ne+\length aux}_{\rightR,0\notin I}${\isacharparenright}{\kern0pt}\ \isanewline
\ \ \ \ \ \ \ \ \ \ \ \ r{\isadigit{2}}{\isacharparenright}{\kern0pt}\isanewline
\ \ \ \ \ \ {\isacharparenright}{\kern0pt}\ aux\ {\isacharbrackleft}{\kern0pt}ne{\isachardot}{\kern0pt}{\isachardot}{\kern0pt}{\isacharless}{\kern0pt}ne{\isacharplus}{\kern0pt}length\ aux{\isacharbrackright}{\kern0pt}{\isacharparenright}{\kern0pt}{\isachardoublequoteclose}
\end{isabellebody}

\paragraph{Example traces.} Finally, we show the formalisation of the traces 
displayed as tables in \S\ref{sec-examples}. We also provide an abbreviated 
version of the monitor's output via Isabelle/HOL's command \isa{\isacommand{value}}
that call's its code generator~\cite{HaftmannN10}, executes the generated
code and displays the final result. The trace for the quality assessment example
is the next one.

\begin{isabellebody}\scriptsize\isanewline
\isacommand{definition}\isamarkupfalse%
\ {\isachardoublequoteopen}mbest\ {\isasymequiv}\ minit\ best{\isachardoublequoteclose}\isanewline
\isacommand{definition}\isamarkupfalse%
\ {\isachardoublequoteopen}mbest{\isadigit{0}}\ {\isasymequiv}\ mstep\ {\isacharparenleft}{\kern0pt}{\isacharbraceleft}{\kern0pt}{\isacharparenleft}$p_1${\isacharcomma}{\kern0pt}{\isacharbrackleft}{\kern0pt}{\isadigit{0}}{\isacharbrackright}{\kern0pt}{\isacharparenright}{\kern0pt}{\isacharcomma}{\kern0pt}\ {\isacharparenleft}$p_1${\isacharcomma}{\kern0pt}{\isacharbrackleft}{\kern0pt}{\isadigit{1}}{\isacharbrackright}{\kern0pt}{\isacharparenright}{\kern0pt}{\isacharcomma}{\kern0pt}\ {\isacharparenleft}$p_1${\isacharcomma}{\kern0pt}{\isacharbrackleft}{\kern0pt}{\isadigit{2}}{\isacharbrackright}{\kern0pt}{\isacharparenright}{\kern0pt}{\isacharcomma}{\kern0pt}\ {\isacharparenleft}$p_1${\isacharcomma}{\kern0pt}{\isacharbrackleft}{\kern0pt}{\isadigit{3}}{\isacharbrackright}{\kern0pt}{\isacharparenright}{\kern0pt}{\isacharbraceright}{\kern0pt}{\isacharcomma}{\kern0pt}\ {\isadigit{0}}{\isacharparenright}{\kern0pt}\ mbest{\isachardoublequoteclose}\isanewline
\isacommand{definition}\isamarkupfalse%
\ {\isachardoublequoteopen}mbest{\isadigit{1}}\ {\isasymequiv}\ mstep\ {\isacharparenleft}{\kern0pt}{\isacharbraceleft}{\kern0pt}{\isacharparenleft}$p_1${\isacharcomma}{\kern0pt}{\isacharbrackleft}{\kern0pt}{\isadigit{0}}{\isacharbrackright}{\kern0pt}{\isacharparenright}{\kern0pt}{\isacharcomma}{\kern0pt}\ {\isacharparenleft}$p_1${\isacharcomma}{\kern0pt}{\isacharbrackleft}{\kern0pt}{\isadigit{1}}{\isacharbrackright}{\kern0pt}{\isacharparenright}{\kern0pt}{\isacharcomma}{\kern0pt}\ {\isacharparenleft}$p_1${\isacharcomma}{\kern0pt}{\isacharbrackleft}{\kern0pt}{\isadigit{2}}{\isacharbrackright}{\kern0pt}{\isacharparenright}{\kern0pt}{\isacharcomma}{\kern0pt}\ {\isacharparenleft}$p_1${\isacharcomma}{\kern0pt}{\isacharbrackleft}{\kern0pt}{\isadigit{3}}{\isacharbrackright}{\kern0pt}{\isacharparenright}{\kern0pt}{\isacharbraceright}{\kern0pt}{\isacharcomma}{\kern0pt}\ {\isadigit{1}}{\isacharparenright}{\kern0pt}\ {\isacharparenleft}{\kern0pt}snd\ mbest{\isadigit{0}}{\isacharparenright}{\kern0pt}{\isachardoublequoteclose}\isanewline
\isacommand{definition}\isamarkupfalse%
\ {\isachardoublequoteopen}mbest{\isadigit{2}}\ {\isasymequiv}\ mstep\ {\isacharparenleft}{\kern0pt}{\isacharbraceleft}{\kern0pt}{\isacharparenleft}$p_2${\isacharcomma}{\kern0pt}{\isacharbrackleft}{\kern0pt}{\isadigit{0}}{\isacharbrackright}{\kern0pt}{\isacharparenright}{\kern0pt}{\isacharcomma}{\kern0pt}\ {\isacharparenleft}$p_2${\isacharcomma}{\kern0pt}{\isacharbrackleft}{\kern0pt}{\isadigit{1}}{\isacharbrackright}{\kern0pt}{\isacharparenright}{\kern0pt}{\isacharcomma}{\kern0pt}\ {\isacharparenleft}$p_1${\isacharcomma}{\kern0pt}{\isacharbrackleft}{\kern0pt}{\isadigit{2}}{\isacharbrackright}{\kern0pt}{\isacharparenright}{\kern0pt}{\isacharcomma}{\kern0pt}\ {\isacharparenleft}$p_2${\isacharcomma}{\kern0pt}{\isacharbrackleft}{\kern0pt}{\isadigit{3}}{\isacharbrackright}{\kern0pt}{\isacharparenright}{\kern0pt}{\isacharbraceright}{\kern0pt}{\isacharcomma}{\kern0pt}\ {\isadigit{2}}{\isacharparenright}{\kern0pt}\ {\isacharparenleft}{\kern0pt}snd\ mbest{\isadigit{1}}{\isacharparenright}{\kern0pt}{\isachardoublequoteclose}\isanewline
\isacommand{definition}\isamarkupfalse%
\ {\isachardoublequoteopen}mbest{\isadigit{3}}\ {\isasymequiv}\ mstep\ {\isacharparenleft}{\kern0pt}{\isacharbraceleft}{\kern0pt}{\isacharparenleft}$p_2${\isacharcomma}{\kern0pt}{\isacharbrackleft}{\kern0pt}{\isadigit{0}}{\isacharbrackright}{\kern0pt}{\isacharparenright}{\kern0pt}{\isacharcomma}{\kern0pt}\ {\isacharparenleft}$p_2${\isacharcomma}{\kern0pt}{\isacharbrackleft}{\kern0pt}{\isadigit{1}}{\isacharbrackright}{\kern0pt}{\isacharparenright}{\kern0pt}{\isacharcomma}{\kern0pt}\ {\isacharparenleft}$p_2${\isacharcomma}{\kern0pt}{\isacharbrackleft}{\kern0pt}{\isadigit{2}}{\isacharbrackright}{\kern0pt}{\isacharparenright}{\kern0pt}{\isacharcomma}{\kern0pt}\ {\isacharparenleft}$p_2${\isacharcomma}{\kern0pt}{\isacharbrackleft}{\kern0pt}{\isadigit{3}}{\isacharbrackright}{\kern0pt}{\isacharparenright}{\kern0pt}{\isacharbraceright}{\kern0pt}{\isacharcomma}{\kern0pt}\ {\isadigit{3}}{\isacharparenright}{\kern0pt}\ {\isacharparenleft}{\kern0pt}snd\ mbest{\isadigit{2}}{\isacharparenright}{\kern0pt}{\isachardoublequoteclose}\isanewline
\isacommand{definition}\isamarkupfalse%
\ {\isachardoublequoteopen}mbest{\isadigit{4}}\ {\isasymequiv}\ mstep\ {\isacharparenleft}{\kern0pt}{\isacharbraceleft}{\kern0pt}{\isacharparenleft}$p_3${\isacharcomma}{\kern0pt}{\isacharbrackleft}{\kern0pt}{\isadigit{0}}{\isacharbrackright}{\kern0pt}{\isacharparenright}{\kern0pt}{\isacharcomma}{\kern0pt}\ {\isacharparenleft}$p_2${\isacharcomma}{\kern0pt}{\isacharbrackleft}{\kern0pt}{\isadigit{1}}{\isacharbrackright}{\kern0pt}{\isacharparenright}{\kern0pt}{\isacharcomma}{\kern0pt}\ {\isacharparenleft}$p_2${\isacharcomma}{\kern0pt}{\isacharbrackleft}{\kern0pt}{\isadigit{2}}{\isacharbrackright}{\kern0pt}{\isacharparenright}{\kern0pt}{\isacharcomma}{\kern0pt}\ {\isacharparenleft}$p_3${\isacharcomma}{\kern0pt}{\isacharbrackleft}{\kern0pt}{\isadigit{3}}{\isacharbrackright}{\kern0pt}{\isacharparenright}{\kern0pt}{\isacharbraceright}{\kern0pt}{\isacharcomma}{\kern0pt}\ {\isadigit{4}}{\isacharparenright}{\kern0pt}\ {\isacharparenleft}{\kern0pt}snd\ mbest{\isadigit{3}}{\isacharparenright}{\kern0pt}{\isachardoublequoteclose}\isanewline
\isacommand{definition}\isamarkupfalse%
\ {\isachardoublequoteopen}mbest{\isadigit{5}}\ {\isasymequiv}\ mstep\ {\isacharparenleft}{\kern0pt}{\isacharbraceleft}{\kern0pt}{\isacharparenleft}$p_3${\isacharcomma}{\kern0pt}{\isacharbrackleft}{\kern0pt}{\isadigit{0}}{\isacharbrackright}{\kern0pt}{\isacharparenright}{\kern0pt}{\isacharcomma}{\kern0pt}\ {\isacharparenleft}$p_3${\isacharcomma}{\kern0pt}{\isacharbrackleft}{\kern0pt}{\isadigit{1}}{\isacharbrackright}{\kern0pt}{\isacharparenright}{\kern0pt}{\isacharcomma}{\kern0pt}\ {\isacharparenleft}$p_2${\isacharcomma}{\kern0pt}{\isacharbrackleft}{\kern0pt}{\isadigit{2}}{\isacharbrackright}{\kern0pt}{\isacharparenright}{\kern0pt}{\isacharcomma}{\kern0pt}\ {\isacharparenleft}$p_3${\isacharcomma}{\kern0pt}{\isacharbrackleft}{\kern0pt}{\isadigit{3}}{\isacharbrackright}{\kern0pt}{\isacharparenright}{\kern0pt}{\isacharbraceright}{\kern0pt}{\isacharcomma}{\kern0pt}\ {\isadigit{5}}{\isacharparenright}{\kern0pt}\ {\isacharparenleft}{\kern0pt}snd\ mbest{\isadigit{4}}{\isacharparenright}{\kern0pt}{\isachardoublequoteclose}\isanewline
\isacommand{definition}\isamarkupfalse%
\ {\isachardoublequoteopen}mbest{\isadigit{6}}\ {\isasymequiv}\ mstep\ {\isacharparenleft}{\kern0pt}{\isacharbraceleft}{\kern0pt}{\isacharparenleft}$p_1${\isacharcomma}{\kern0pt}{\isacharbrackleft}{\kern0pt}{\isadigit{4}}{\isacharbrackright}{\kern0pt}{\isacharparenright}{\kern0pt}{\isacharcomma}{\kern0pt}\ {\isacharparenleft}$p_3${\isacharcomma}{\kern0pt}{\isacharbrackleft}{\kern0pt}{\isadigit{1}}{\isacharbrackright}{\kern0pt}{\isacharparenright}{\kern0pt}{\isacharcomma}{\kern0pt}\ {\isacharparenleft}$p_3${\isacharcomma}{\kern0pt}{\isacharbrackleft}{\kern0pt}{\isadigit{2}}{\isacharbrackright}{\kern0pt}{\isacharparenright}{\kern0pt}{\isacharcomma}{\kern0pt}\ {\isacharparenleft}$p_1${\isacharcomma}{\kern0pt}{\isacharbrackleft}{\kern0pt}{\isadigit{5}}{\isacharbrackright}{\kern0pt}{\isacharparenright}{\kern0pt}{\isacharbraceright}{\kern0pt}{\isacharcomma}{\kern0pt}\ {\isadigit{6}}{\isacharparenright}{\kern0pt}\ {\isacharparenleft}{\kern0pt}snd\ mbest{\isadigit{5}}{\isacharparenright}{\kern0pt}{\isachardoublequoteclose}\isanewline
\end{isabellebody}

Below, we do not show the full output for the second argument in the monitor's state 
because it is long and difficult to parse. For a shorter version, see the next example. 
The monitor correctly identifies the best quality products to have IDs $0$ 
and $3$. 

\begin{isabellebody}\scriptsize\isanewline
\isacommand{value}\isamarkupfalse%
\ {\isachardoublequoteopen}mbest{\isadigit{6}}{\isachardoublequoteclose}\ %
\isamarkupcmt{\isa{{\isacharparenleft}{\kern0pt}{\isacharbraceleft}{\kern0pt}{\isacharparenleft}{\kern0pt}{\isadigit{0}}{\isacharcomma}{\kern0pt}{\isacharbrackleft}{\kern0pt}Some\ {\isadigit{0}}{\isacharbrackright}{\kern0pt}{\isacharparenright}{\kern0pt}{\isacharcomma}{\kern0pt}\ {\isacharparenleft}{\kern0pt}{\isadigit{0}}{\isacharcomma}{\kern0pt}\ {\isacharbrackleft}{\kern0pt}Some\ {\isadigit{3}}{\isacharbrackright}{\kern0pt}{\isacharparenright}{\kern0pt}{\isacharbraceright}{\kern0pt}{\isacharcomma}{\kern0pt}\isanewline
\ \ \ \ {\isasymlparr}mstate{\isacharunderscore}{\kern0pt}i\ {\isacharequal}{\kern0pt}\ {\isadigit{1}}{\isacharcomma}{\kern0pt}\ mstate{\isacharunderscore}{\kern0pt}m\ {\isacharequal}{\kern0pt}\ best$_M${\isacharcomma}{\kern0pt}\ mstate{\isacharunderscore}{\kern0pt}n\ {\isacharequal}{\kern0pt}\ {\isadigit{1}}{\isasymrparr}{\isacharparenright}{\kern0pt}}%
}\isanewline
\end{isabellebody}

The piracy trace is formalised with functions \isa{minit} and \isa{mstep} as shown below.
\begin{isabellebody}\scriptsize\isanewline
\isacommand{definition}\isamarkupfalse%
\ {\isachardoublequoteopen}mpira\ {\isasymequiv}\ minit\ pirated{\isachardoublequoteclose}\isanewline
\isacommand{definition}\isamarkupfalse%
\ {\isachardoublequoteopen}mpira{\isadigit{0}}\ {\isasymequiv}\ mstep\ {\isacharparenleft}{\kern0pt}{\isacharbraceleft}{\kern0pt}{\isacharparenleft}{\kern0pt}no{\isacharunderscore}{\kern0pt}sign{\isacharcomma}{\kern0pt}{\isacharbrackleft}{\kern0pt}{\isadigit{1}}{\isacharbrackright}{\kern0pt}{\isacharparenright}{\kern0pt}{\isacharcomma}{\kern0pt}\ {\isacharparenleft}{\kern0pt}no{\isacharunderscore}{\kern0pt}sign{\isacharcomma}{\kern0pt}{\isacharbrackleft}{\kern0pt}{\isadigit{2}}{\isacharbrackright}{\kern0pt}{\isacharparenright}{\kern0pt}{\isacharcomma}{\kern0pt}\ {\isacharparenleft}{\kern0pt}sign{\isacharcomma}{\kern0pt}{\isacharbrackleft}{\kern0pt}{\isadigit{3}}{\isacharbrackright}{\kern0pt}{\isacharparenright}{\kern0pt}{\isacharbraceright}{\kern0pt}{\isacharcomma}{\kern0pt}\ {\isadigit{0}}{\isacharparenright}{\kern0pt}\ mpira{\isachardoublequoteclose}\isanewline
\isacommand{definition}\isamarkupfalse%
\ {\isachardoublequoteopen}mpira{\isadigit{1}}\ {\isasymequiv}\ mstep\ {\isacharparenleft}{\kern0pt}{\isacharbraceleft}{\kern0pt}{\isacharparenleft}{\kern0pt}no{\isacharunderscore}{\kern0pt}sign{\isacharcomma}{\kern0pt}{\isacharbrackleft}{\kern0pt}{\isadigit{1}}{\isacharbrackright}{\kern0pt}{\isacharparenright}{\kern0pt}{\isacharcomma}{\kern0pt}\ {\isacharparenleft}{\kern0pt}no{\isacharunderscore}{\kern0pt}sign{\isacharcomma}{\kern0pt}{\isacharbrackleft}{\kern0pt}{\isadigit{2}}{\isacharbrackright}{\kern0pt}{\isacharparenright}{\kern0pt}{\isacharcomma}{\kern0pt}\ {\isacharparenleft}{\kern0pt}sign{\isacharcomma}{\kern0pt}{\isacharbrackleft}{\kern0pt}{\isadigit{3}}{\isacharbrackright}{\kern0pt}{\isacharparenright}{\kern0pt}{\isacharbraceright}{\kern0pt}{\isacharcomma}{\kern0pt}\ {\isadigit{1}}{\isacharparenright}{\kern0pt}\ {\isacharparenleft}{\kern0pt}snd\ mpira{\isadigit{0}}{\isacharparenright}{\kern0pt}{\isachardoublequoteclose}\isanewline
\isacommand{definition}\isamarkupfalse%
\ {\isachardoublequoteopen}mpira{\isadigit{2}}\ {\isasymequiv}\ mstep\ {\isacharparenleft}{\kern0pt}{\isacharbraceleft}{\kern0pt}{\isacharparenleft}{\kern0pt}no{\isacharunderscore}{\kern0pt}sign{\isacharcomma}{\kern0pt}{\isacharbrackleft}{\kern0pt}{\isadigit{1}}{\isacharbrackright}{\kern0pt}{\isacharparenright}{\kern0pt}{\isacharcomma}{\kern0pt}\ {\isacharparenleft}{\kern0pt}no{\isacharunderscore}{\kern0pt}sign{\isacharcomma}{\kern0pt}{\isacharbrackleft}{\kern0pt}{\isadigit{2}}{\isacharbrackright}{\kern0pt}{\isacharparenright}{\kern0pt}{\isacharcomma}{\kern0pt}\ {\isacharparenleft}{\kern0pt}sign{\isacharcomma}{\kern0pt}{\isacharbrackleft}{\kern0pt}{\isadigit{3}}{\isacharbrackright}{\kern0pt}{\isacharparenright}{\kern0pt}{\isacharbraceright}{\kern0pt}{\isacharcomma}{\kern0pt}\ {\isadigit{2}}{\isacharparenright}{\kern0pt}\ {\isacharparenleft}{\kern0pt}snd\ mpira{\isadigit{1}}{\isacharparenright}{\kern0pt}{\isachardoublequoteclose}\isanewline
\isacommand{definition}\isamarkupfalse%
\ {\isachardoublequoteopen}mpira{\isadigit{3}}\ {\isasymequiv}\ mstep\ {\isacharparenleft}{\kern0pt}{\isacharbraceleft}{\kern0pt}{\isacharparenleft}{\kern0pt}off{\isacharunderscore}{\kern0pt}route{\isacharcomma}{\kern0pt}{\isacharbrackleft}{\kern0pt}{\isadigit{1}}{\isacharbrackright}{\kern0pt}{\isacharparenright}{\kern0pt}{\isacharcomma}{\kern0pt}\ {\isacharparenleft}{\kern0pt}no{\isacharunderscore}{\kern0pt}sign{\isacharcomma}{\kern0pt}{\isacharbrackleft}{\kern0pt}{\isadigit{2}}{\isacharbrackright}{\kern0pt}{\isacharparenright}{\kern0pt}{\isacharcomma}{\kern0pt}\ {\isacharparenleft}{\kern0pt}sign{\isacharcomma}{\kern0pt}{\isacharbrackleft}{\kern0pt}{\isadigit{3}}{\isacharbrackright}{\kern0pt}{\isacharparenright}{\kern0pt}{\isacharbraceright}{\kern0pt}{\isacharcomma}{\kern0pt}\ {\isadigit{3}}{\isacharparenright}{\kern0pt}\ {\isacharparenleft}{\kern0pt}snd\ mpira{\isadigit{2}}{\isacharparenright}{\kern0pt}{\isachardoublequoteclose}\isanewline
\end{isabellebody}

We provide the monitor's output at time-point $3$ and show its full state. 

\begin{isabellebody}\scriptsize\isanewline
\isacommand{value}\isamarkupfalse%
\ mpiracy{\isadigit{3}}\ %
\isamarkupcmt{\isa{{\isacharparenleft}{\kern0pt}{\isacharbraceleft}{\kern0pt}{\isacharparenleft}{\kern0pt}{\isadigit{0}}{\isacharcomma}{\kern0pt}\ {\isacharbrackleft}{\kern0pt}Some\ {\isadigit{1}}{\isacharbrackright}{\kern0pt}{\isacharparenright}{\kern0pt}{\isacharcomma}{\kern0pt}\ {\isacharparenleft}{\kern0pt}{\isadigit{0}}{\isacharcomma}{\kern0pt}\ {\isacharbrackleft}{\kern0pt}Some\ {\isadigit{2}}{\isacharbrackright}{\kern0pt}{\isacharparenright}{\kern0pt}{\isacharbraceright}{\kern0pt}{\isacharcomma}{\kern0pt}\isanewline
\ \ \ \ {\isasymlparr}mstate{\isacharunderscore}{\kern0pt}i\ {\isacharequal}{\kern0pt}\ {\isadigit{1}}{\isacharcomma}{\kern0pt}\isanewline
\ \ \ \ mstate{\isacharunderscore}{\kern0pt}m\ {\isacharequal}{\kern0pt}\ MRelease\ True\ {\isacharparenleft}{\kern0pt}MPred\ {\isacharprime}{\kern0pt}{\isacharprime}{\kern0pt}off{\isacharunderscore}{\kern0pt}route{\isacharprime}{\kern0pt}{\isacharprime}{\kern0pt}\ {\isacharbrackleft}{\kern0pt}\isactrlbold v\ {\isadigit{0}}{\isacharbrackright}{\kern0pt}{\isacharparenright}{\kern0pt}\ True\ {\isacharparenleft}{\kern0pt}Abs{\isacharunderscore}{\kern0pt}{\isasymI}\ {\isacharparenleft}{\kern0pt}{\isacharunderscore}{\kern0pt}{\isacharcomma}{\kern0pt}\ {\isacharunderscore}{\kern0pt}{\isacharcomma}{\kern0pt}\ True{\isacharparenright}{\kern0pt}{\isacharparenright}{\kern0pt}\ {\isacharparenleft}{\kern0pt}MPred\ {\isacharprime}{\kern0pt}{\isacharprime}{\kern0pt}no{\isacharunderscore}{\kern0pt}signal{\isacharprime}{\kern0pt}{\isacharprime}{\kern0pt}\ {\isacharbrackleft}{\kern0pt}\isactrlbold v\ {\isadigit{0}}{\isacharbrackright}{\kern0pt}{\isacharparenright}{\kern0pt}\ {\isacharparenleft}{\kern0pt}{\isacharbrackleft}{\kern0pt}{\isacharbrackright}{\kern0pt}{\isacharcomma}{\kern0pt}\ {\isacharbrackleft}{\kern0pt}{\isacharbrackright}{\kern0pt}{\isacharparenright}{\kern0pt}\ {\isacharbrackleft}{\kern0pt}{\isacharbrackright}{\kern0pt}\ {\isacharbrackleft}{\kern0pt}{\isacharparenleft}{\kern0pt}{\isadigit{1}}{\isacharcomma}{\kern0pt}\ {\isacharbraceleft}{\kern0pt}{\isacharbraceright}{\kern0pt}{\isacharcomma}{\kern0pt}\ {\isacharbraceleft}{\kern0pt}{\isacharbrackleft}{\kern0pt}Some\ {\isadigit{2}}{\isacharbrackright}{\kern0pt}{\isacharbraceright}{\kern0pt}{\isacharparenright}{\kern0pt}{\isacharcomma}{\kern0pt}\ {\isacharparenleft}{\kern0pt}{\isadigit{2}}{\isacharcomma}{\kern0pt}\ {\isacharbraceleft}{\kern0pt}{\isacharbraceright}{\kern0pt}{\isacharcomma}{\kern0pt}\ {\isacharbraceleft}{\kern0pt}{\isacharbrackleft}{\kern0pt}Some\ {\isadigit{2}}{\isacharbrackright}{\kern0pt}{\isacharbraceright}{\kern0pt}{\isacharparenright}{\kern0pt}{\isacharcomma}{\kern0pt}\ {\isacharparenleft}{\kern0pt}{\isadigit{3}}{\isacharcomma}{\kern0pt}\ {\isacharbraceleft}{\kern0pt}{\isacharbraceright}{\kern0pt}{\isacharcomma}{\kern0pt}\ {\isacharbraceleft}{\kern0pt}{\isacharbrackleft}{\kern0pt}Some\ {\isadigit{2}}{\isacharbrackright}{\kern0pt}{\isacharbraceright}{\kern0pt}{\isacharparenright}{\kern0pt}{\isacharbrackright}{\kern0pt}{\isacharcomma}{\kern0pt}\isanewline
\ \ \ \ mstate{\isacharunderscore}{\kern0pt}n\ {\isacharequal}{\kern0pt}\ {\isadigit{1}}{\isasymrparr}{\isacharparenright}{\kern0pt}}%
}\isanewline

\end{isabellebody}

\end{document}